%% file: paper.tex
\documentclass[traditabstract,longauth]{aa}  

\usepackage{amsmath}
\usepackage{amssymb}
\usepackage{fixltx2e}
\usepackage[english]{babel}
\usepackage{graphicx}
\usepackage{epstopdf}
\usepackage{epsf,color}
\usepackage[mathscr]{eucal}
\usepackage{amsmath}
\usepackage{amssymb,amsfonts}
\usepackage{natbib}
\usepackage{graphicx}
\usepackage{txfonts}
\usepackage{dsfont}
\definecolor{Mygreen}{rgb}{0.75, 0.0, 0.0}
\definecolor{Mypink}{rgb}{1.0, 0.0, 0.5}
\definecolor{Myred}{rgb}{0.7, 0.0, 0.0}
\usepackage[breaklinks, citecolor=blue, linkcolor=Myred, urlcolor=Myred, colorlinks=true, debug, baseurl=' ']{hyperref}
\usepackage{float} 
\usepackage{color}
\usepackage{ulem}
\usepackage{hyperref}
\usepackage{mwe,tikz}
\usepackage[percent]{overpic}

\bibpunct{(}{)}{;}{a}{}{,}
\defcitealias{pierre_xxl_2015}{XXL~Paper~I}
\defcitealias{pacaud_xxl_2015}{XXL~Paper~II}
\defcitealias{giles_xxl_2015}{XXL~Paper~III}
\defcitealias{2016A&A...592A...4L}{XXL~Paper~IV}
\defcitealias{2016A&A...592A..12E}{XXL~Paper~XIII}
\defcitealias{adami_2018}{XXL~Paper~XX}
\defcitealias{guglielmo_xxl_2017}{XXL~Paper~XXII}
\defcitealias{2018A&A...620A..12C}{XXL~Paper~XXVII}
\defcitealias{Ricci_LF}{XXL~Paper~XXVIII}
\defcitealias{2018A&A...620A..20K}{XXL~Paper~XXXV}
\defcitealias{2010A&A...517A..92A}{A10}
\defcitealias{Planck2014XX}{Planck XX}
\defcitealias{2020ApJ...890..148U}{Umetsu~et~al.~2019}

\begin{document}

\title{The XXL Survey\\
XLIV. Sunyaev-Zel'dovich mapping of a low-mass cluster at $z \sim 1$:\\ 
a multi-wavelength approach
\thanks{Based on observations carried out under project number 179-17 and 094-18, with the NIKA2 camera at the IRAM 30 m Telescope. IRAM is supported by INSU/CNRS (France), MPG (Germany) and IGN (Spain).}, \thanks{Based on observations obtained with XMM-Newton, an ESA science mission with instruments and contributions directly funded by ESA Member States and NASA.}}
\input{listeauthors}

\date{Received \today \ / Accepted --}
\abstract {
{High-mass clusters at low redshifts have been intensively studied at various wavelengths. However, while more distant objects at lower masses constitute the bulk population of future surveys, their physical state remain poorly explored to date.
In this paper, we present resolved observations of the Sunyaev-Zel'dovich (SZ) effect, obtained with the NIKA2 camera, towards the cluster of galaxies XLSSC~102, a relatively low-mass system ($M_{500} \sim 2 \times 10^{14}$ M$_{\sun}$) at $z = 0.97$ detected from the XXL survey. We combine NIKA2 SZ data, \textit{XMM-Newton} X-ray data, and Megacam optical data to explore, respectively, the spatial distribution of the gas electron pressure, the gas density, and the galaxies themselves.
We find significant offsets between the X-ray peak, the SZ peak, the brightest cluster galaxy, and the peak of galaxy density. Additionally, the galaxy distribution and the gas present elongated morphologies.  This is interpreted as the sign of a recent major merging event, which induced a local boost of the gas pressure towards the north of XLSSC~102 and stripped the gas out of the galaxy group.
The NIKA2 data are also combined with XXL data to construct the thermodynamic profiles of XLSSC~102, obtaining relatively tight constraints up to about $\sim r_{500}$, and revealing properties that are typical of disturbed systems. We also explore the impact of the cluster centre definition and the implication of local pressure substructure on the recovered profiles.
Finally, we derive the global properties of XLSSC~102 and compare them to those of high-mass-and-low-redshift systems, finding no strong evidence for non-standard evolution. We also use scaling relations to obtain alternative mass estimates from our profiles. The variation between these different mass estimates reflects the difficulty to accurately measure the mass of low-mass clusters at z$\sim$1, especially with low signal-to-noise ratio (S/N) data and for a disturbed system.
However, it also highlights the strength of resolved SZ observations alone and in combination with survey-like X-ray data. This is promising for the study of high redshift clusters from the combination of \textit{eROSITA} and high resolution SZ instruments and will complement the new generation of optical surveys from facilities such as LSST and \textit{Euclid}.
}}

\titlerunning{The XXL Survey XLIV. Sunyaev-Zel'dovich mapping of a low-mass cluster at $z \sim 1$}
\authorrunning{M. Ricci, R. Adam, D. Eckert et al.}
\maketitle

\section{Introduction}
\label{sec:intro}

In our current cosmological paradigm, galaxy clusters form at the intersection of filaments in the cosmic web and trace the peaks of matter density in the Universe. They can thus be used to test cosmological models and address questions in fundamental physics, in particular by measuring their abundance as a function of mass and redshift, which is sensitive to both the expansion of the Universe and the growth history of large scale structures \citep[see e.g.][]{2006astro.ph..9591A,Allen2011,2013PhR...530...87W,2015APh....63...23H}.
The cosmological use of clusters requires the precise measurement of their masses and the control of any bias related to their formation and evolution. 

Galaxy clusters are made of dark matter, hot gas (the intra cluster medium, ICM), and galaxies that interact together. They are, therefore, key environments for the study of the co-evolution of dark and baryonic matter.
They can be studied at different wavelengths, as their galaxies principally emit in the optical and infrared (IR) and the ICM shines in X-ray and leaves an imprint in the cosmic microwave background, known as the Sunyaev-Zel'dovich (SZ) effect \citep{Sunyaev1972,Sunyaev1980}, observable at millimetre wavelengths. 
In surveys, the cluster total mass can be inferred from those observables, through the use of scaling relations. This is derived from the idea that at first order, clusters can be seen as self-similar objects, whose formation is driven by gravitation only \citep{1986MNRAS.222..323K}.

In reality, clusters are affected by complex physical processes, due to the interplay between their gas, galaxies and dark matter (merger induced shocks, active galactic nuclei (AGN) feedback, etc), which introduce bias and scatter in the scaling relations and affect cluster detection \citep[see][for a review]{2019SSRv..215...25P}. The correct modelling of cluster physics constitutes the main current limitation of cluster cosmological analyses \citep[e.g.][hereafter \textit{Planck} XX]{2016A&A...594A..24P, Planck2014XX}. 
Thus, to fully exploit the statistical power of surveys, precise individual measurements, obtained through pointed observations, are also required. Indeed, they allow for an in-depth characterisation of the astrophysical processes at place in clusters \citep[e.g.][]{Maughan2007, Eckert2017} and serve as references to test the hypotheses used in surveys or calibrate their mass scale.
Moreover, the characterisation of clusters benefits from multi-wavelength approaches. It permits a better understanding of their formation and evolution and allows for a more realistic modelling of their physics. 

So far, most of the cosmological analyses have been limited to massive clusters up to intermediate redshifts \citepalias[e.g.][]{Planck2014XX}, whereas distant clusters have higher cosmological constraining power, and low-mass clusters are by far more numerous.
Due to their shallow potential wells, low-mass clusters may be more affected by gas stripping, shock heating, or turbulence that are caused by merging events as well as AGN feedback \citep[e.g.][]{2011MNRAS.412.1965M}. Therefore, deviations from self-similar scaling relations are expected to be enhanced in this regime, in particular at high redshift, where such effects are more efficient \citep[see e.g.][]{2017MNRAS.466.4442L}. 
Understanding in depth the physical properties of low-mass, high-redshift clusters is also crucial to calibrate numerical simulations that are later used when comparing cosmological models to observations \citep[see e.g. the impact of baryonic processes on the halo mass function,][or the effect of cluster properties in AGN activity \citealt{2018A&A...620A..20K} (XXL~Paper~XXXV); \citealt{2019A&A...623L..10K}]{2016MNRAS.456.2361B}. 

Despite its importance, the low-mass and high-redshift region is still unexplored using high resolution SZ observations, due to the lack of dedicated instruments, while such data would provide unique insight into the physical properties of these objects. Indeed, provided that sufficiently deep and resolved observations are available, the thermal SZ effect (hereafter referred to as 'SZ effect') is very useful to study the dynamical state of clusters, since merging events cause over-pressure in the ICM \citep[see e.g.][]{1999ApJ...519L.115P,2001PASJ...53...57K}. Moreover, the integrated SZ flux, $Y$, directly probes the overall thermal energy of clusters, and has been shown to closely track the clusters' total masses \citepalias[see e.g.][]{Planck2014XX}, with a low intrinsic scatter. Finally, the SZ surface brightness is independent of redshift. Consequently, resolved SZ observations, combined with optical/NIR and X-ray data, offer a unique opportunity to study in depth the physics of distant clusters. 

\begin{figure}
		  \includegraphics[width=0.485\textwidth]{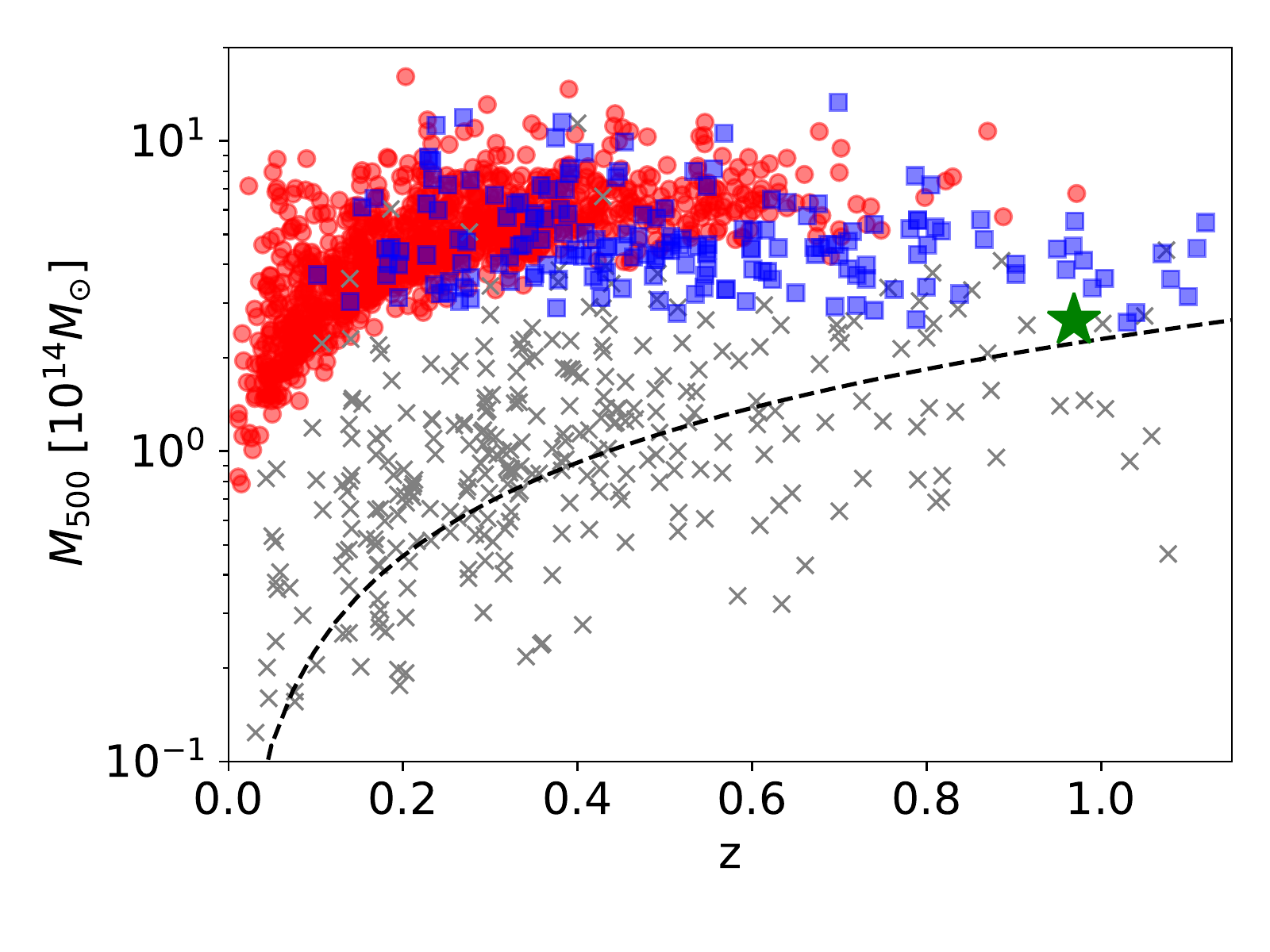}
		 \caption{\label{fig:Mz_plane} 
		 \footnotesize{
Location of XLSSC~102 (indicated by the green star) in the mass-redshift plane, and comparison to the distribution of X-ray and SZ cluster samples. The C1 and C2 XXL clusters with spectroscopic redshift measurements \citepalias{adami_2018} are shown by the grey crosses. Their masses are derived from internal scaling relations ($M_{500,scal}$). The Planck SZ sample \citep{PlanckXXVII2015} and the ACT sample \citep{2018ApJS..235...20H} are shown by the red circles and the blue squares, respectively. The dashed line indicates the expected eROSITA limiting mass corresponding to a detection limit of 50 photons in the 0.5 -- 2.0 keV band and an exposure time of 1.6 ks \citep[see][]{2016MNRAS.456.2361B}.}}
		\end{figure}

In this paper, we present the multi-wavelength analysis of the galaxy cluster XLSSC~102 at z=0.97. This object was independently detected in X-rays by the XXL survey \citep[][hereafter \citetalias{pacaud_xxl_2015}]{pacaud_xxl_2015}, via the SZ effect by the ACT survey \citep{2018ApJS..235...20H} and in the optical by, for instance, the WaZP and CAMIRA cluster finder algorithms \citep{2014becs.confE...8B,2018PASJ...70S..20O}. Recently, we observed XLSSC~102 with the NIKA2 high resolution millimetre camera \citep{adam_nika2_2018}.
The main properties of XLSSC~102 derived from previous analyses are presented in Table \ref{tab:prop_summary}. As shown in Figure \ref{fig:Mz_plane}, the mass of this cluster is estimated to be $M_{500} \sim 2 \times 10^{14} M_{\odot}$ (see Table \ref{tab:prop_summary}), and it thus resides in the high mass tail of XXL detections but in the low-mass tail of SZ samples. Our goal is to combine SZ, X-ray, and optical data in order to fully characterise the physics of this cluster and study its impact on the thermodynamic profiles and integrated quantities; in particular, those that drive clusters detection and mass estimation in surveys. The final aim is then to investigate and quantify the systematics and biases that may affect cosmological samples that extend to unexplored mass and redshift regimes, but for which the detection and the mass estimation are usually done by 'blindly' applying methods that pertain to massive, low redshift relaxed clusters.

This paper is organised as follows: in Section \ref{sec:Data} we present the NIKA2 SZ, Megacam optical/NIR, and \textit{XMM-Newton} X-ray data; in Section \ref{sec:Morphology}, we use these data to study the morphology of XLSSC~102 and assess its dynamical state; in Section \ref{sec:thermo_profiles}, we compute the thermodynamic profiles of XLSSC~102 and investigate their stability under different centring definitions and with respect to its internal structure. 
We then derive the global properties of XLSSC~102 and compare them to that of low redshift massive systems and expectations from scaling relations in Section \ref{sec:global_prop}. Finally, we give our conclusions in Section \ref{sec:Summary_and_conclusions}.

Throughout this paper, we assume a flat $\Lambda$CDM cosmology with $H_0 = 70$ km s$^{-1}$ Mpc$^{-1}$, $\Omega_{\rm M} = 0.3$, and $\Omega_{\Lambda} = 0.7$. 
We refer to overdensity scaled radii as $r_{\Delta}$, and other quantities $Q$ computed within these radii as $Q_{\Delta}$. The overdensity contrast $\Delta$ is defined with respect to the critical density of the Universe at the cluster's redshift (see Eq. \ref{eq_icmtool_rdelta}). 
At the cluster redshift, 1 arcmin corresponds to 477 kpc.

		\begin{table*}
	\begin{center}
\caption{Main properties of XLSSC~102. Columns 1 to 5 : XXL ID, coordinates, spectroscopic redshift and X-ray temperature from \citealt{adami_2018} (hereafter \citetalias{adami_2018}), column 6 : optical richness from \citealt{Ricci_LF} (hereafter \citetalias{Ricci_LF}), columns 7 to 9 : mass estimates from XXL scaling relations \citepalias[see][for col 7, 8 and 9 respectively]{adami_2018,pacaud_xxl_2015,2020ApJ...890..148U}, columns 10 and 11 : mass estimates from the ACT survey  \citep{2018ApJS..235...20H}. See section \ref{subsec:mass_compare} for details.}
	\label{tab:prop_summary}
 
	\begin{tabular}{c|cccccccc|cc}
	\hline\hline
	ID & R.A. & Dec. & $z$  & $T_{300 \ {\rm kpc}}$ & $\lambda_{0.5 {\rm Mpc}}$ & $M^{\rm XXL}_{500,{\rm scal}}$ & $M^{\rm XXL}_{500,{\rm MT\star}}$ & $M^{\rm XXL}_{500,{\rm MT\star,\star}}$ &  $M^{\rm ACT, UPP}_{500c}$& $M^{\rm ACT, cal}_{500c}$ \\

	 (---) & (deg) &  (deg) &  (---) &  (keV) & (---) &  ($10^{14}$ M$_{\odot}$) & ($10^{14}$ M$_{\odot}$) & ($10^{14}$ M$_{\odot}$)& ($10^{14}$ M$_{\odot}$) & ($10^{14}$ M$_{\odot}$)   \\

	\hline
	XLSSC~102 &  $31.322$ & $-4.652$ & $0.969$  & $3.9^{+0.8}_{-0.8}$ &  $25 \pm 8$ & $2.6 \pm 1.1$ & $1.9\pm1.1$ & $1.17^{+1.16}_{-0.60}$ &  $3.1^{+0.5}_{-0.4}$ & $4.6_{-1.0}^{+1.1} $ \\   
	\end{tabular}
	\end{center}
	\end{table*}
	

\section{Data}
\label{sec:Data}

	\begin{figure*}[h]
	\centering
	\includegraphics[trim={0cm 0cm 0cm 0cm}, clip, height=5.4cm]{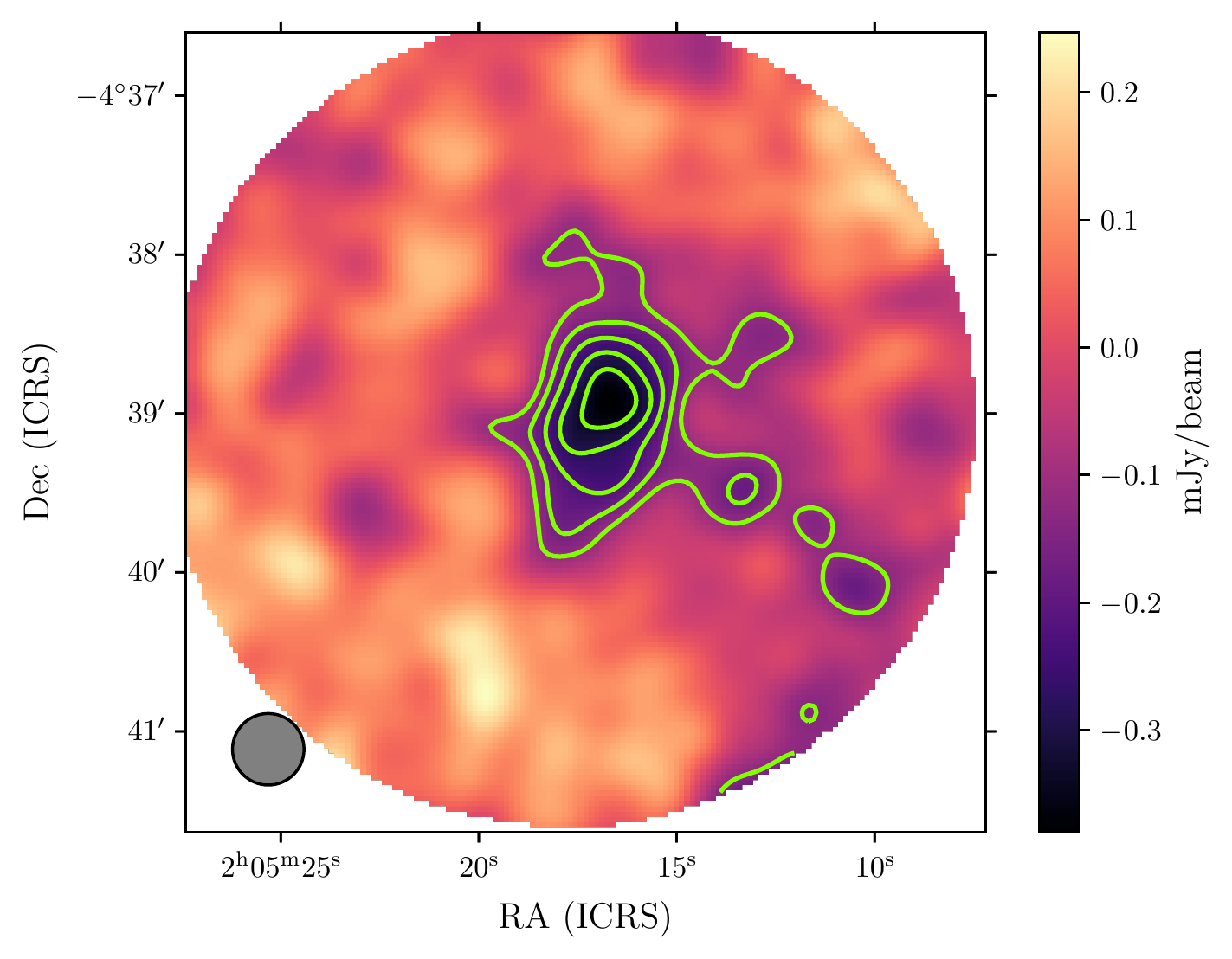}
	\includegraphics[trim={1.95cm 0cm 0cm 0cm}, clip, height=5.4cm]{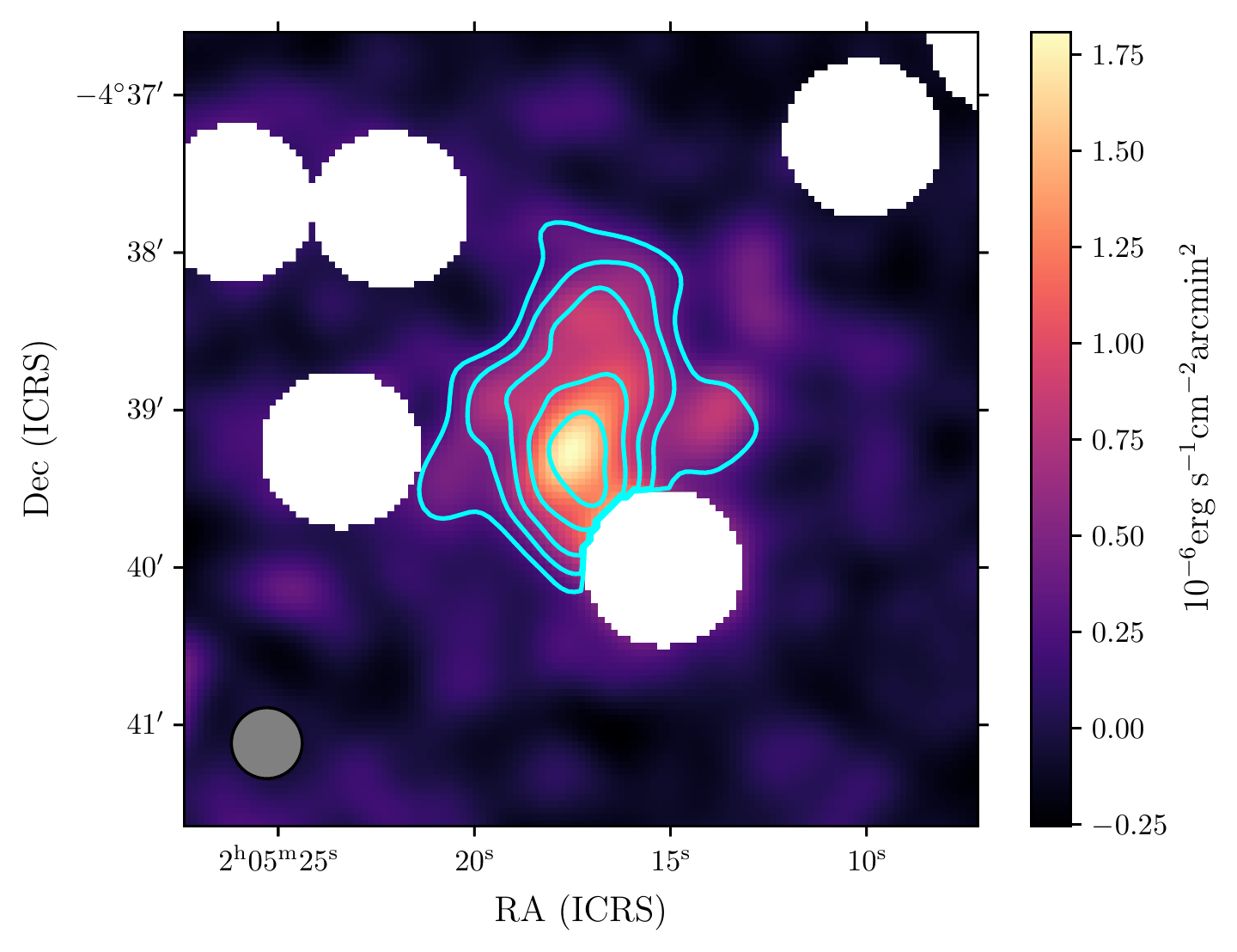}
	\includegraphics[trim={1.95cm 0cm 0cm 0cm}, clip, height=5.4cm]{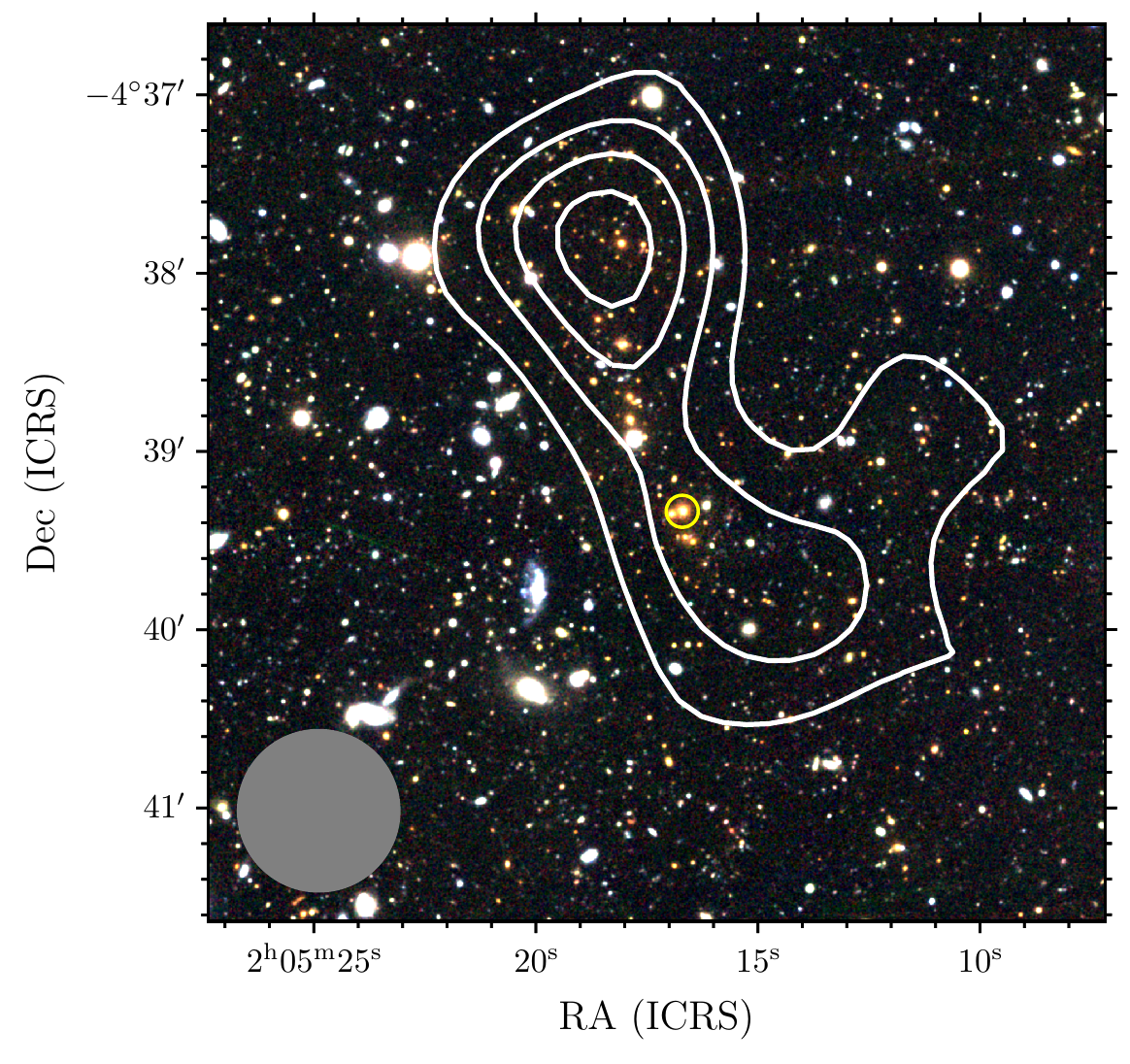}
	
	\caption{\footnotesize{Multi-wavelength view of XLSSC~102. The size of the images is 2.4$\times$2.4 Mpc$^2$ and the S/N contours start at S/N = 2 and increase by step of one (left and middle) and two (right). The maps are smoothed to an effective resolution of 27 arcsec (left, middle) and 54 arcsec (right), as indicated by the circles on the bottom left corners. The projection of the maps is given in the international celestial reference system (ICRS). Left: map in the 150 GHz NIKA2 band. The S/N contours are shown in green. We masked the regions where the noise is higher than 25\% with respect to the centre.  Middle: surface brightness map in the [0.5-2] keV \textit{XMM-Newton} band. The S/N contours are shown in cyan and the masked white regions correspond to AGN. Right: composite optical image of XLSSC~102 made of Hyper Suprime-Cam images taken in the R, I and Z filters. The galaxy density map S/B contours are over-plotted in white and the BCG is indicated by the yellow circle.}}
	\label{fig:multi_images}
	\end{figure*}

In this section we present the NIKA2, \textit{XMM-Newton} and Megacam data used for the multi-wavelength analysis of XLSSC~102.

	\subsection{NIKA2 Sunyaev-Zel'dovich data}
	\label{sec:Data_NIKA2}

The SZ data were obtained with the  dual band NIKA2 millimetre camera \citep[][]{adam_nika2_2018,2019arXiv191002038P}, installed at the IRAM 30m telescope. NIKA2 is operating with kinetic inductance detectors \citep[KIDs,][]{monfardini_dual-band_2011,Bourrion2016,calvo_nika2_2016} disposed in three arrays: two with a maximum transmission around 260 GHz (A1 and A3) and one at 150 GHz (A2). 
Its large field of view (6.5 arcmin) coupled with its high angular resolution ($17.7''$ beam at 150 GHz and $11.2''$ at 260 GHz) makes it particularly well-suited to observe the inner structure of galaxy clusters. 

The data presented in this study correspond to $\sim$ 6.6 hours (23.8 ks) of un-flagged observations, obtained during the winter pool 2018, with average weather conditions, that is, with a typical opacity of $\tau_{150\rm{GHz}}\sim 0.15$. 
The scanning strategy consisted of a series of $8\times4$ on-the fly scans, oriented in the RA-Dec coordinate system along six different directions, to minimise residual striping patterns in the maps. We defined the reference pointing coordinate as the position of the X-ray centroid measured from the XXL survey (RA : 33.322 deg.; Dec : -4.652 deg., J2000).
We verified the calibration using Uranus scans taken around the same time as our target cluster, and thus reflecting the effective performance at that time. For each scan we measured the source flux and the size of the beam for each matrix. We obtain beam FWHM of $12.1''\pm0.4''$ for A1, $12.1''\pm0.4''$ for A3 and $18.0''\pm0.3''$ for A2, absolute calibration uncertainties of 7\%, 8\% and 4\% for A1, A3 and A2 and root mean square pointing errors of $2.2''$ per scan for A1 and A3 and $2.3''$ for A2.
The performance that we measure is in agreement with that expected from the science verification run \citep[][]{adam_nika2_2018,2019arXiv191002038P}.

The time ordered data of each individual scan were processed to remove the contribution of atmospheric and electronic correlated noise and to obtain surface brightness maps following the procedure described in \citet{Catalano2014} and \citet{Adam2014,Adam2015}. The maps were latter inspected to flag and remove bad quality scans, presenting excessive noise spatial correlation. Finally, we combined the scans to obtain co-added surface brightness maps and jackknife maps. The latter were used to generate Monte Carlo realisations of the noise and compute the noise covariance matrix, following the procedure described in \citet{adam_high_2016}.
The transfer function linked to the data reduction procedure was computed by comparing the input and the output images of mock realisations of simulated data processed through the pipeline, as described in \citet{Adam2015}. This showed that the large scales, especially beyond the field of view (6.5'), are filtered out (see Figure \ref{fig:TF} in Annexe \ref{annexe}). The transfer function, interpolated using a parametric function (polynomial plus exponential cutoff), is taken into account is the rest of the analyse.

While NIKA2 provides imaging at both 260 and 150 GHz, we focus in this paper on the 150 GHz data because it is the channel in which the SZ signal is recovered with sufficient significance. The 260 GHz channel is only used to control contamination from point sources and no SZ emission is detected there, as expected given the noise amplitude and the signal strength. 
The strength of the kinetic SZ effect expected for even the most extreme objects \cite[see e.g.][]{Mroczkowski2012,Sayers2013,Adam2016b} is well below the sensitivity of our measurements. In the following we will thus assume that the SZ signal is dominated by its thermal component.

	\subsection{\textit{XMM-Newton} X-ray data}
	
The X-ray data used in this study come from the XXL survey \citep[][XXL~Paper~I]{pierre_xxl_2015}, which is an \textit{XMM-Newton} project designed to provide a well defined sample of galaxy clusters out to redshift above unity, suitable for precision cosmology, and for the analysis of galaxy evolution and active galactic nuclei \citep[see][]{pierre_precision_2011}. XLSSC~102 was detected as one of the hundred brightest clusters in the XXL survey \citepalias{pacaud_xxl_2015} and its redshift has been confirmed using member galaxy spectra \citepalias[see][for the procedure]{adami_2018}. We reduced the \textit{XMM-Newton} data with \texttt{XMMSAS} v16.1 and the pipeline developed in the framework of the \textit{XMM-Newton Cluster Outskirts Project} \citep[X-COP,][]{Eckert2017}. The procedure for data reduction is described in detail in \citet[see their Sect. 2]{Ghirardini2019}. After performing the standard data reduction steps for the three EPIC detectors, we filtered out time periods of flaring soft proton activity with the \texttt{mos-filter} and \texttt{pn-filter} tools. We extracted X-ray count maps in the [0.5-2] keV band and from the cleaned event files and used the \texttt{XMMSAS} tool \texttt{eexpmap} to extract exposure maps including vignetting effects. We used a collection of filter-wheel-closed data to model the spatial distribution of the non X-ray background, and rescaled the filter-wheel-closed dataset such that the high-energy count rates matches the count rates measured in the corners of the detectors, which are located outside of the field of view of the telescopes. The resulting count maps, exposure maps and background maps were co-added to create combined EPIC maps. We repeated the same procedure for the two XXL pointings where the source was included and combined the images to create a mosaic image of the field. For more details on the analysis procedure, we refer the reader to \citet{Ghirardini2019}.
On the first pointing, the exposure time is 10.4 ks and the cluster position is separated by an angle of 7.0 arcmin from the pointing centre. On the second pointing, the exposure time is 7.7 ks and the off-axis angle of XLSSC102 is 13.3 arcmin. After correcting for vignetting effect, this leads to an effective exposure time of 6.2 ks for the co-added map.

	\subsection{Megacam optical/NIR data}

We used data taken with the optical and near infrared wide field imager MegaCam\footnote{See \url{http://www.cfht.hawaii.edu/Instruments/Imaging/Megacam/}} in the five pass-bands: $u^*$, $g’$, $r’$, $i’$ and $z’$, from approximately $350$ to $1000$ nm, as part of the Canada-France-Hawai Telescope Legacy Survey \citep[CFHTLS,][]{2012AJ....143...38G}. The source detection and characterisation were performed upstream and we thus directly used the galaxy photometric catalogue.

XLSSC~102 falls in the W1 field of the CFHTLS, where the photometry of extended sources reaches a 80\% completeness of $24.0\pm0.1$ mag in the $r'$ band \citep{hudelot_vizier_2012}. We used the most recent version of the data release, T007, for which a catalogue of photometric redshifts, obtained with the {\sc{LePhare}} code, is also available  \citep[see][]{ilbert_accurate_2006,coupon_photometric_2009}. The set of SED templates used for their computation was constructed using elliptical, spiral (SBc and Scd) and irregular galaxy templates from \cite{coleman_colors_1980}, and two star-forming galaxy templates from \cite{kinney_template_1996}.
The statistical choice to get discrete photometric redshift values from their probability distribution function was to take the median value. Only the objects with reliable photometric redshift were included in the final catalogue. This catalogue was then cut at a magnitude of $i'=24$ to remove unreliable detections, and other criteria were applied to select the member galaxies (see Section \ref{subsec:maps} for details). 

In addition to the CFHTLS galaxy catalogue we also used for visualisation purposes the images taken with the Hyper Suprime-Cam (HSC) Subaru telescope as part of the Subaru Strategic Program \citep[SSP, ][]{2018PASJ...70S...4A} \footnote{The fits files were taken from the HSC public database, \url{http://hsc.mtk.nao.ac.jp/ssp/data-release/}}. This is motivated by the fact that the HSC-SSP images are deeper than those of the CFHTLS, but that their photometric redshifts are of comparable quality to that of the CFHTLS at z$\sim$1, due to their lack of u$^*$ band data \citep[][]{2018PASJ...70S...9T}.

The brightest cluster galaxy (BCG) of XLSSC~102 was identified by \citetalias{Ricci_LF}. It has a magnitude of $ 20.82 \pm 0.03 $ in the $i'$ band and its coordinates are RA : 31.3196 deg, Dec: -4.6556 deg.

\section{Multi-wavelength morphological analysis}
\label{sec:Morphology}
In order to characterise the physical state of XLSSC~102, at such low-mass, high redshift, and given the moderate S/N of our data, we use and compare different morphological estimators at different wavelengths.
In this Section we present our method to determine the morphology of XLSSC~102 from SZ, optical and X-ray data and estimate its associated uncertainties. We then perform a multi-wavelength comparison and discuss its interpretation in terms of cluster dynamics. 
	\subsection{Image processing} 
	\label{subsec:maps}

					\paragraph{{\bf{In SZ:}}}
Point sources, such as dusty and radio galaxies, are known to be a major contaminant for the SZ signal at millimetre wavelengths, and may affect the recovery of cluster morphologies. We thus checked the FIRST \citep{1995ApJ...450..559B} 1.4 GHz data, and we found that no radio sources are detected in our field. Assuming galaxies with standard radio spectral indices \citep[typically $\sim -0.7 \pm 0.2$, see][]{1979AJ.....84..942W}, we expect any radio source to be negligible in the NIKA2 bands around our cluster. We then used the 260 GHz map to search for dusty galaxies and we found two sources with a S/N greater than four within 2 arcmin of the cluster centre. We removed their contribution in the 150 GHz map by fitting their flux following the procedure given in \cite{adam_high_2016}. We do not expect any contamination of the cluster SZ signal as the sources are located at more than 1.2 arcmin from the cluster centre.
The 150 GHz map was then smoothed to reach an effective resolution $\theta_{\rm eff}=27$ arcsec and the signal to noise was estimated by applying the same filtering to the Monte Carlo realisations of the noise discussed in Section \ref{sec:Data} \citep[see][for the procedure]{adam_high_2016}. The resulting map and S/N contours are presented in the left panel of Figure \ref{fig:multi_images}. At this resolution, the cluster signal reaches an S/N of 6.9 at the peak. 

					\paragraph{{\bf{In X-ray:}}}
The surface brightness map of XLSSC~102 was created by dividing the background subtracted co-added [0.5-2] keV count rate map by the corresponding exposure map.
The image was then smoothed to reach the same mean effective resolution $\theta_{\rm eff}$ as that of the SZ map and the S/N was defined by propagating Poisson errors.
Several bright point sources are present in the field \citep[][hereafter \citetalias{2018A&A...620A..12C}]{2018A&A...620A..12C} and one in particular may contaminate the cluster signal. To reduce their contribution, we measured the median pixel values defined in rings of 20-25 arcsec from their centres and used these to fill the regions enclosed within a radius of 20 arcsec. After smoothing the map, the point sources were finally masked by discs of 30 arcsec radii. The resulting surface brightness image is shown in the middle panel of Figure \ref{fig:multi_images}. We can see that at this resolution we reach a S/N of 6.6 at the peak. 

					\paragraph{{\bf{In the optical:}}}
In order to analyse the distribution of galaxies we created a background corrected galaxy density map, using Gaussian filtering with FWHM = 54 arcsec (2 $\times \theta_{\rm eff}$ for SZ and X-ray images). 
For this purpose, we first selected galaxies within a photometric redshift slice centred on the spectroscopic cluster redshift, after accounting for the photometric redshift bias \citepalias[see][for the procedure]{Ricci_LF}. The width of the slice depends on the galaxy magnitude and has been defined to select 68\% of objects, based on the photometric-spectroscopic redshift distribution. 
We then subtracted the density measured in the background, taken here as the $10\times10$ Mpc$^2$ field area centred on the cluster, after we discarded the regions containing structures\footnote{Structures were identified as the regions where the galaxy density filtered at $4 \times \theta_{\rm eff}$ is above two times the standard deviation of the map.}.
We found that the cluster signal was more enhanced with respect to the background when taking only galaxies with elliptical SEDs (as determined by {\sc{LePhare}}), fainter than the BCG and brighter than $m^* + 1$ \footnote{With $m^*$ the characteristic magnitude at $z$=0.97 expected from the evolution of an elliptical galaxy that formed its stars at $z=3$.  The $m^*$ model was computed with {\scshape{LePhare}} using the elliptical galaxy SED template {\scshape{burst\_sc86\_zo.sed}} from the {\scshape{PEGASE2}} library \citep{1997A&A...326..950F}. We normalised the model using $K^*$ values from \cite{lin_evolution_2006} corrected to AB system. }, that is, with $20.821<i'< 23.78$. We kept that selection in the following.
Instead of S/N we defined here the signal-to-background ratio (S/B)  of the background corrected density map $M_{\Sigma}$ as :
	\begin{equation}S/B~(M_{\Sigma}) = \frac{M_{\Sigma} - N_{\rm bkg}/A_{\rm bkg}}{{N_{\rm bkg}}/A_{\rm bkg}},
	\end{equation}
with $N_{\rm bkg}$ the number of galaxies in the effective background area $A_{\rm bkg}$. The S/B contours of the optical density map are over-plotted on the HSC R,I,Z colour image of the XLSSC~102 field in the right panel of Figure \ref{fig:multi_images}. At the peak, the map reaches a S/B of 9.2.

					\paragraph{{\bf{Mock realisations:}}}
The multi-wavelength comparison is only meaningful if the confidence intervals of our morphological estimators are well measured. For this purpose, we simulated 1000 Monte-Carlo realisations of our three data sets:
in SZ, we modelled the cluster signal and we added noise realisations that reproduce the statistics of the noise seen in the real data (see Section \ref{sec:Data_NIKA2});
in the X-ray, we computed Poissonian realisations of idealised maps, composed of a model for the cluster emission injected in the background maps;
in the optical, we also computed Poissonian realisations of idealised maps, composed of a model for the cluster emission and a constant background. The value of the background in the optical was measured as the mean galaxy density in a $10\times10$ Mpc$^2$ field where the structures were masked. 

We tested two types of model for the cluster signal: either constructed from the real data or from the best 2D Gaussian fit of the real data. We found that the two classes of models give compatible results and error bars. In the following we use the model constructed from the data themselves.

	\begin{figure}[h]
	\centering
	\includegraphics[width=0.5\textwidth]{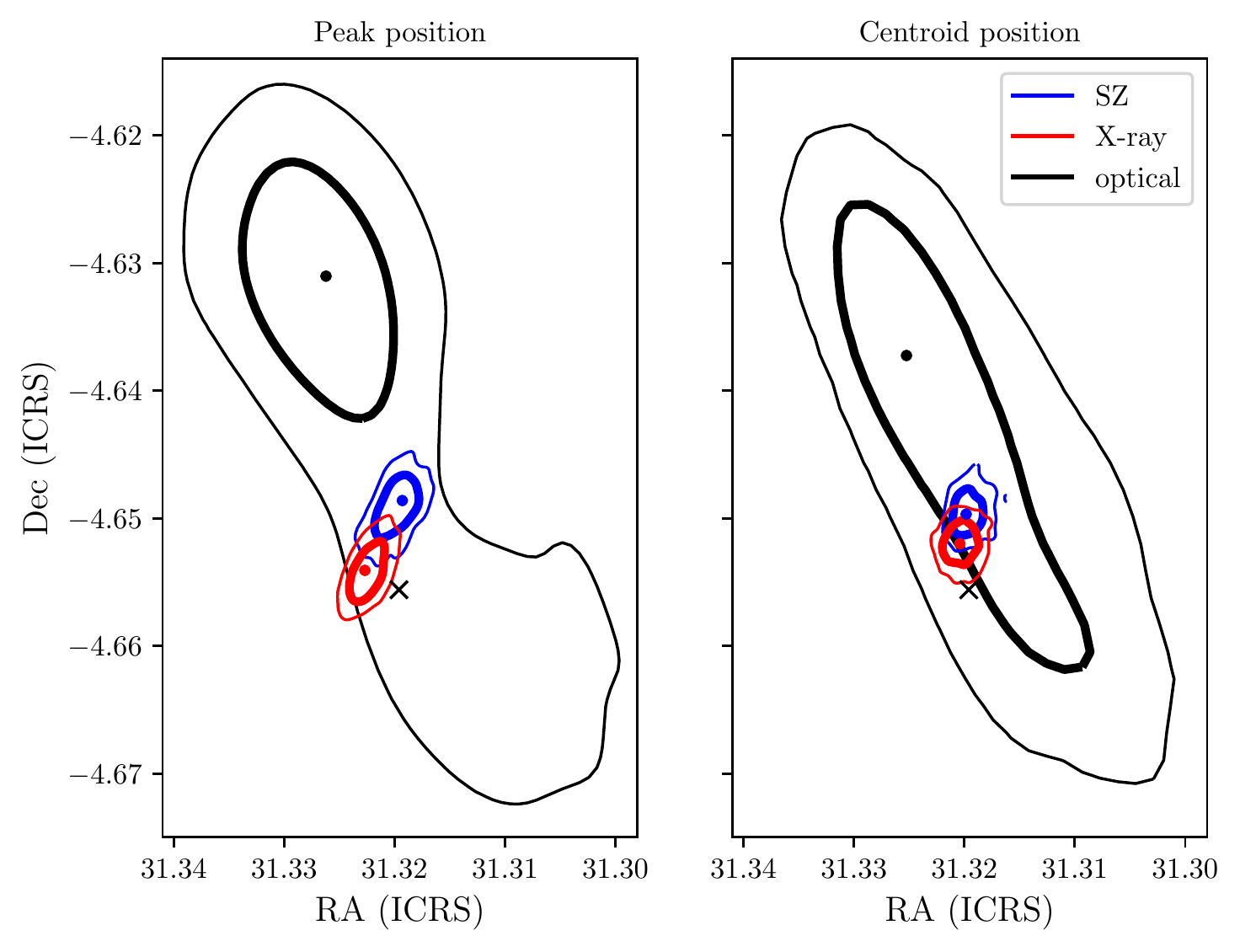}
	\caption{\footnotesize{Position of the cluster signal peaks (left) and centroids (right) in the optical density map (black), the SZ map (blue) and the X-ray map (red). The best-fit coordinates are indicated by the points and the 68\% and 95\% confidence regions are shown by the contours. The position of the BCG is shown by the black crosses.}}
	\label{fig:peak_centroid}
	\end{figure}

	\subsection{Signal peak position}

Our goal is to assess the morphology of the gas and galaxy distributions in XLSSC~102, as revealed by the optical, X-ray and SZ data. The first pertinent morphological indicator when investigating the signal at different wavelengths is to compare the peaks of emission, as their possible offsets from one another is a sign of perturbed dynamical state \citep[see e.g.][]{1999ApJ...519L.115P, 2003ApJ...585..687K, 2010A&A...513A..37H,2016MNRAS.457.4515R, 2017A&A...599A.138Z}. We also use the position of the BCG, which is expected to coincide with the bottom of the cluster potential well in relaxed clusters \citep[see e.g.][]{0004-637X-617-2-879}.

We defined the positions of the SZ, X-ray and optical peaks as the coordinates of the pixels with maximum signal on the smoothed surface brightness and galaxy density maps. The simulated maps were processed in the same way as the true data maps, and we measured the positions of the peaks on the 1000 Monte-Carlo mock realisations to evaluate the uncertainties.

The positions of the peaks in the three maps and their confidence contours are shown in the left panel of Figure \ref{fig:peak_centroid}. 
We can see that the three 95\% confidence regions are in agreement but not the 68\% ones. The BCG position is excluded from the X-ray and SZ confidence regions at more than 95\%, and from the galaxy density peak confidence region at more than 68\%.

	\begin{figure}[h]
	\centering
	\includegraphics[width=0.5\textwidth]{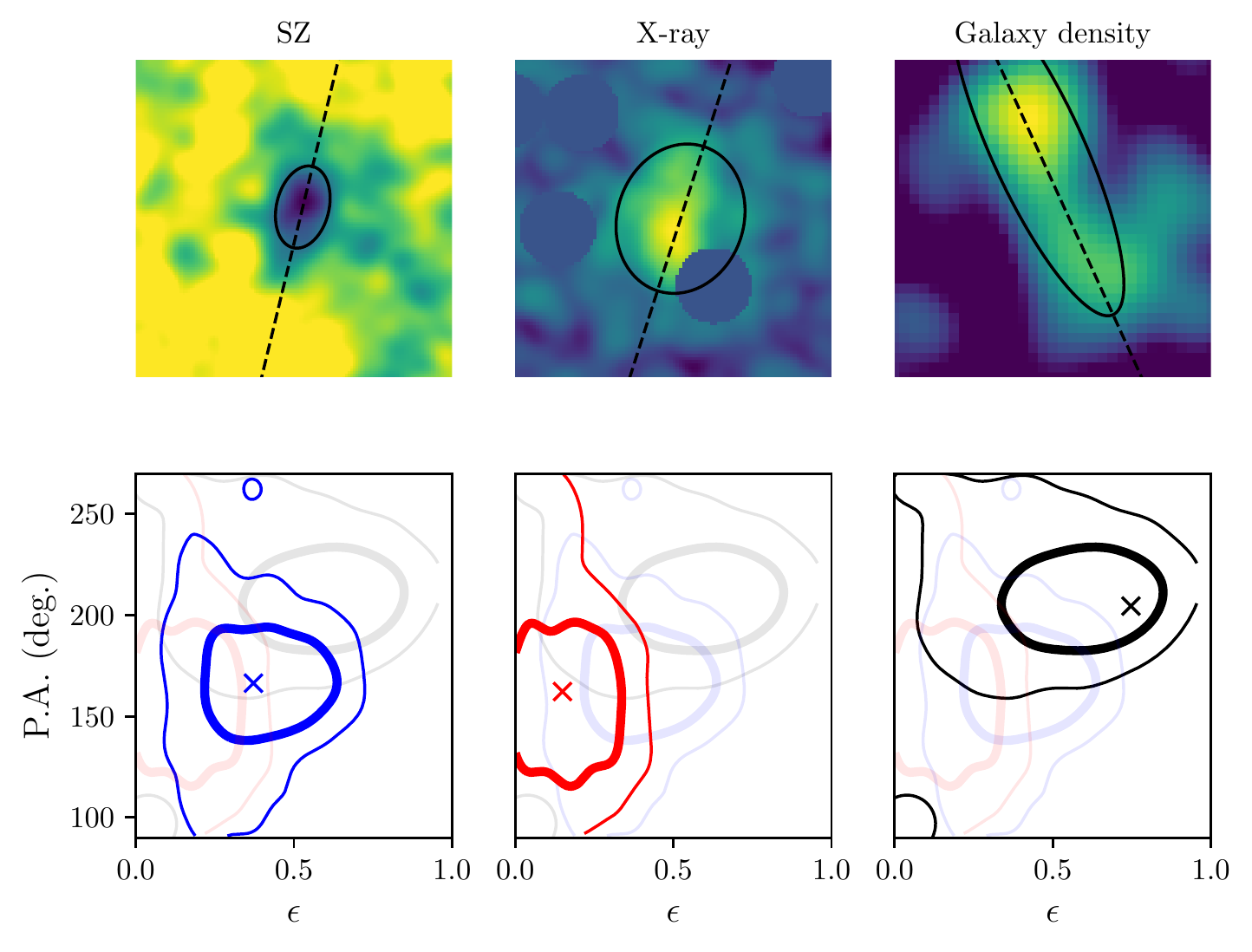}
	\caption{\footnotesize{Upper panel: best 2D Gaussian fit over-plotted on the SZ map (left), the X-ray surface brightness map (middle) and the optical over-density map (right). The position of the major axis is indicated by the dashed line. The sizes of the ellipses correspond to the 2D Gaussian FWHM.  Bottom panel: confidence contours of the position angle and ellipticities, for the SZ image (left, blue), the X-ray map (middle, red) and the galaxy density (right, black). The crosses indicate the best-fit parameters.}}
	\label{fig:ellipticity_PA}
	\end{figure}

	\subsection{Large scale morphology}

Various morphological estimators have been used in the literature to precisely assess cluster morphology from X-ray, SZ or optical data \citep[see e.g.][]{2016ApJ...819...36D}. However, those are often applied on high signal to noise data and optimised for a certain wavelength. 

Here we chose to estimate the centroid, ellipticity and position angle of the signal in the different maps by fitting 2D Gaussian of centre $(x_0, y_0)$, rotation angle $\theta$ and widths ($\sigma_{\rm min},\sigma_{\rm maj}$). 
The ellipticity is then given by: 
	\begin{equation}
	\epsilon = 1 - \frac{\sigma_{\rm min}}{\sigma_{\rm maj}}.
	\end{equation}
 Although the profiles are usually parametrised by more complex functions, we found that the 2D Gaussian fit was appropriate for a meaningful multi-wavelength comparison at our relatively low signal to noise. This choice also allows us to make use of all the signal and not to define a special region, as needed for, e.g. isophotal measurements. 

The simulated maps were processed as the true data maps and we measured the morphological estimators (ellipticity, position angle and centroid positions) in the 1000 Monte-Carlo realisations to evaluate their uncertainties.

The positions of the centroids (the centre of the fitted 2D Gaussian) and their confidence contours are shown in the right panel of Figure \ref{fig:peak_centroid}. 
We can see that the three centroid positions are compatible at the 68\% level. The peak and centroid for each wavelength are compatible at the 68\% level. The positions of the peaks and centroids are summarised in Table \ref{table:ellipticity_PA}. The BCG position is also excluded from the X-ray and SZ centroid at more than 95\% confidence, and from the galaxy density centroid at at least 68\% confidence.

The morphological measurements extracted from the 2D Gaussian fits are shown in Figure \ref{fig:ellipticity_PA} for the three wavelengths. The upper panels show the 2D Gaussian best fits over-plotted on the the SZ map (left), the X-ray surface brightness map (middle) and the galaxy density map (right). The bottom panels show the corresponding position angles and ellipticity confidence ellipsoids. The values of the ellipticities and major axis position angles are given in Table \ref{table:ellipticity_PA}. 
We can see that a null ellipticity of the galaxy density and the SZ signal is excluded at more than 95\%. For the X-ray emission, a null ellipticity is only excluded at 68\%. We can also note that the galaxy density map of XLSSC~102 is more elliptical than that of the gas and that the SZ signal is more elliptical than the X-ray emission. The position angles of the SZ and X-ray emission are almost co-linear but tilted by $\sim 40$ degrees with respect to the galaxy density axis.
As a complementary dynamical state estimator, we computed the concentration of the X-ray signal as the ratio of X-ray flux between aperture radii of 40 and 400 kpc, following the method described, for example, in \citet{2017MNRAS.468.1917R}. We find a value of $c = 0.044^{+0.019}_{-0.011}$, compatible with that of non cool core clusters, following the classification found in the literature \citep[e.g using the threshold of $c < 0.075$ as][]{2008A&A...483...35S}.

	\begin{figure}[h]
 \begin{flushright}
    \includegraphics[width=0.5\textwidth]{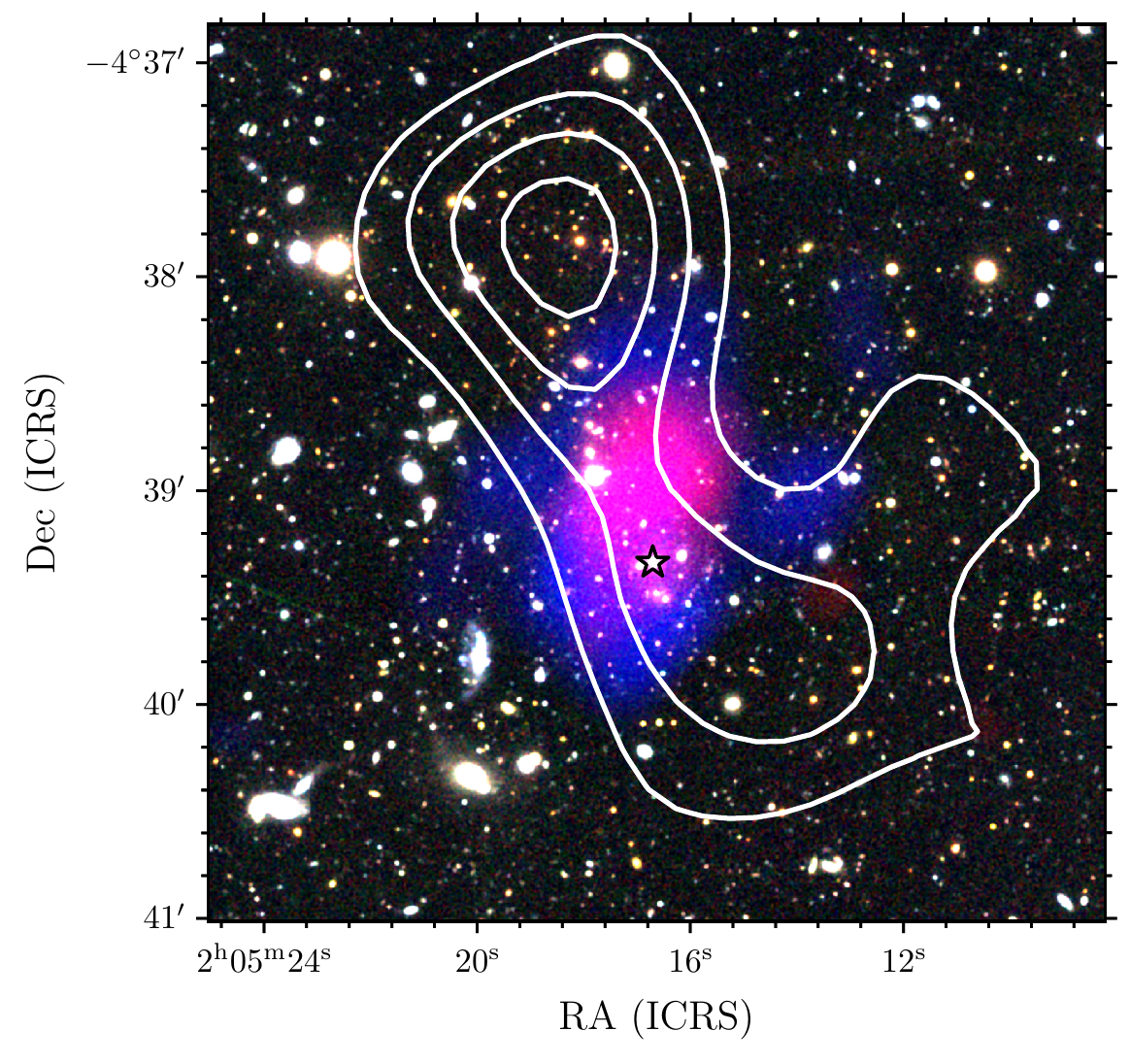}  
	\includegraphics[width=7.5cm]{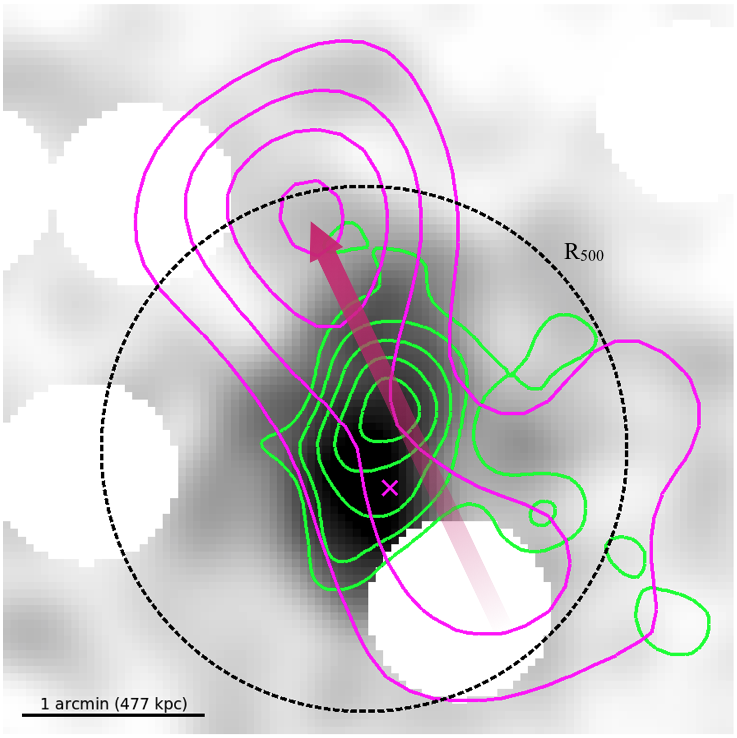}
	\end{flushright}
	\caption{\footnotesize{Multi-wavelength view of XLSSC~102. Top: composite image of XLSSC~102. The background optical image is as in the right panel of Figure \ref{fig:multi_images}, the SZ signal is shown in red and the X-ray signal is shown in blue. Bottom: The SZ S/N contours are over-plotted in green and the galaxy isodensity S/B contours are over-plotted in magenta on the X-ray surface brightness map. 
	The smoothing is the same as in Figure \ref{fig:multi_images}. The BCG position is shown by the magenta cross. The scale is indicated by the black line and the radius $r_{500}$ is given by the black dashed circle. The pink arrow represents the possible merger axis.}}
	\label{fig:multi_lambda}
	\end{figure}

\begin{table*}
\center
\caption{\footnotesize{Morphological measurements in the optical, SZ and X-ray observations of XLSSC~102. Errors are estimated from the marginalised probability distribution function (PDF) of the parameters, see Figures \ref{fig:peak_centroid} and \ref{fig:ellipticity_PA} for the parameters confidence regions.}}
\begin{tabular}{c||c|c|c}

      				 & Optical & SZ & X-ray  \\
\hline
Ellipticity ($\epsilon$) & $0.75^{+0.17}_{-0.27}$ & $0.37^{+0.13}_{-0.15}$ & $0.15^{+0.11}_{-0.12}$ \\
Major axis position angle (PA) [deg] & $205^{+20}_{-19}$ & $166^{+22}_{-20}$ & $162^{+29}_{-27}$ \\
Peak position [deg] & ($31.326^{+0.006}_{-0.011}$, $-4.631^{+0.004}_{-0.027}$) & ($31.319^{+0.001}_{-0.001}$, $-4.649^{+0.002}_{-0.002}$) & ($31.323^{+0.001}_{-0.001}$, $-4.654^{+0.001}_{-0.001}$)\\
Centroid position [deg] & ($31.325^{+0.006}_{-0.009}$, $-4.637^{+0.009}_{-0.016}$) & ($31.320^{+0.001}_{-0.001}$, $-4.650^{+0.001}_{-0.001}$) & ($31.320^{+0.001}_{-0.001}$, $-4.652^{+0.001}_{-0.001}$)
\end{tabular}
\label{table:ellipticity_PA}
\end{table*}

	\subsection{Interpretation by a merging scenario}

A multi-wavelength view of XLSSC~102 is shown in Figure \ref{fig:multi_lambda}. The SZ and optical S/N contours are over-plotted in green and magenta, respectively, on the X-ray surface brightness map. The maps and contours are produced as explained in Section \ref{subsec:maps} and shown in Figure \ref{fig:multi_images}. The BCG is indicated by the magenta cross.  
On a large scale ($\sim$ 0.5 Mpc), we can see that the galaxy over-density region is elongated along a NE-SW axis, while the SZ signal and X-ray emission in the inner part are elongated along a NW-SE axis. 

In Figure \ref{fig:multi_lambda} we can see that the main peak of the optical density is highly offset from the BCG ($d \sim 1.5'$ or $\sim 0.7$ Mpc), the X-ray centroid ($d \sim 1.3'$ or $\sim 0.6$ Mpc), the X-ray peak ($d \sim 1.4'$ or $\sim 0.7$ Mpc), and the SZ peak ($d \sim 1.1'$ or $\sim 0.5$ Mpc). 
If the tracers of the gas are in good agreement at large scale, at smaller scales we see that the SZ and X-ray peaks are also offset from one another ($d \sim 0.4'$ or $\sim 0.2$ Mpc). 
The BCG is also offset from the X-ray peak ($d \sim 0.2$ or $\sim 0.1$ Mpc) and the SZ peak ($d \sim 0.4$ or $\sim 0.2$ Mpc).

The ellipticity of XLSSC~102 SZ, X-ray or optical emission alone cannot be used to distinguish between the effect of a disturbed dynamical state or triaxiality \citep[see e.g.][]{2018MNRAS.477..139C}. However, taken together and with the other morphological indicators it suggests a perturbed morphology due to a merging event. The fact that the ellipticity is higher for the galaxy distribution than for the gas and that the SZ signal is more elongated than the X-ray one is in agreement with a post-merger scenario.
The offset between the SZ peak, tracing the peak of gas pressure, and the X-ray peak, tracing the gas density peak, indicates the presence of an over-pressure region and a local boost in the gas temperature away from the cluster potential well centre. 
The offset between the X-ray and SZ peaks has been investigated in hydro-dynamical simulations by \cite{2014ApJ...796..138Z}. Based on their results, we evaluated that the probability to have an offset of $\sim 0.2$ Mpc or larger at $z=1$ for a cluster with mass higher than $M=1.4\times10^{14}$ M$_{\odot}$ is around 0.15, and thus not uncommon. 
For a high redshift comparison, \citet{2018A&A...620A...2M} (XXL~Paper~XVII) found in their $z=1.99$ galaxy cluster an offset of $\sim 35''$ ($\sim 0.3$ Mpc at their redshift) between the SZ peak and the X-ray peak (which coincides with the BCG) albeit with lower resolution SZ data. 
The offset between the BCG and the X-ray centroid is known to be a dynamical state indicator \citep[see e.g.][]{2010A&A...513A..37H,2017A&A...599A.138Z,2003ApJ...585..687K}. As the gas density and the BCG trace the cluster potential well, their offset should be small in relaxed clusters \citep[see e.g.][]{0004-637X-617-2-879}, but may be high in perturbed clusters.

While the offsets between the BCG, X-ray peak and SZ peak can be understood in the context of a merger, the location of the optical density peak with respect to the BCG is more puzzling.
However, we can see that a bright galaxy - possibly a second BCG - can be seen at the location of the optical density peak. We note that the galaxies belonging to the northern and southern groups have compatible photometric redshift and $i'$ band magnitude distributions (according to Kolmogorov-Smirnov tests).

A plausible explanation, illustrated on the composite image shown in Figure \ref{fig:multi_lambda}, is that a group of galaxies passed through the main cluster (located near the X-ray centroid), causing over-pressure and disturbances in the gas emission and shifting the optical density peak to the North. The group of galaxies may have been stripped of its gas during this process, which would explain the X-ray and SZ signal elongation toward the North. The enhanced galaxy density in the group, with respect to the main cluster, could be due to the projection of the merger axis along the line of sight. However, in absence of enough spectroscopic members and without gravitational lensing measurements, such scenario remains speculative. The merging scenario would indicate that the cluster is in a post-merger state after a collision with a smaller unit.
 
\section{Profiles of the thermodynamic variables} 
\label{sec:thermo_profiles}

In the previous section, we show that XLSSC~102 is a perturbed cluster, likely in a post-merging phase. Hence, it presents a local over-pressure region and its centre of mass is not well determined. 
In this section we combine the SZ and X-ray data to construct  the thermodynamic profiles of the ICM in XLSSC~102. We then measure and discuss the effects of the merging event and the centre definition. 

	\subsection{Construction of the thermodynamic profiles}
		\begin{figure*}
	\resizebox{0.5\hsize}{!}{\begin{overpic}[scale=1]{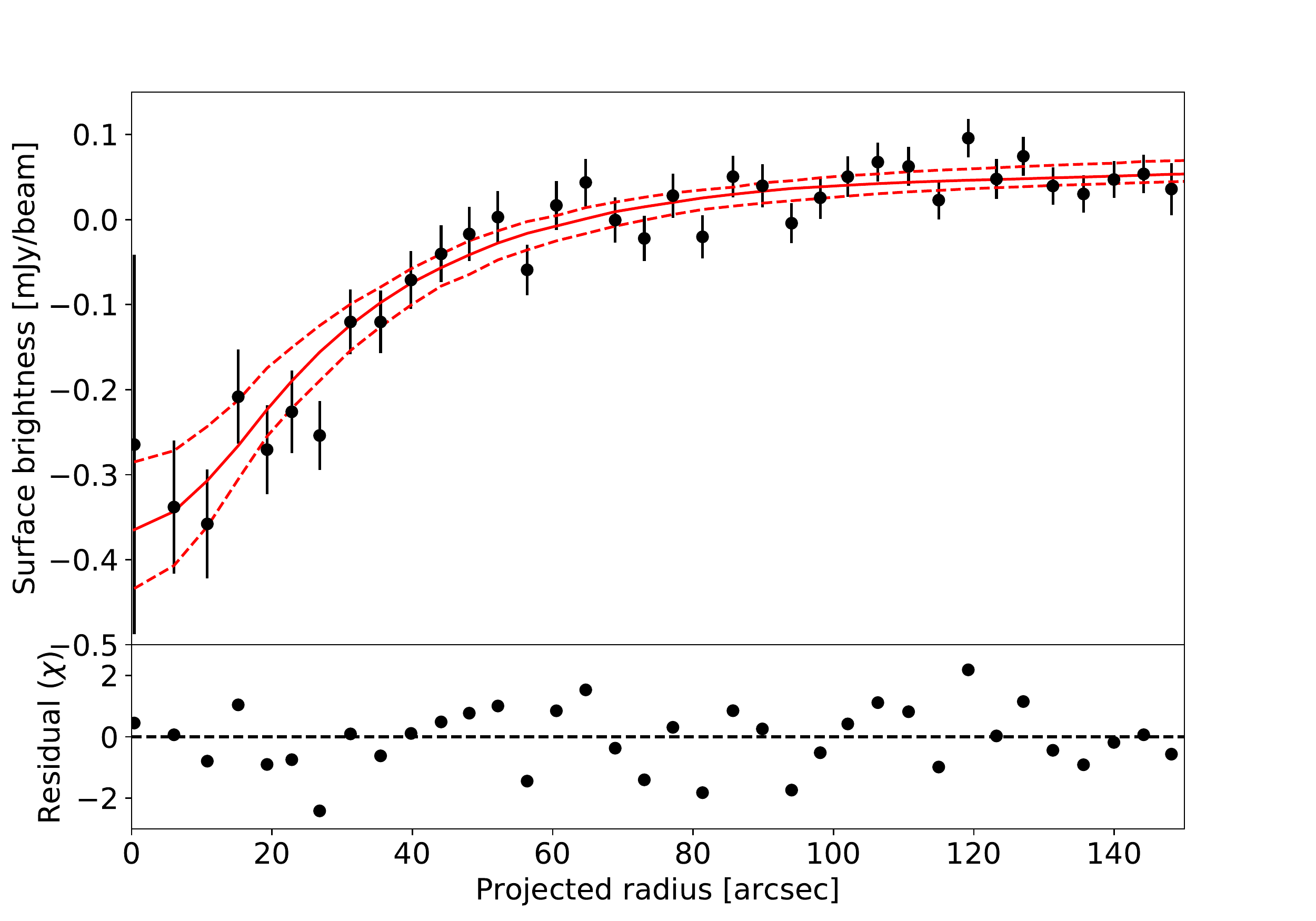}
     \put(54,22){\includegraphics[scale=0.52]{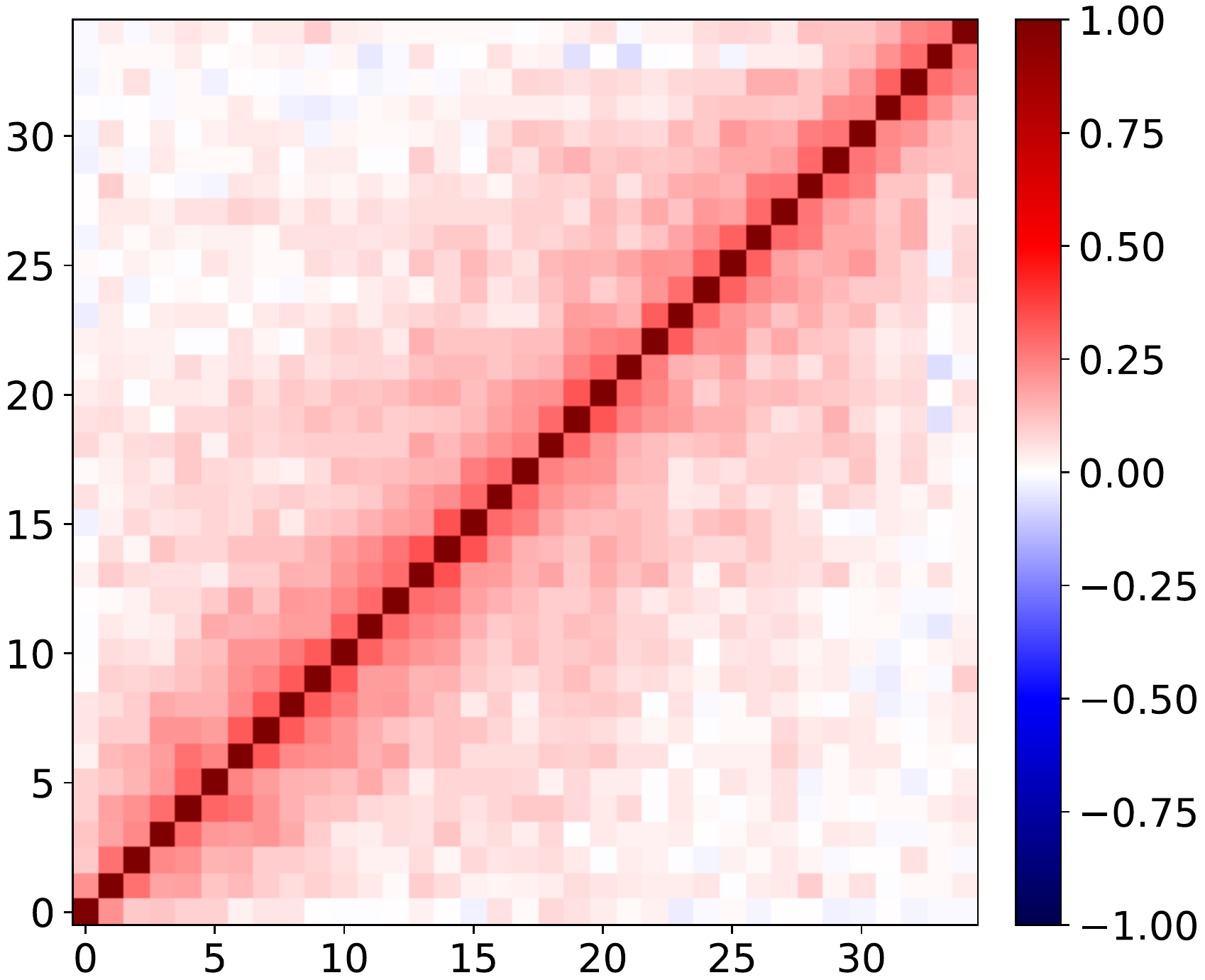}}  
  \end{overpic}}
	\resizebox{0.5\hsize}{!}{\begin{overpic}[scale=1]{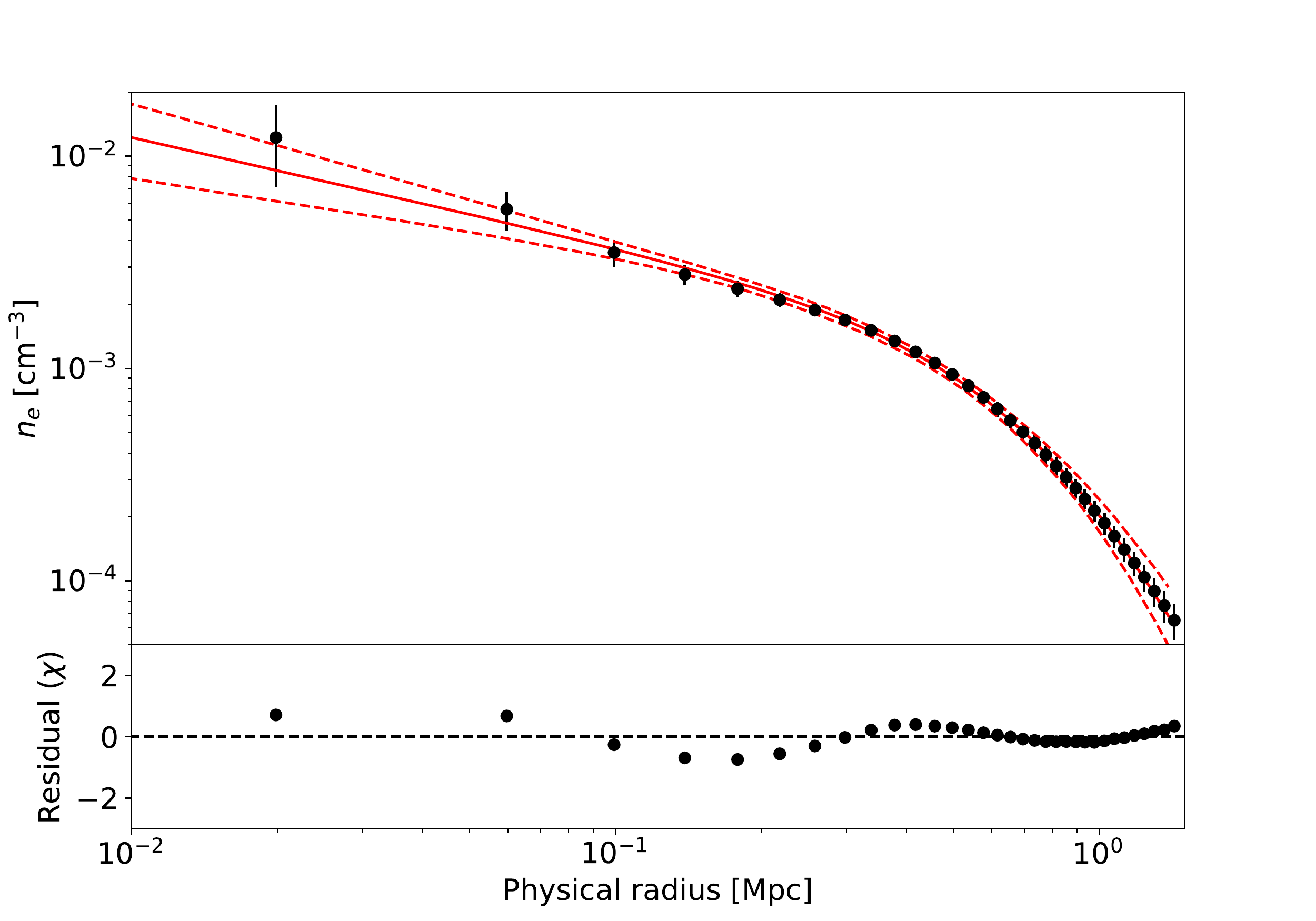}
     \put(10.5,22){\includegraphics[scale=0.52]{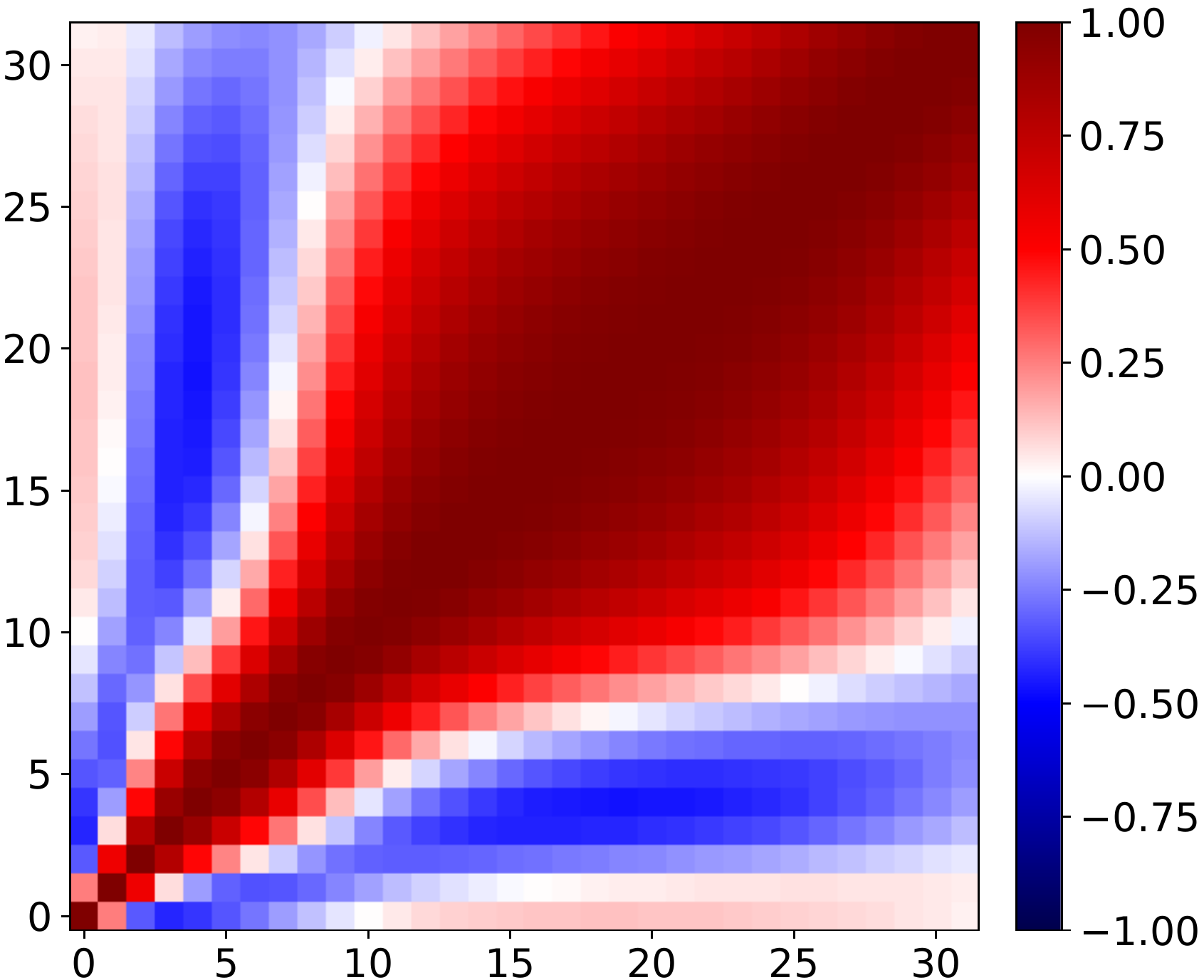}}  
  \end{overpic}}  
		 \caption{\label{profile_SB_and_X}\footnotesize{SZ and X-ray visualisation of the input data for the combined analysis. The maximum likelihood model profiles and their 68\% confidence intervals are shown by the red lines, and the bottom plots show the residuals. The error bars indicate the 68\% c.i., but they only represent the diagonal of the covariance matrix as errors are correlated. The bin-to-bin error correlation matrices are shown by the inset figure in each plot. {\it Left:} SZ surface brightness profile as a function of projected radius, centred on the X-ray peak position. The black points show the inverse variance weighted mean surface brightness in each annulus. The slightly positive value at large projected radius is due to the unconstrained zero level of the map. As detailed in the text, in practice we use the map and not just the profile.{\it Right:} de-projected X-ray electron density profile, centred on the X-ray peak position. The errors in each bin are highly correlated because of the de-projection scheme.}}
		\end{figure*}
		\begin{figure*}[h]
		  \includegraphics[trim={0cm 0cm 4.5cm 0cm}, clip, height=5.8cm]{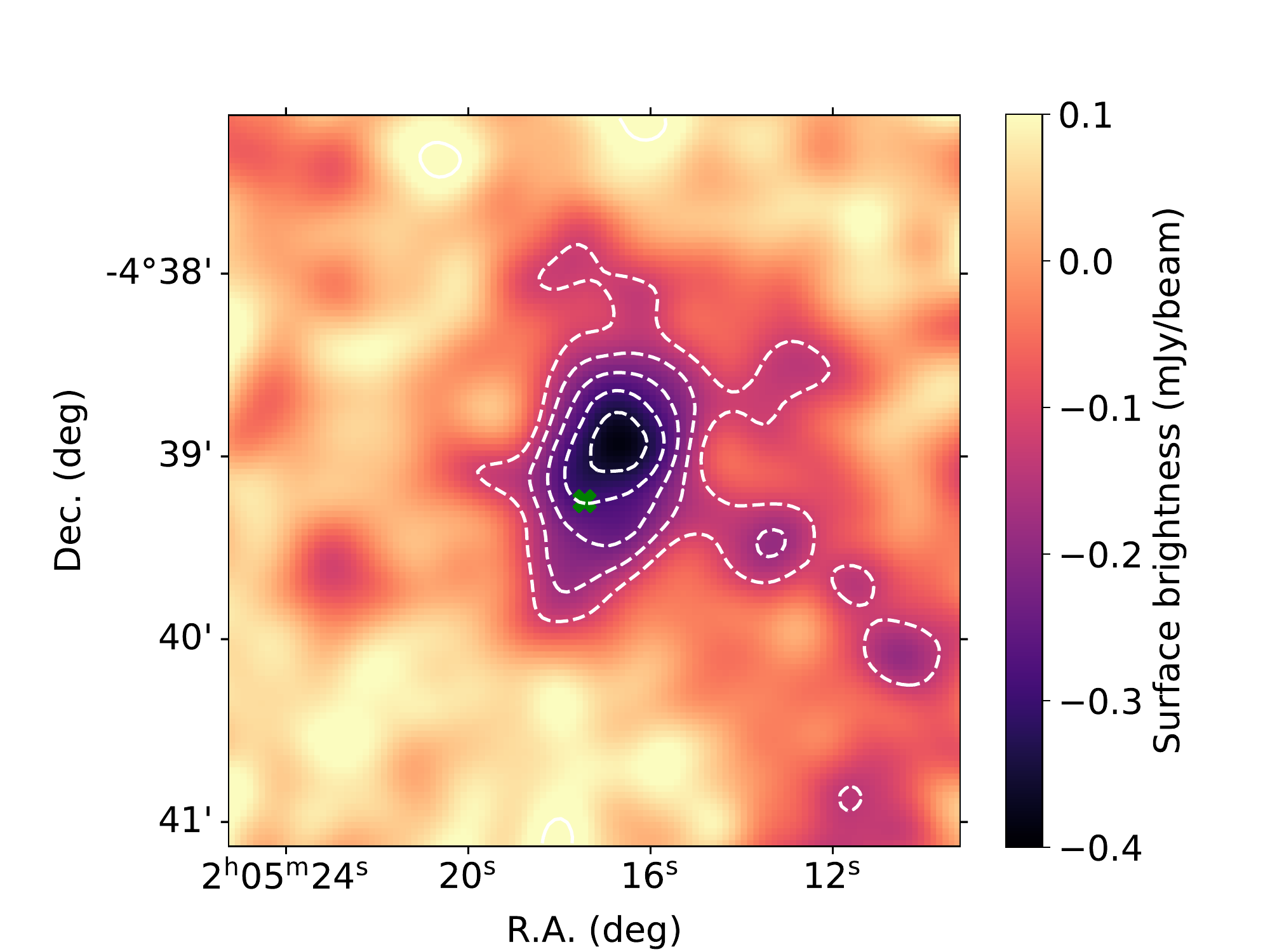}
		  \includegraphics[trim={1.7cm 0cm 4.5cm 0cm},clip, height=5.8cm]{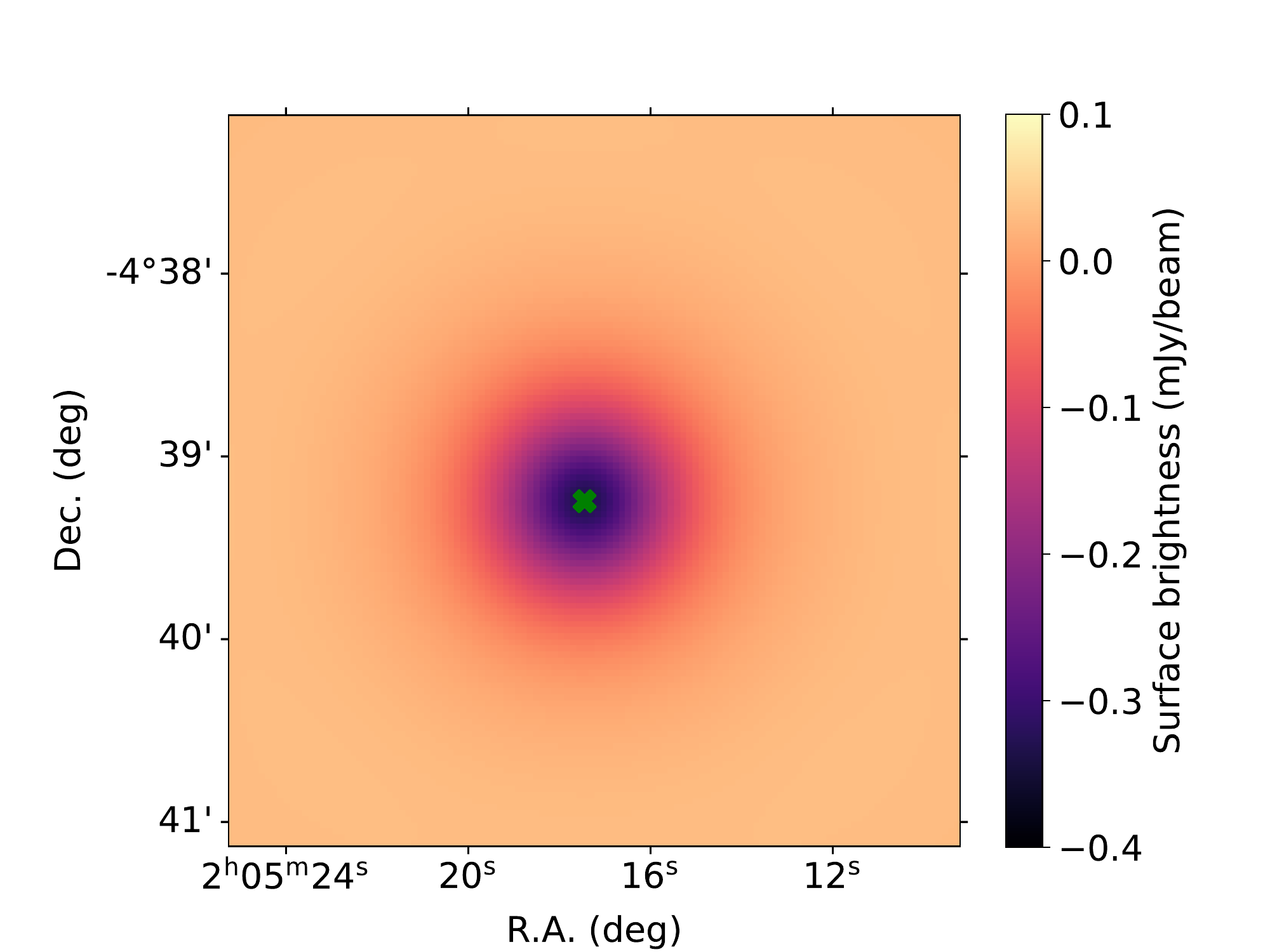}
		  \includegraphics[trim={1.7cm 0cm 0cm 0cm},clip, height=5.8cm]{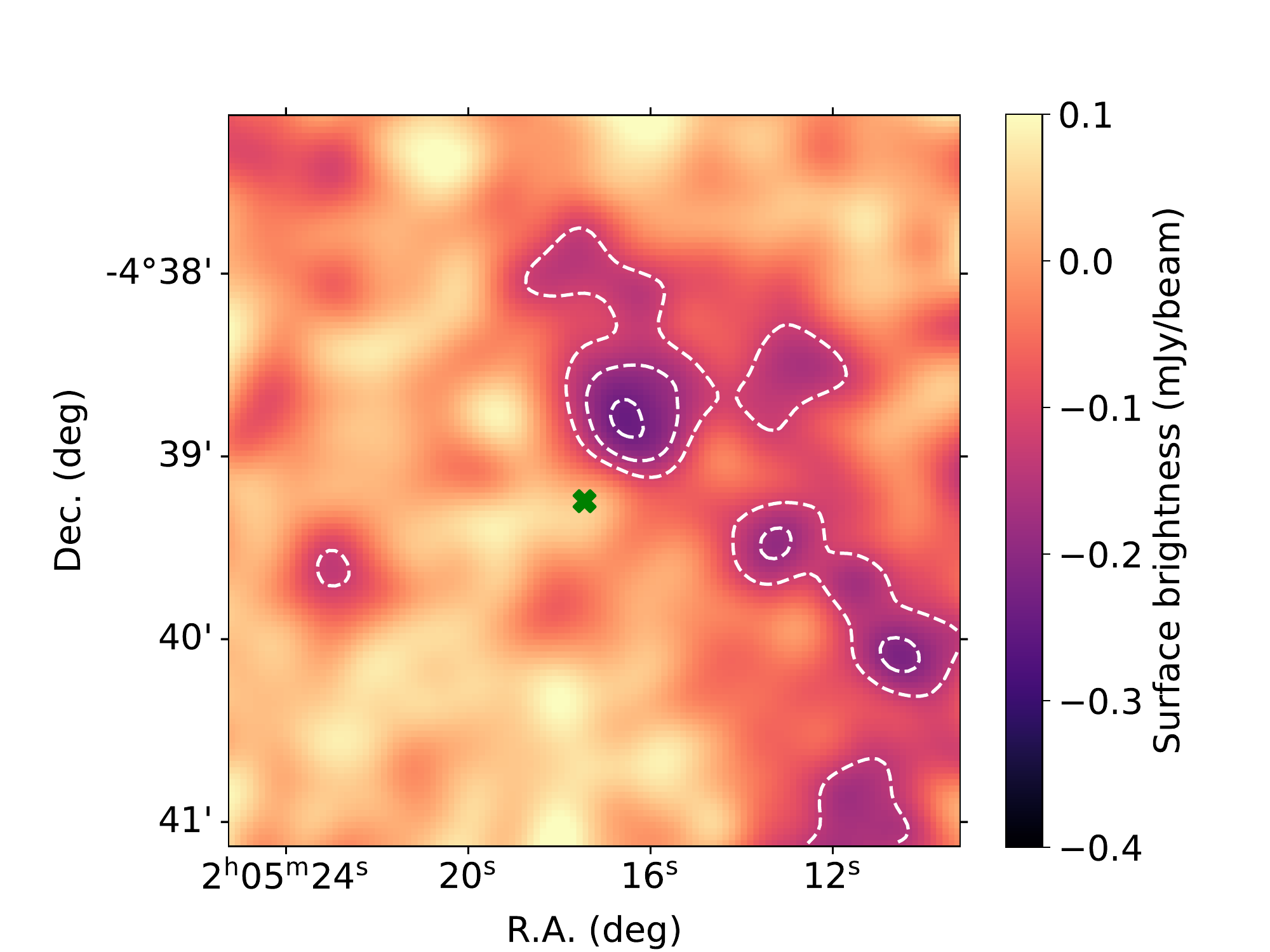}
		 \caption{\label{SZ_model_residuals} 
		 \footnotesize{Comparison between the NIKA2 data and the best-fit model evaluated at the X-ray peak (green cross). {\it Left:} input SZ map. {\it Middle:} maximum likelihood model map. {\it Right:} residuals. The white contours indicate the S/N levels, starting at S/N = 2 and linearly increasing by step of one. The green crosses indicate the position chosen as the centre of the model (here, the X-ray peak).}}
		\end{figure*}

We combine NIKA2 and XXL data to derive the 3D thermodynamic profiles of XLSSC~102 ICM. The methodology is described in the following. 

		\subsubsection{Physical description of the ICM} 
		\label{subsubsec:eq}
The SZ surface brightness $S_{\rm SZ}$ is related to the ICM electron pressure $P_e$ via \citep[see][for a review]{Birkinshaw1999}:
\begin{equation}
	S_{\rm SZ}(\nu) = \frac{\sigma_{\rm T}}{m_e c^2} \int f(\nu) (1+\delta(T)) \ P_e \ d\ell,
	\label{eq_icmtool_Ssz_to_P}
\end{equation}
where $\sigma_{\mathrm{T}}$ is the Thomson cross section, $m_e c^2$ the electron rest mass energy, $f(\nu)$ the SZ frequency spectrum, $T$ the gas temperature, and $\delta(T)$ gives relativistic corrections \citep{1998ApJ...502....7I}. We use a mean temperature of 3.9 keV, using the XXL X-ray temperature estimates (see Table \ref{tab:prop_summary}) to compute an average relativistic correction. However, we neglect internal temperature variation relative to the mean value, as the corresponding change in the relativistic correction is expected to be about two to three percents.

The X-ray surface brightness $S_{\rm X}$ is related to the gas density $n_e$ through \citep[see e.g.][for a review]{bohringer2010}:
\begin{equation}
	S_{\rm X} = \frac{1}{4\pi (1+z)^4} \int n_e^2 \Lambda(T,Z) d\ell,
	\label{eq_icmtool_Sx_to_ne}
\end{equation}
where $ \Lambda(T,Z)$ is the cooling function, which weakly depends on temperature and on the ICM metallicity $Z$.

By combining the electron pressure $P_e$ obtained from SZ observations with the gas electron density $n_e$ obtained from X-ray observations, and under the assumption of spherical symmetry, it is possible to derive the 3-D radial profiles of other thermodynamic quantities \cite[see e.g.][]{2009ApJ...694.1034M,adam_pressure_2015,ruppin_first_2017}. 

The electron temperature is given by:
\begin{equation}
	k_{\rm B} \ T(r) = P_e(r) / n_e(r),
	\label{eq_icmtool_temperature}
\end{equation}
under the ideal gas assumption, with $k_{\rm B}$ the Boltzmann constant. It is connected to the depth of the cluster potential well, but also to its dynamical and thermal states. In the following we will assume thermal equipartition between ions and electrons and we will thus refer to $T$ as the ICM temperature.

The electron entropy index $K_e$, which records the thermal history of the cluster, can be defined as \citep[see][]{2005RvMP...77..207V}:
\begin{equation}
	K_e(r) =  \frac{P_e(r)}{n_e(r)^{5/3}}.
	\label{eq_icmtool_entropy}
\end{equation}

The SZ flux, which is proportional to the cluster total thermal energy, can be given by the spherically integrated Compton parameter, expressed as a function of the pressure as:
\begin{equation}
	Y_{\rm sph}(R) = 4 \pi \frac{\sigma_{\rm T}}{m_e c^2} \int_0^R P_e(r) r^2 dr.
	\label{eq_icmtool_ysph}
\end{equation}

The matter distribution can also be constrained, as the gas mass profile is given by:
\begin{equation}
	M_{\rm gas}(R) = 4 \pi \int_0^R \mu_e m_p n_e(r) r^2 dr,
	\label{eq_icmtool_gas_mass}
\end{equation}
with $\mu_e = 1.15$ the mean molecular weight per electron and $m_p$ the mass of the proton.

The total hydrostatic mass can be computed as: 
\begin{equation}
	M_{\rm HSE}(r) = \frac{- r^2}{G \mu_{\rm gas} m_p n_e(r)} \frac{dP_e(r)}{dr},
	\label{eq_icmtool_masse_hse}
\end{equation}
with $\mu_{\rm gas}=0.61$ the mean molecular weight per gas particle, computed from primordial abundances from \cite{1989GeCoA..53..197A}, and $G$ the gravitational constant. Hydrostatic masses are known to be biased low with respect to true masses because of the contribution of non thermal pressure support and bulk motions in the gas \citep[see][for a review]{2013SSRv..177..119E}. The total true mass is thus related to the hydrostatic mass via the hydrostatic bias $b_{\rm HSE}$ as:
\begin{equation}
	M_{\rm HSE}(R) = (1-b_{\rm HSE}) M_{\rm tot}(R),
	\label{eq_icmtool_hse_bias}
\end{equation}
where $b_{\rm HSE}$ may also depend on the radius, the mass scale or other cluster properties, such as the dynamical state. By combining Eq. \ref{eq_icmtool_gas_mass},  \ref{eq_icmtool_masse_hse} and \ref{eq_icmtool_hse_bias}, we can also define the gas fraction profile as:
\begin{equation}
	f_{\rm gas}(R) = \frac{M_{\rm gas}(R)}{M_{\rm tot}(R)},
	\label{eq_icmtool_gas_frac}
\end{equation}
which provides a probe of the relative spatial distribution of dark matter and gas in clusters.
Finally, the overdensity contrast within radius $R$ can be calculated as:
\begin{equation}
	\Delta(R) = \frac{3 M_{\rm tot}(R)}{4 \pi R^3\rho_c(z) }.
	\label{eq_icmtool_rdelta}
\end{equation}
with $\rho_c(z)$ the critical density of the Universe at the cluster's redshift.

	\begin{table*}[]
	\caption{\label{tab:priors} Priors on the model parameters.}
	\begin{center}
	\begin{tabular}{c||c|c|c|c|c||c|c|c|c}
	\hline
	\hline
	& \multicolumn{5}{c}{Pressure profile }&  \multicolumn{4}{c}{Density profile} \\
	 \hline
	Parameter & $P_0$ & $r_p$ & a & b & c & $n_{e,0}$ & $r_c$ & $\beta$ & $\alpha$ 
\\
	Unit & keV cm$^{-3}$ & kpc & -- & -- & -- & cm$^{-3}$ & kpc & -- & -- \\
	\hline
	Value & $>$ 0 & $>$ 50 & $1.33\pm0.33$ & $4.13\pm1.03$ & $0.31\pm0.08$
& $>$ 0 & [0,2000] & [0,5] & $>$ 0 \\
	Prior type & Flat & Flat & Gaussian & Gaussian & Gaussian & Flat & Flat & Flat & Flat \\
	\hline
	\end{tabular}
	\end{center}
	\end{table*}

		\subsubsection{Input data} 
The following data were used in order to derive the ICM physical profiles : 

{\bf{1), NIKA2 surface brightness map:}}
We used the NIKA2 150 GHz surface brightness map, projected in $5''\times5''$ pixels and a $5'$ field of view in order to lighten the numerical computation. The associated noise covariance matrix and the transfer function of the data reduction were computed as described in Section \ref{sec:Data_NIKA2} and taken into account in the analysis.
We consider a spherical model which contains the same information as the surface brightness profile presented in the left panel of Figure \ref{profile_SB_and_X}.

{\bf{2), XMM X-ray density profile}}
To extract the gas density profile of XLSSC 102, we used the public code \texttt{pyproffit}\footnote{\href{https://github.com/domeckert/pyproffit}{https://github.com/domeckert/pyproffit}}, which is a Python implementation of the popular {\sc Proffit} code \citep{Eckert2011}. The surface brightness profile was accumulated in circular annuli of 5 arcsec width. We conservatively masked circles of 30 arcsec radius around sources detected by the XXL pipeline \citepalias{2018A&A...620A..12C} to avoid contamination by point-like sources. The multiscale decomposition method introduced by \citet[][\citetalias{2016A&A...592A..12E}]{2016A&A...592A..12E} was used to model the gas distribution and deproject the profile, assuming the gas distribution is spherically symmetric. Namely, the surface brightness profile was described as a sparse linear combination of a large number of King functions with fixed radial shape. The model was convolved with the \textit{XMM-Newton} PSF and fitted to the data by optimizing the Poisson likelihood function. Optimization was performed using the Hamiltonian Monte Carlo code \texttt{PyMC3} \citep{pymc3}, which is suitable for high-dimensional optimization problems. The conversion between count rate and emission measure was calculated by assuming that the X-ray spectral distribution follows a single-temperature APEC model with a temperature fixed to the measured spectral temperature \citep[][XXL~Paper~III]{giles_xxl_2015}. The covariance matrix of the resulting gas density profile was computed from the output \texttt{PyMC3} chain and taken into account later on in the joint fitting procedure. The output gas density profile is presented on the right-hand panel of Fig. \ref{profile_SB_and_X}. The error bars shown here are the square root of the diagonal elements of the covariance matrix.

{\bf 3), {\textit{Planck} total SZ flux}}
Because of the large beam of \textit{Planck} \citep[$10'$ for the Compton parameter map,][]{2016A&A...594A..22P} and the faintness of its signal, XLSCC102 is not detected in the \textit{Planck} SZ catalogue \citep{2016A&A...594A..27P} at such mass and redshift.
Nevertheless, we used the value of the total SZ flux $Y^{Planck}_{\rm tot}$ measured in the \textit{Planck} Compton parameter MILCA map as an extra constraint on the cluster total flux to better define the zero level of the NIKA2 map. This was done by extracting the total SZ signal at the XXL coordinates in the \textit{Planck} Compton parameter map \citep{2016A&A...594A..22P,2013A&A...558A.118H}, by fitting 2D Gaussian of FWHM equal to the map resolution. The flux error was estimated by repeating the measurements at random positions around the cluster. This procedure assumes that our targets are point sources with respect to the $10'$ \textit{Planck} beam and provide a constraint on the total SZ flux (equivalent to the $Y_{\rm 5\theta_{500}}$ definition used in \citep{2016A&A...594A..27P}). We found $Y_{\rm tot}^{\rm Planck} = 40 \pm 57$ kpc$^2$.

		\subsubsection{Fitting procedure} 
		\label{sec:fitting}
			\paragraph{{\bf{Profile modelling:}}}
In order to fit the 3-D profiles of the quantities described above we use a parametric modelling of the electron pressure and density profiles.

We model the electron pressure profile by a generalised Navarro, Frenk and White model \citep[gNFW,][]{2007ApJ...668....1N} given by:
\begin{equation}
	P_e(r) = \frac{P_0}{\left(\frac{r}{r_p}\right)^c \left(1+\left(\frac{r}{r_p}\right)^a\right)^{\frac{b-c}{a}}},
\label{eq_gNFW}
\end{equation}
with $P_0$ a normalisation constant, $r_p = R_{\Delta} / c_{\Delta}$ the characteristic radius expressed with $c_{\Delta}$ describing the gas concentration, and $a$, $b$ and $c$ the parameters describing the slope of the profile at radii $r \sim r_p$, $r \gg r_p$ and $r \ll r_p$, respectively.

We model the electron density profile by a simplified Vikhlinin model \citep[SVM,][]{2006ApJ...640..691V}, given by:
\begin{equation}
	n_e(r) = n_{e0} \left[1+\left(\frac{r}{r_c}\right)^2 \right]^{-3 \beta /2 + \alpha/4}  \left(\frac{r}{r_c}\right)^{-\alpha/2}.
\label{eq_density_SVM}
\end{equation}
In this expression, $n_{e0}$ is the normalisation, the first term in bracket corresponds to a $\beta$-model with characteristic core radius $r_c$ and outer slope $\beta$ and the second term allows for modification of the inner slope according to the parameter $\alpha$. The SVM model parametrisation also includes a third term allowing for a change of the outer slope. As we aim at an accurate description of our data with a minimal set of parameters and after testing different parametrisation we found that adding this terms was unnecessary.

			\paragraph{{\bf{Fitting algorithm:}}}
The profile fitting algorithm was adapted from that developed in \cite{adam:tel-01303736}. We briefly summarise its operating principle in the following. The approach consists in sampling the parameter space of equations \ref{eq_gNFW} and \ref{eq_density_SVM} using a Markov chain Monte Carlo algorithm and evaluating at each step the likelihood of the parameter set, given the data. 

The fitted variables are the five parameters of the gNFW model, the four parameters of our SVM model plus the zero level of the NIKA2 surface brightness map $Z_0^{\rm SZ}$, which is a nuisance parameter. The priors chosen for the parameters are presented in Table \ref{tab:priors}. They arise both from physical and numerical considerations. The Gaussian priors on $a$, $b$ and $c$ are centred on the values measured on \textit{Planck} nearby clusters \cite{2013A&A...550A.131P} and have standard deviations  equal to 25\% of their mean values.  We also added a prior on the combination of parameters to force $dM/dr \geqslant 0$, and thus ensuring that the mass increases with radius. 

The likelihood of the proposed set of parameters is then evaluated with respect to the X-ray density profile, the NIKA2 SZ surface brightness map and the \textit{Planck} total SZ flux. While the density profile model is directly compared to the electron density inferred from the X-ray image, the pressure profile model is integrated along the line of sight, accounting for the NIKA2 instrumental response (beam and transfer function), to compute the corresponding SZ surface brightness model. The latter can then be compared to the NIKA2 data. The pressure profile model is also used to compute the total integrated Compton parameter, following equation \ref{eq_icmtool_ysph}. This quantity is then compared to the total flux as measured from \textit{Planck} data.

Finally, the fitting algorithm uses a Metropolis-Hastings method of MCMC sampling to constrain the parameters. Once the chains converge and the burn-in is removed, we obtain the PDF sampling in the 10-dimensional parameter space. The PDF of one parameter can then be obtained by marginalising over the other nine.

			\paragraph{{\bf{Application to our data:}}}
In Figure \ref{profile_SB_and_X}, we show the best-fit parametric models of the NIKA2 SZ surface brightness profile and the X-ray density profile over-plotted on the data. We take as our reference centre the position of the X-ray peak. We can see that the models provide a good description of the data at all radial scales. In the case of NIKA2, we also provide the comparison between the model and the data at the map level in Figure \ref{SZ_model_residuals}. We can see that overall the model provides a good description of the data. Nonetheless, as the morphology of the cluster is elongated and the model centred on the X-ray peak, the residuals in the north-west region of the cluster present a significant excess ($>4 \sigma$). This agrees with XLSSC~102 having a disturbed morphology and is consistent with the merger scenario proposed in Section \ref{sec:Morphology}. We note that the best-fit model SZ total flux is in full agreement with the Planck constraints (58 kpc$^2$ versus $40 \pm 57$ kpc$^2$).

	\begin{figure*}
	\center
	\begin{tabular}{rr}
	\includegraphics[trim=0cm 2.cm 0cm 0cm, width=0.45\textwidth]{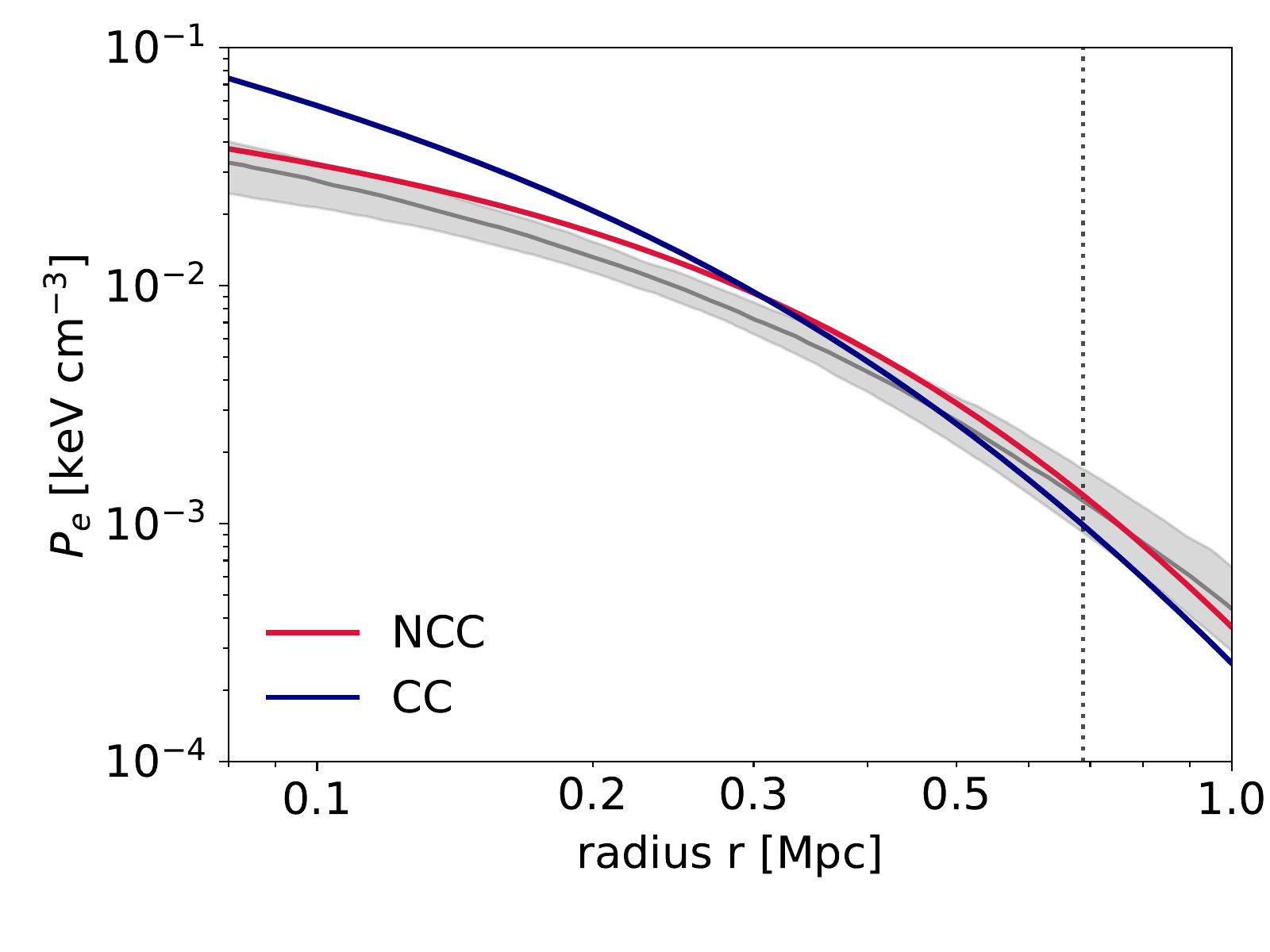} & 
	\includegraphics[trim=0cm 2.cm 0cm 0cm, width=0.45\textwidth]{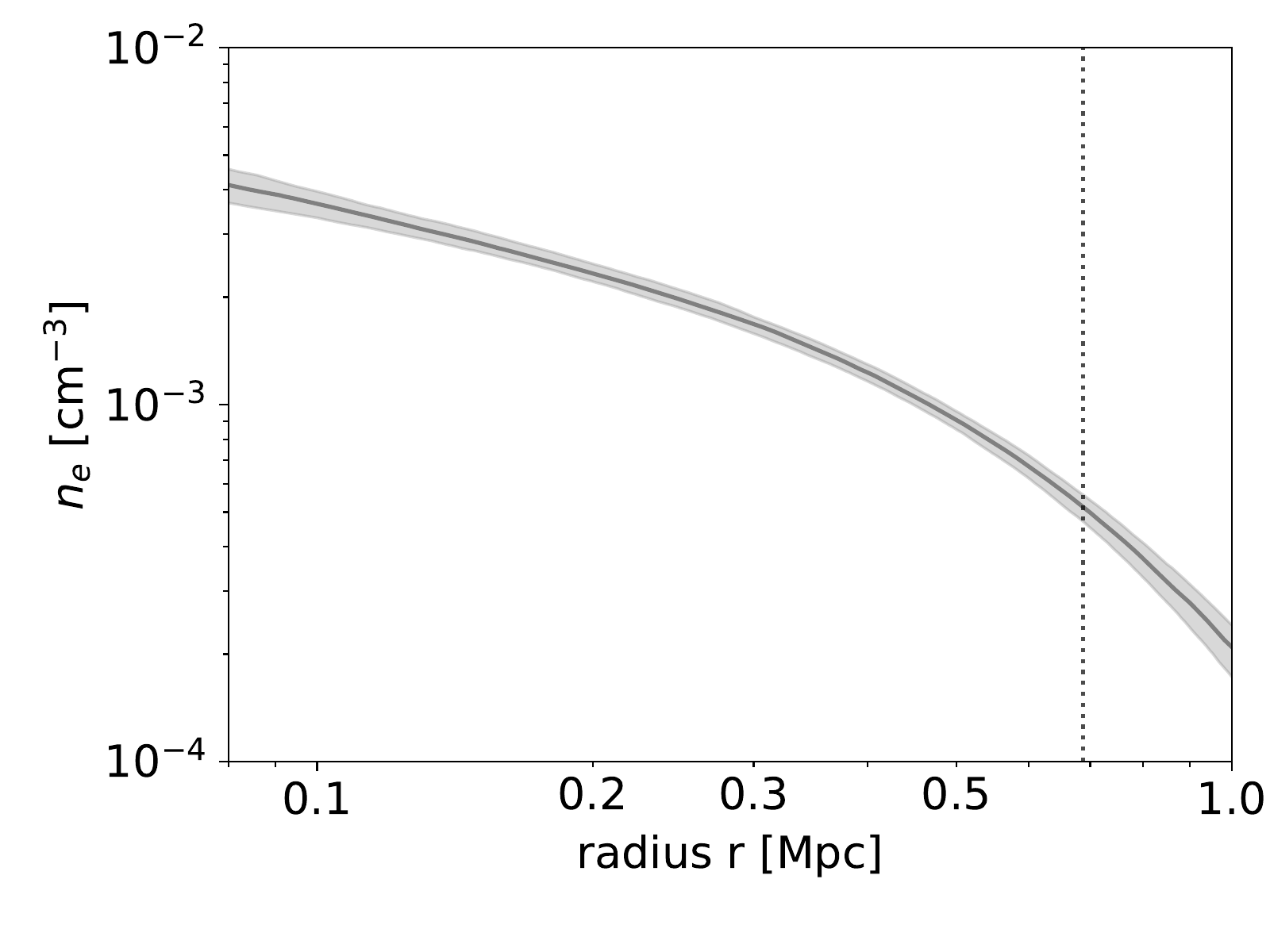} \\
	\includegraphics[trim=0cm 2.cm 0cm 0cm, width=0.45\textwidth]{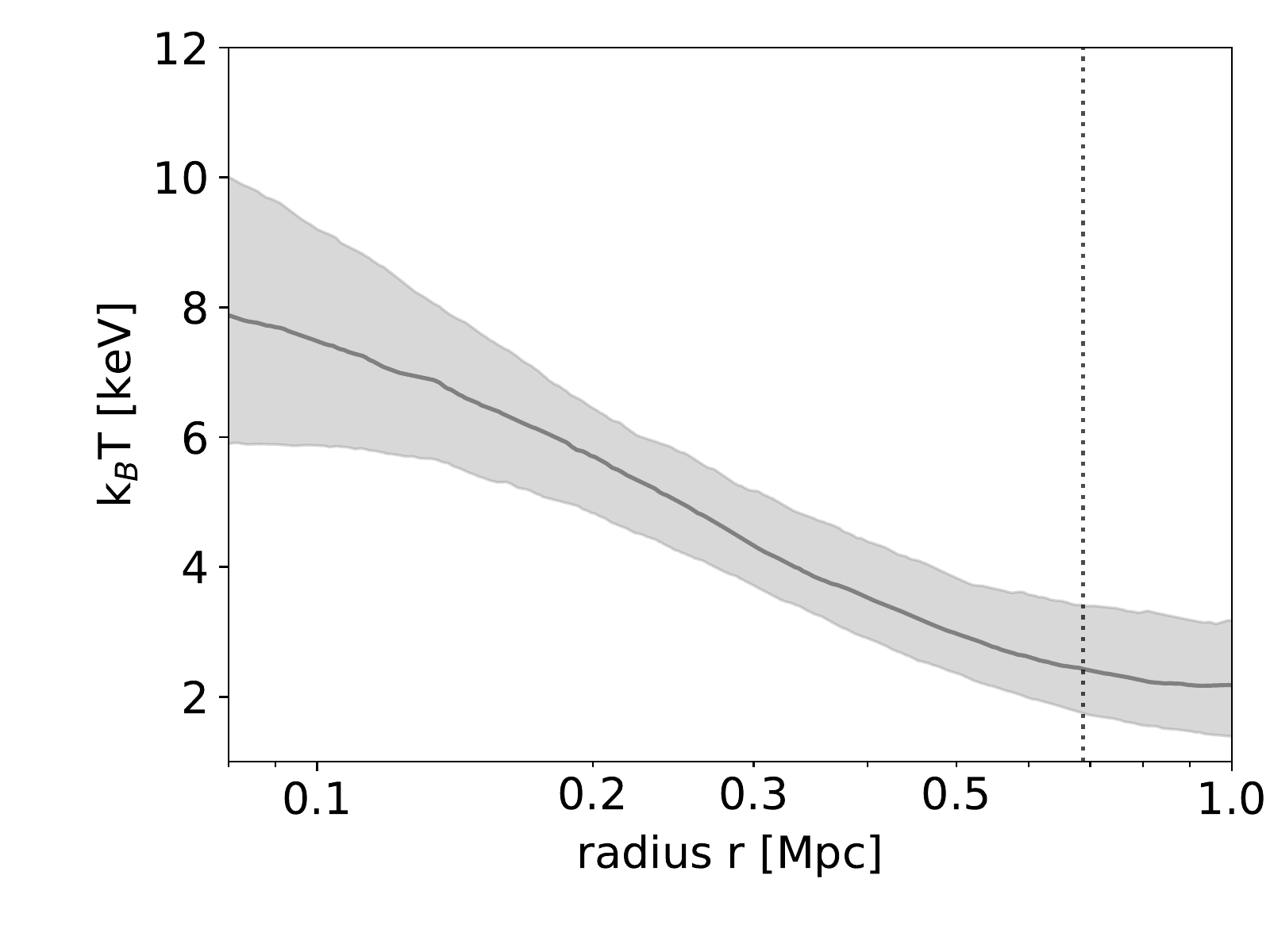} &
	\includegraphics[trim=0cm 2.cm 0cm 0cm, width=0.45\textwidth]{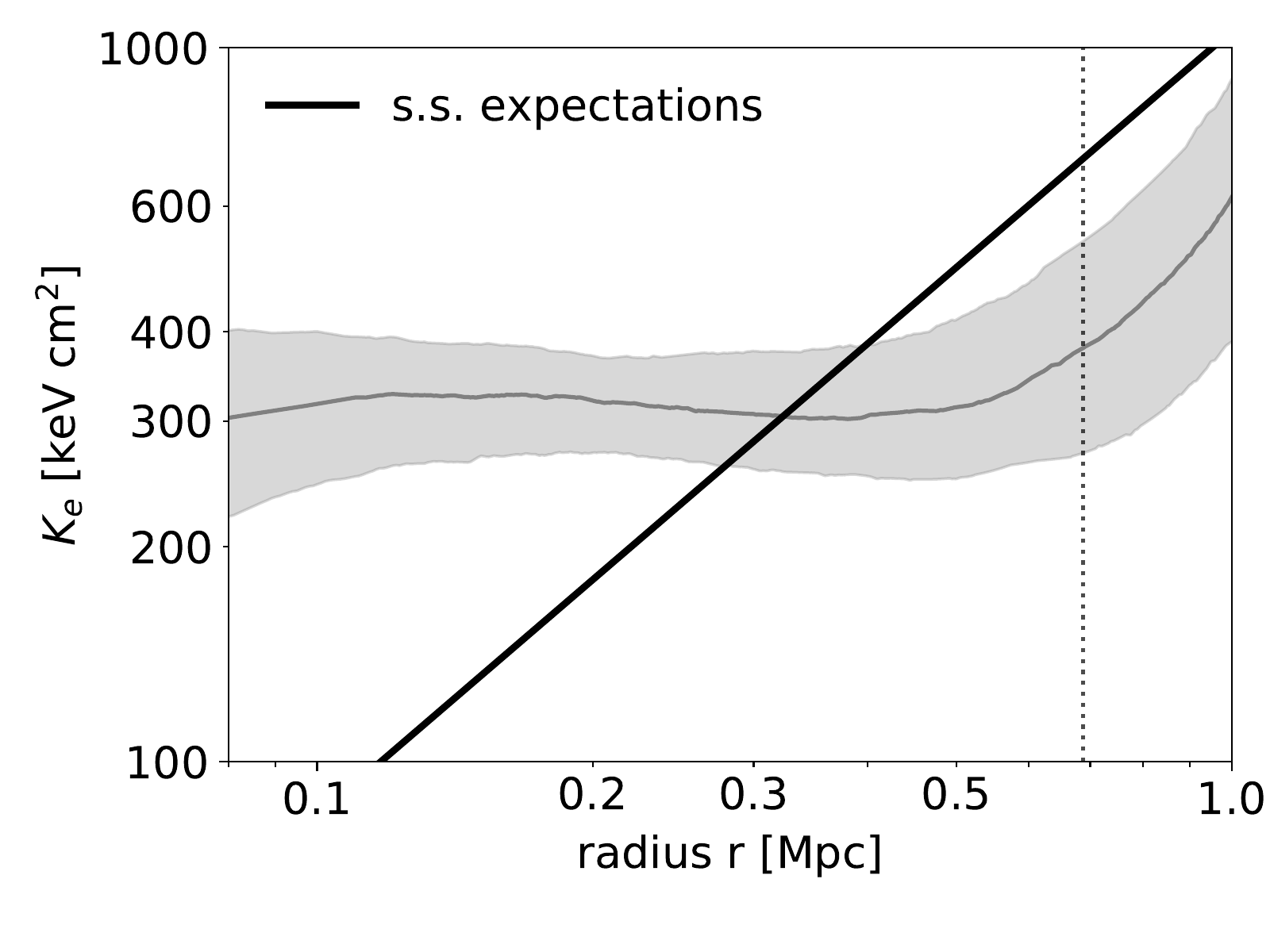} \\
	\includegraphics[trim=0cm 2.cm 0cm 0cm, width=0.45\textwidth]{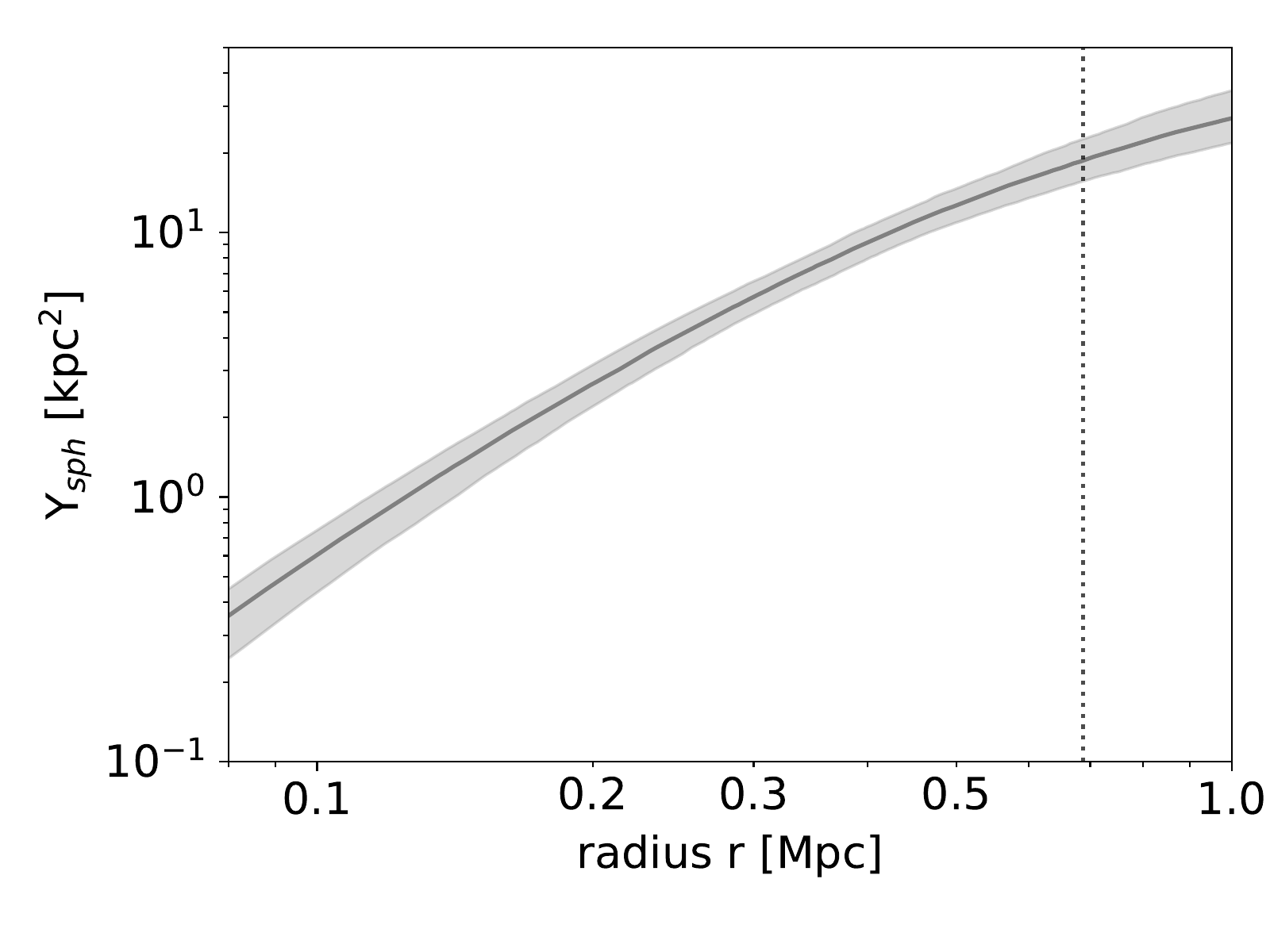} &
	\includegraphics[trim=0cm 2.cm 0cm 0cm, width=0.45\textwidth]{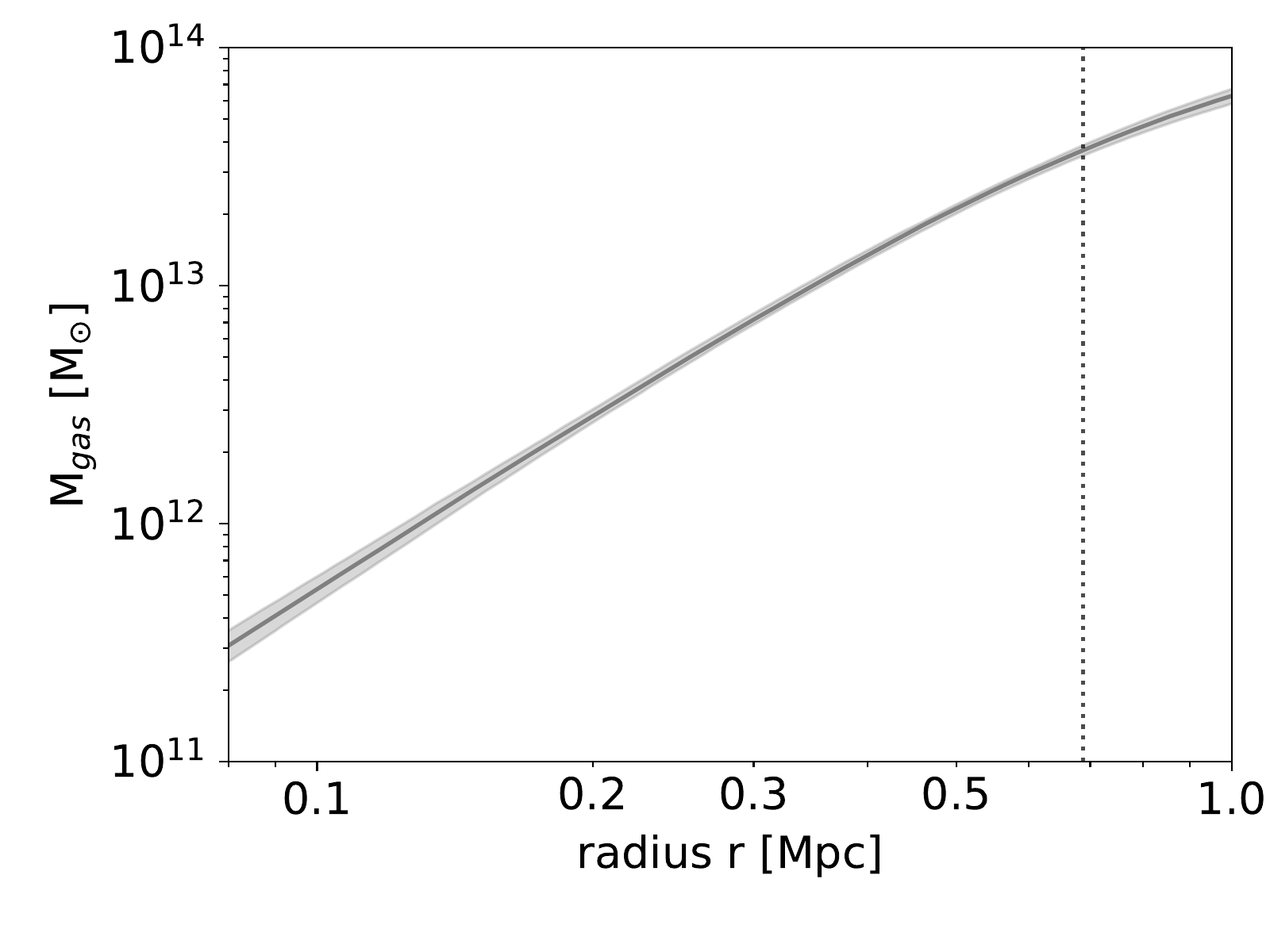} \\
	\includegraphics[trim=0cm 0.0cm 0cm 0cm, width=0.45\textwidth]{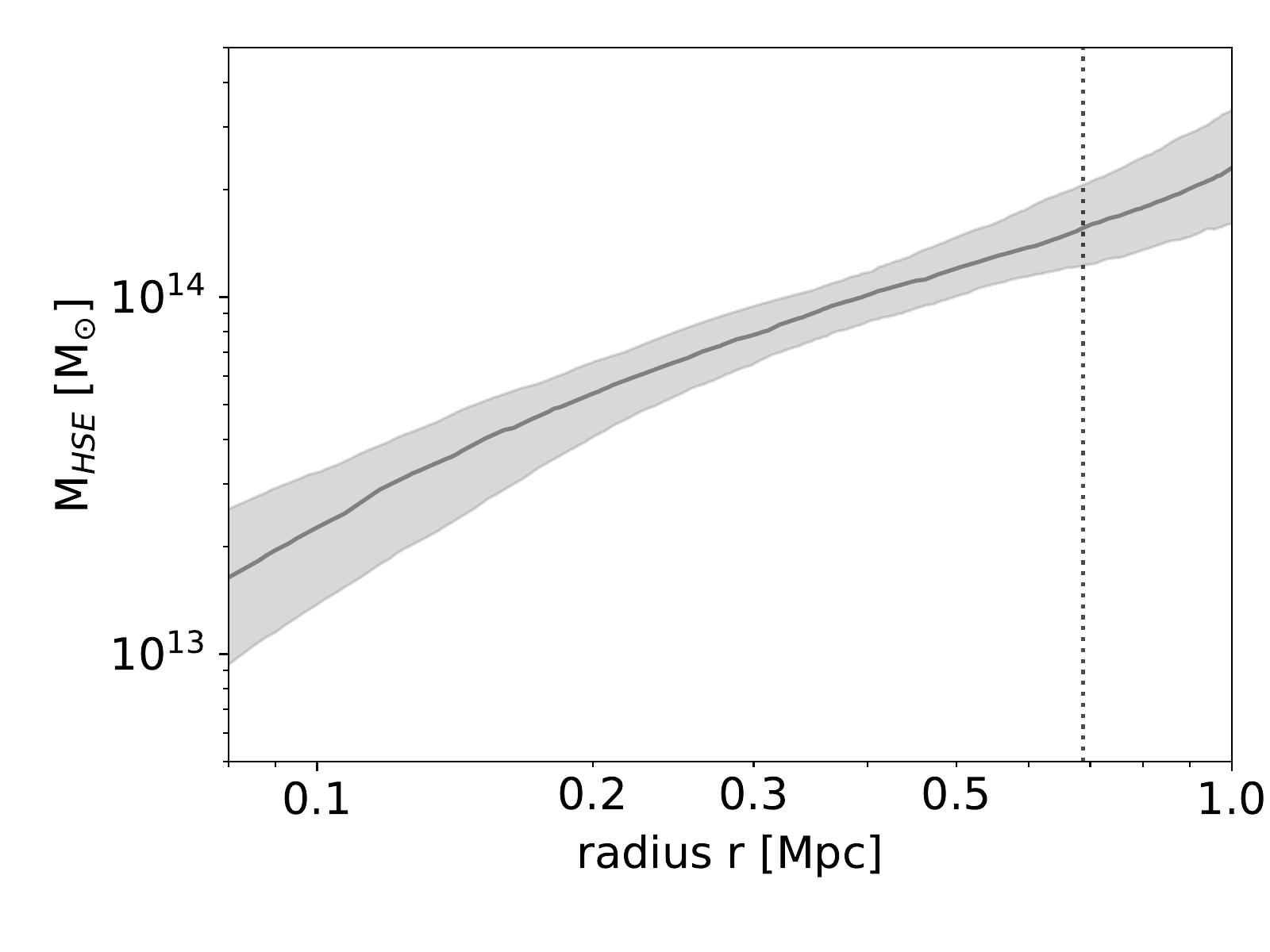} & 
	\includegraphics[trim=0cm 0.0cm 0cm 0cm, width=0.45\textwidth]{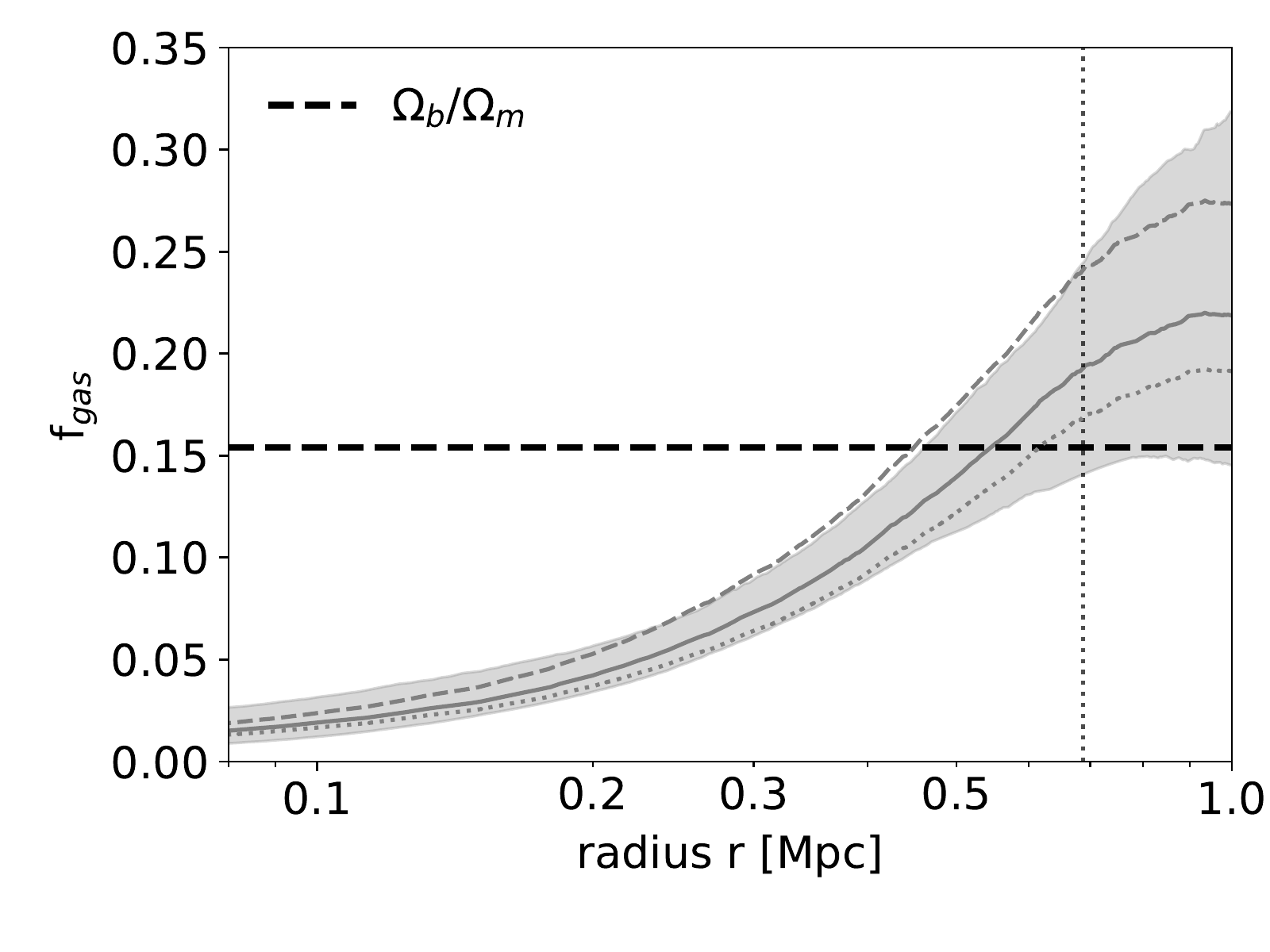}
	\end{tabular}
	\caption{\label{fig:thermo_profiles_Xp} 
	\footnotesize{Radial physical profiles of XLSSC~102. From left to right and top to bottom: electron pressure, electron density, temperature, electron entropy index, spherically integrated SZ flux, gas mass, hydrostatic mass, and gas fraction profiles. The grey shaded regions shows 68\% c.i. around the median profiles. 
The thin dotted black line indicates the value of $r_{500}$ from XXL.
For comparison, the mean pressure profiles of cool core and disturbed clusters from the REXCESS sample \citep[][hereafter, A10]{2010A&A...517A..92A}, computed with the mass estimated from XXL scaling relations $M^{{\rm XXL}}_{500,{\rm scal}}$ (see Table \ref{tab:prop_summary}), are shown by the dark blue and red curves.
The self-similar entropy index expectation from \cite{2005RvMP...77..207V} computed for a mass $M^{{\rm XXL}}_{500,{\rm scal}}$ (see Table \ref{tab:prop_summary}) and the value of the cosmic gas fraction are shown by the dashed black lines. 
The latter is computed as $\frac{\Omega_b}{\Omega_m}$ under the assumption that all the baryons are in the hot gas phase, with baryonic and matter density taken from \cite{2018arXiv180706209P}.
The gas fraction profile is computed assuming $b_{\rm HSE}$ = 0.2. The median profiles computed assuming $b_{\rm HSE}$ = 0 and $b_{\rm HSE}$ = 0.3 are shown by the dotted and dashed grey curves respectively.
}}
	\end{figure*}

	\begin{figure*}
	\center
	\begin{tabular}{rr}
	\includegraphics[trim=0cm 2cm 0cm 0cm, width=0.45\textwidth]{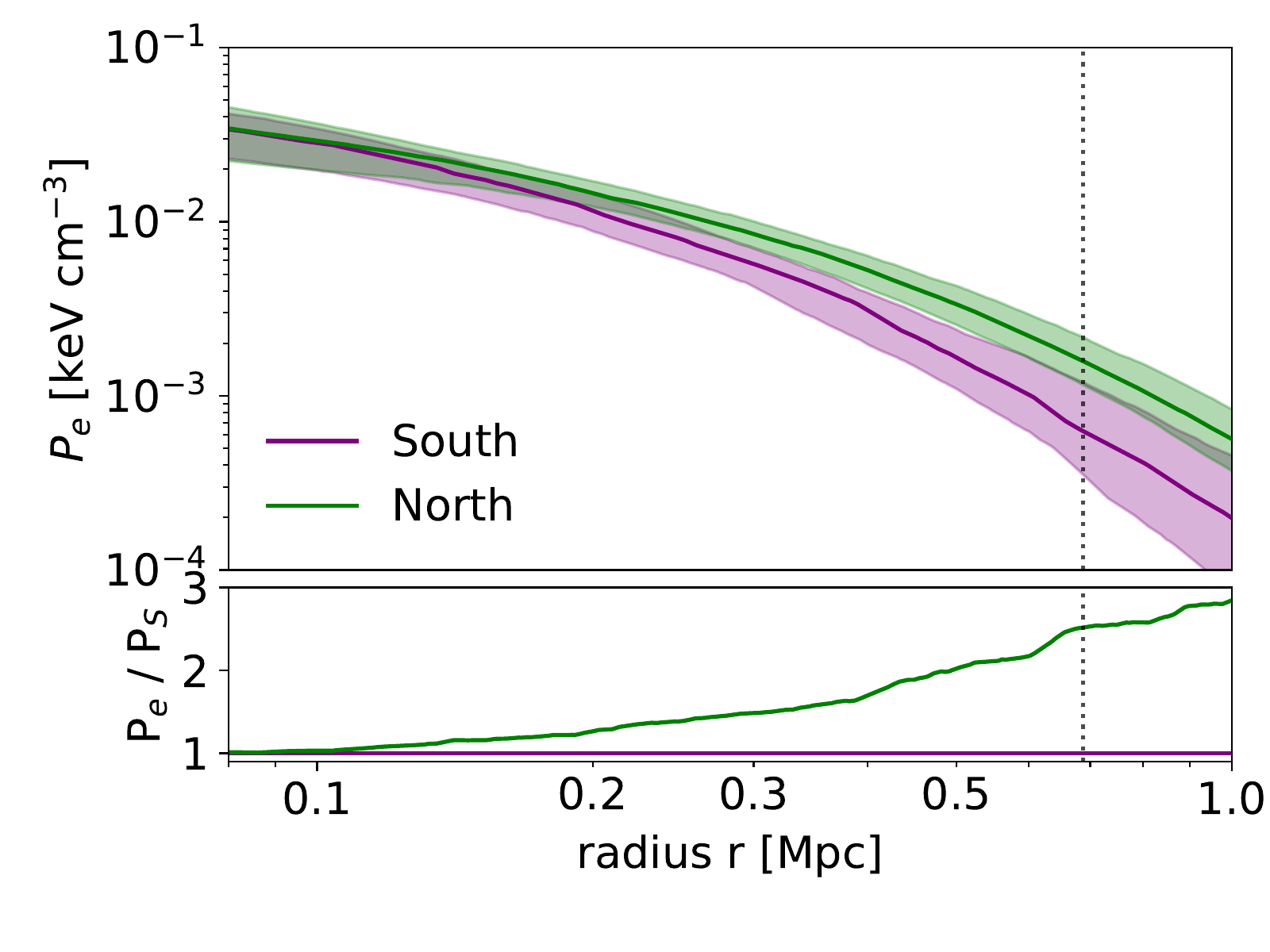} & 
	\includegraphics[trim=0cm 2cm 0cm 0cm, width=0.45\textwidth]{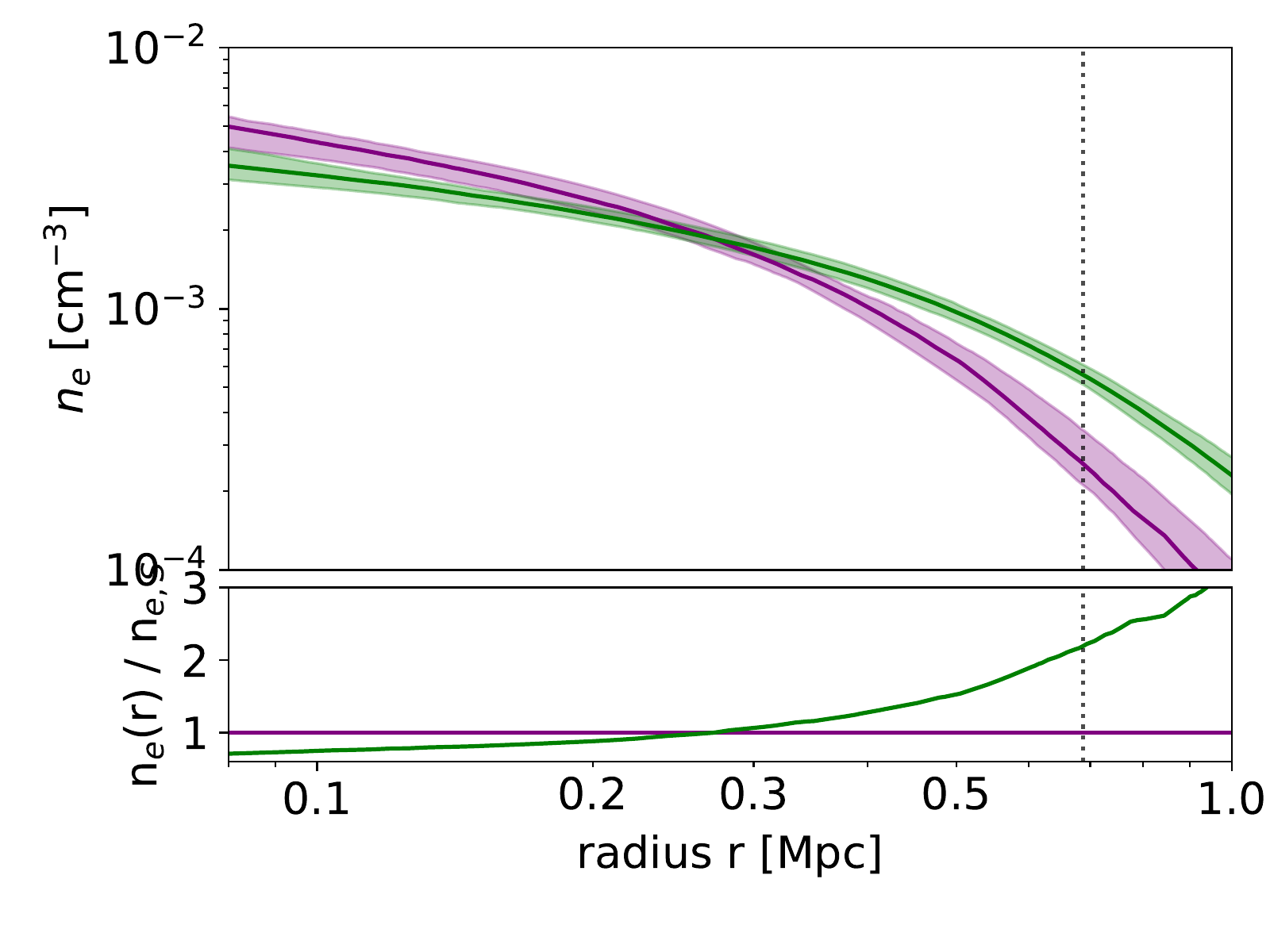} \\
	\includegraphics[trim=0cm 2cm 0cm 0cm, width=0.45\textwidth]{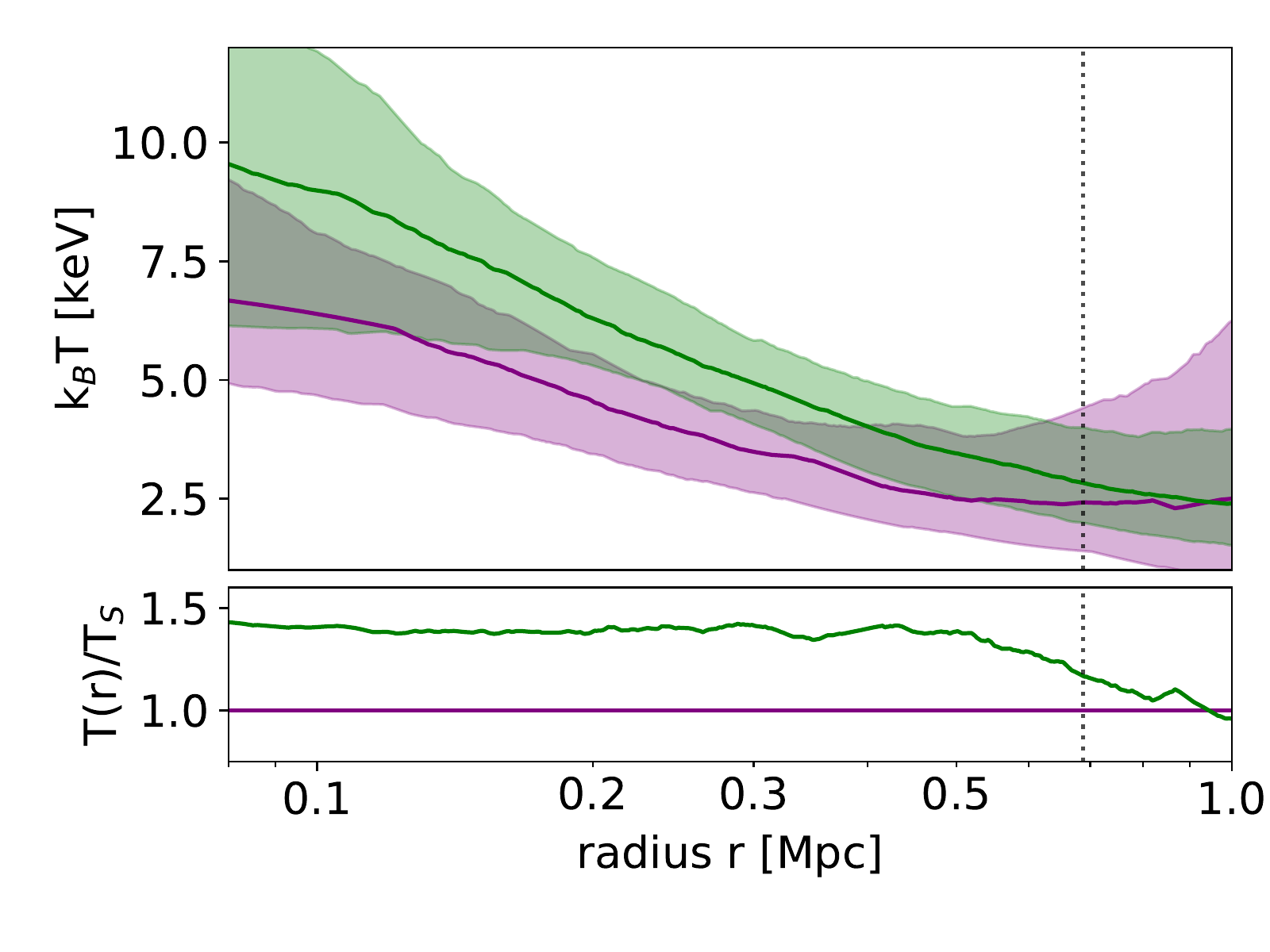} &
	\includegraphics[trim=0cm 2cm 0cm 0cm, width=0.45\textwidth]{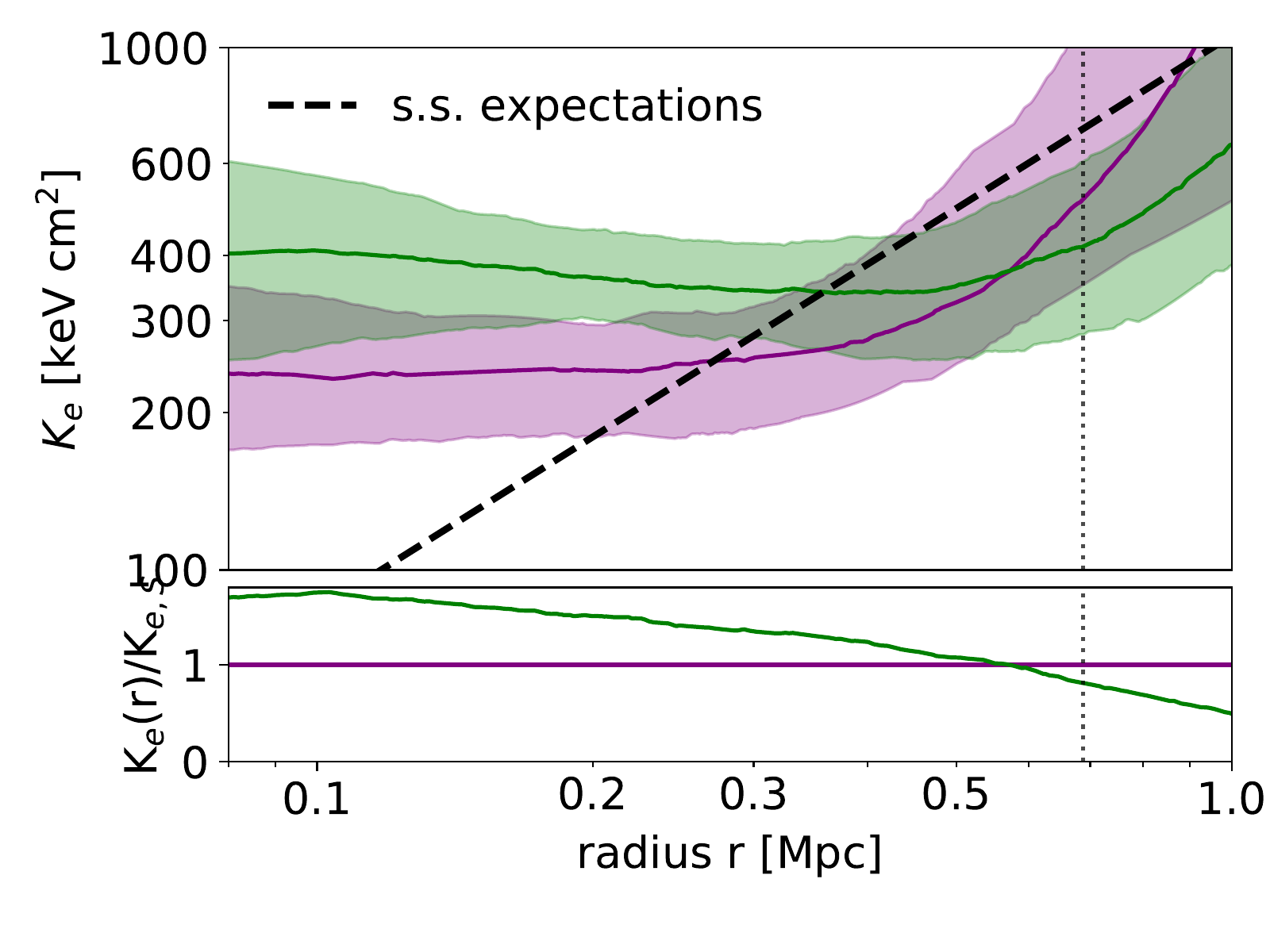} \\
	\includegraphics[trim=0cm 2cm 0cm 0cm, width=0.45\textwidth]{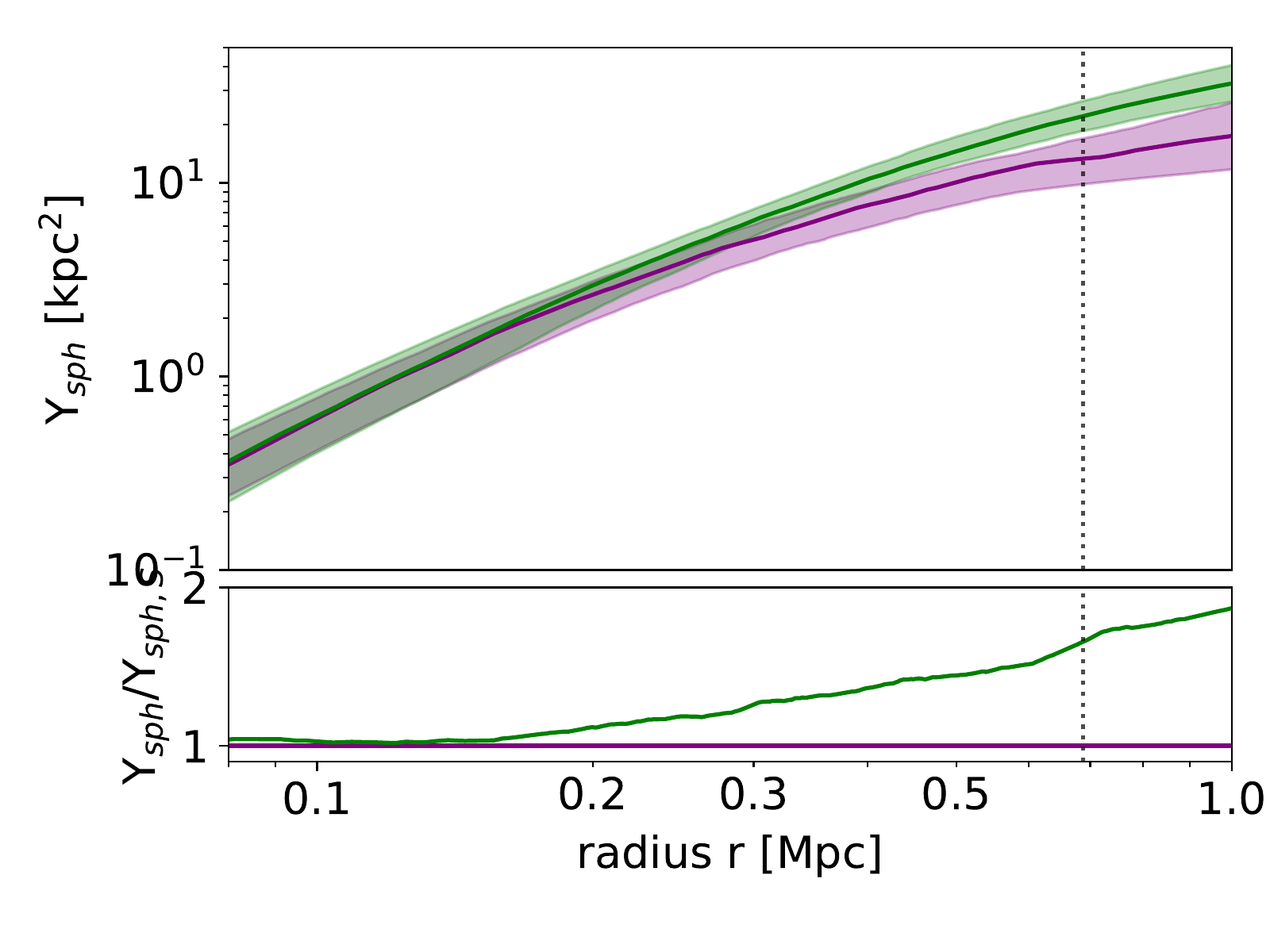} &
	\includegraphics[trim=0cm 2cm 0cm 0cm, width=0.45\textwidth]{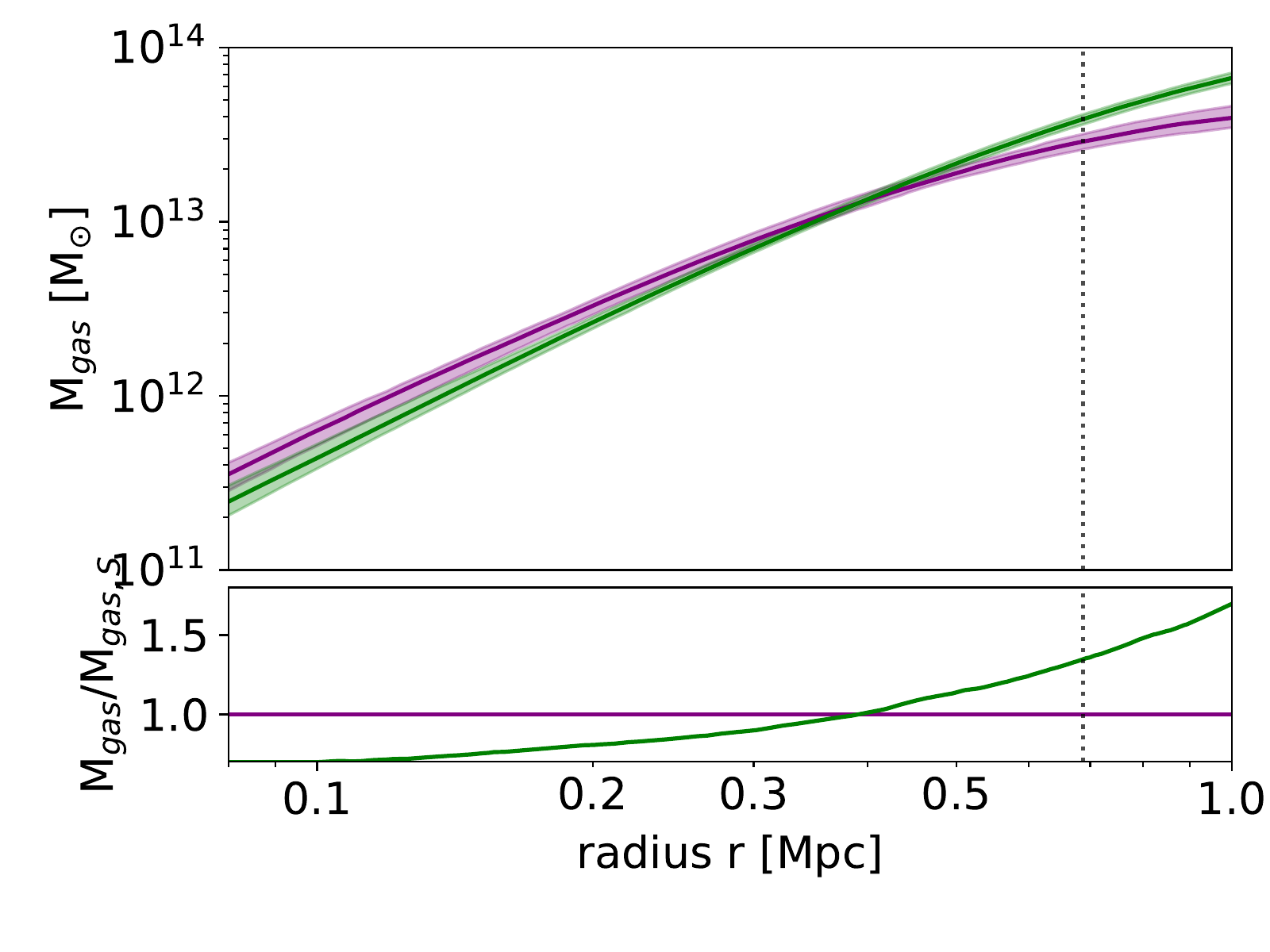} \\
	\includegraphics[trim=0cm 0.0cm 0cm 0cm, width=0.45\textwidth]{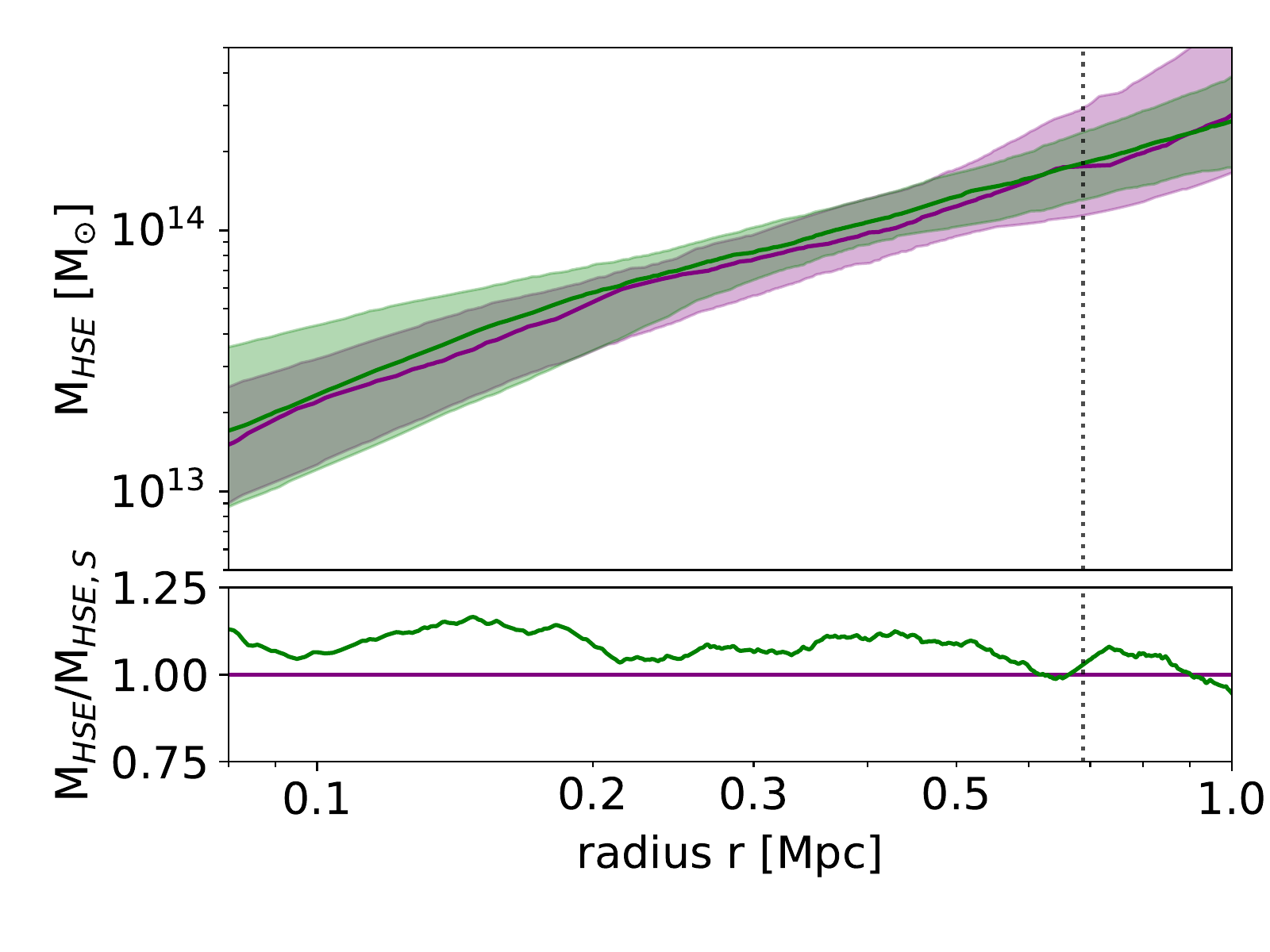} & 
	\includegraphics[trim=0cm 0.0cm 0cm 0cm, width=0.45\textwidth]{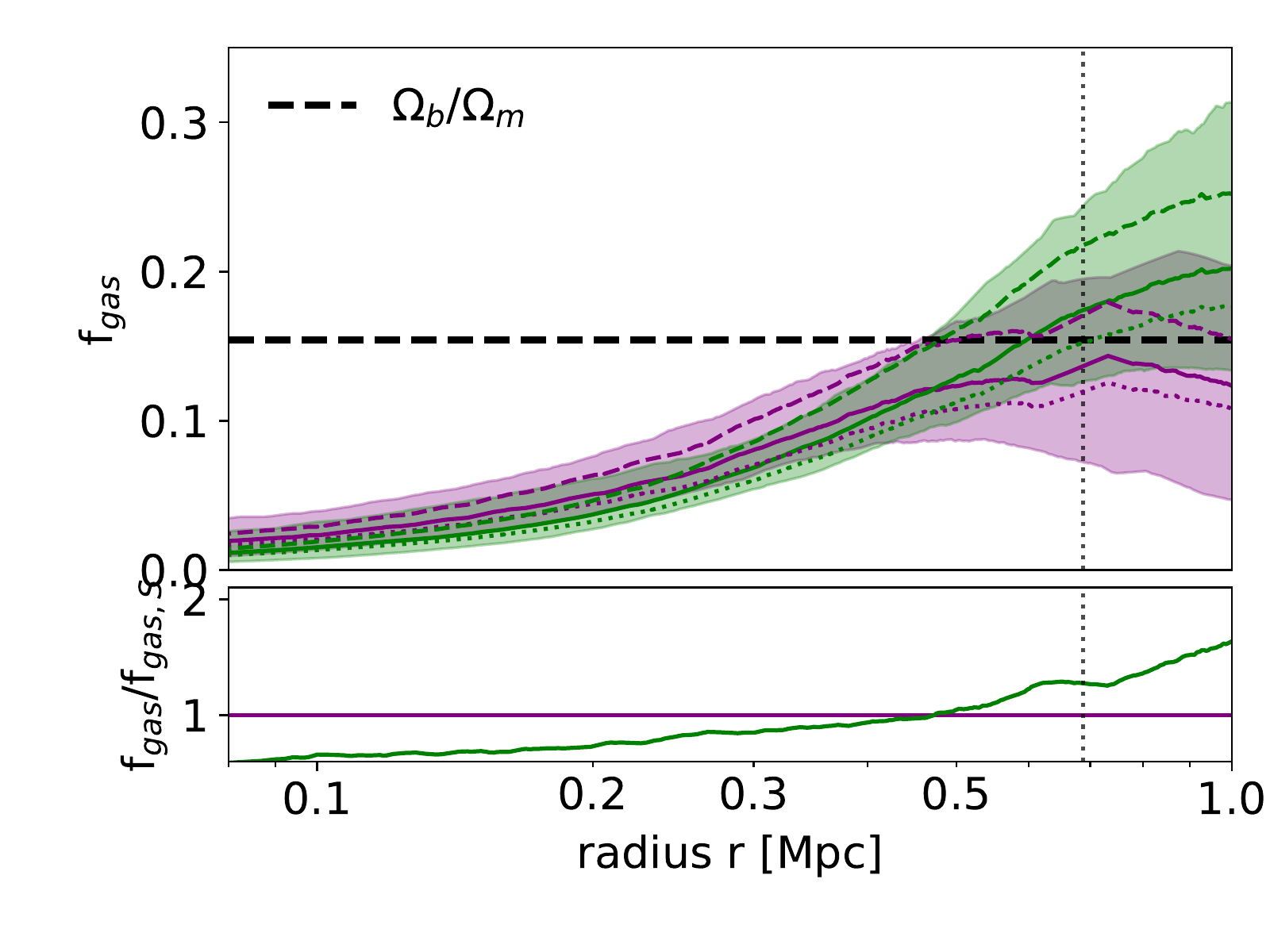}
	\end{tabular}
	
	\caption{\label{fig:thermo_profiles_regions}\footnotesize{Radial physical profiles of XLSSC~102 computed in the northern region (green) and the southern region (purple).
	The legend is the same as in Figure \ref{fig:thermo_profiles_Xp}. The bottom inset panel in each plot shows the ratio of the profiles  computed in the northern and southern regions.
   \textcolor{white}{This sentence is here to overpass a bug in latex.}
    }}

	\end{figure*}


	\subsection{Results and impact of the complex morphology of XLSSC~102}
With the MCMC chains in hand, we construct the thermodynamic and mass profiles using the relations presented in Section \ref{subsubsec:eq}. To do so, we compute the median and the 68\% confidence intervals of the profiles of the derived quantities at each radius. In the following we present 3-D profiles of thermodynamic variables for XLSSC~102 and discuss the impact of the different centre definition and of the overpressure region.

		\subsubsection{Thermodynamic state characterisation}

The thermodynamic and mass 3D profiles of XLSCC102, computed at the X-ray peak position are presented in Figure \ref{fig:thermo_profiles_Xp}. From left to right and top to bottom the quantities are the electron pressure $P_e$, electron density $n_e$, temperature $k_BT$, electron entropy index $K_e$, spherically integrated Compton parameter $Y_{\rm sph}$, gas mass $M_{\rm gas},$ hydrostatic mass $M_{\rm HSE}$, and gas fraction $f_{\rm gas}$ profiles. The gas fraction is computed by assuming a constant hydrostatic mass bias such as $b_{\rm HSE} = 0.2$. 
We display the profiles in the typical radius range where they are constrained by the SZ and X-ray data : from 80 kpc to $\sim 1$ Mpc, which corresponds to half the NIKA2 150GHz beam ($\sim 10$ arcsec) and to the maximum radius at which the SZ and X-ray signal is detected ($\sim 2$ arcmin). 
We note that we expect $r_{500} \sim 0.7$ Mpc from XXL scaling law measurements, which roughly correspond to $r_{200}\sim 1$ Mpc (we verify that with the over-density profile computed from Eq. \ref{eq_icmtool_rdelta}). We thus probe the thermodynamic and mass profiles of XLSSC~102 up to $\sim r_{200}$. However, as we use a parametric model, the shape of the profiles at large radii ($r \gtrsim r_{500}$), where the signal is low, might be predominantly driven by the data in the inner region. 

We can see that we obtain relatively tight constraints, considering the moderate depth of the XXL and NIKA2 data and the redshift and mass of the source. 
The constraints are globally tighter at intermediate scales, where NIKA2 is the most sensitive. Indeed, on small and large scales, the beam and the transfer function filter out the signal. Additionally, the noise correlations and the uncertainties in the zero level boost the error bars on large scales.

As expected, we find that the pressure and density decrease with radius. The profiles are relatively flat in the inner part, not as seen for cool core clusters.
The pressure profile of XLSSC~102 is similar to that of low redshift perturbed clusters from the REXCESS sample \citepalias{2010A&A...517A..92A}. 

The temperature profile reaches about 8 keV at the X-ray peak (at r $\sim$ 80 kpc) and drops by a factor of $\sim4$ towards the outskirts. This steep decrease is larger than expectations from studies of more nearby cluster samples \cite[typically a temperature drop by a factor of $2.5-3$ at $r_{200}$ with respect to the inner regions, see][]{2013A&A...551A..22E,Reiprich2013}. The temperature profile we measure is consistent with that of a morphologicaly disturbed cluster, but not with that of a cool-core. We note that the temperature profile is consistent with the average XMM cluster temperature measured within a projected aperture of 300 kpc (see Table \ref{tab:prop_summary}). 

The entropy index profile of XLSCC102 is flat and the central value is high. We see an excess of entropy in the inner region with respect to the pure gravitational collapse expectation indicated by the black dotted line \citep[see][for its derivation]{2005RvMP...77..207V,2010A&A...511A..85P}. This behaviour is expected for disturbed clusters \citep[e.g.][]{2010A&A...511A..85P,2011ApJ...728...54Z}. 
More surprisingly, we observe that the entropy is lower than the pure gravitational collapse expectation beyond 350 kpc ($\sim 0.5 \times r_{500}$), although the deviation is of low significance. This phenomenon has been observed in un-relaxed clusters \citep[see e.g.][]{2013A&A...551A..22E,2016A&A...595A..42T} and may be due to gas inhomogeneities that bias high the density measurement or to turbulence and non non-thermal pressure support.

As expected, the spherically integrated Compton parameter smoothly increases with radius. Similarly, the gas mass increases with radius and it is nearly consistent with a power law at r $< r_{500}$. The hydrostatic mass profile is relatively steep in the centre and flattens at $r \sim$ 200 kpc. 
These profiles will be used in Section \ref{subsec:scaling_relations} when studying scaling relations. 

Finally, the gas fraction increases with the distance from the centre. The median profile becomes larger than the cosmic gas fraction at $r \gtrsim 500$ kpc, however, the cosmic value is still within the 68\% confidence interval. We note that a high hydrostatic bias would diminish the gas fraction, but an excessively large value would be necessary to avoid exceeding the cosmic fraction. This behaviour is also seen in non-relaxed clusters \citep[see e.g.][]{2013A&A...551A..23E} and, as for the entropy behaviour, could be attributed to gas inhomogeneities biasing high the density measurement or to non thermal pressure support.  

All the recovered physical radial profiles of XLSSC~102 up to $r_{200}$ are in agreement with the cluster being un-relaxed and support the post-merger scenario discussed in Section \ref{sec:Morphology}. Additionally, based on the definitions of Section \ref{subsubsec:eq}, several quantities indicate that the gas density deduced from the X-ray data is too high compared to expectations from the pressure at radii larger than $\sim$ 350 kpc given the expected mass (see e.g. Eqs \ref{eq_icmtool_entropy},  \ref{eq_icmtool_masse_hse} and \ref{eq_icmtool_gas_mass}). This is in apparent contradiction to the presence of an overpressure region identified in Section \ref{sec:Morphology}. As discussed above, several physical reasons could be invoked, and in particular the presence of gas inhomogeneities, which can bias high the recovered density, and are common in merging systems.

		\subsubsection{Impact of the internal structure}
		\label{subsec:internal_structure}

As the profiles are computed assuming spherical symmetry, it can be interesting to compute them in different sectors to link the thermodynamic state and the inner structure \citep[as done in][]{ruppin_first_2018}. Here we split the cluster in two halves to investigate the effect of the overpressure region identified in Section \ref{sec:Morphology}. The two halves are then alternately masked to derive the quantities corresponding to an entire cluster.
To define our two regions, we chose the X-ray peak as the centre, and we divided the cluster perpendicularly to the elongation axis of the X-ray and SZ emission (following a North-West to South-East direction). The overpressure region is thus localised in the northwestern part ('North' in the following), whereas the southeastern part ('South' in the following) is expected to be less perturbed.

The thermodynamic and mass profiles for the two regions are presented in Figure \ref{fig:thermo_profiles_regions}. The order of the plots is the same as in Figure  \ref{fig:thermo_profiles_Xp}. The profiles of the northern and southern regions are shown in green and purple, respectively. The bottom inset panel in each plot shows the ratio of the profiles computed in the northern region to that computed in the southern region. 

We can remark that the pressure and density profiles in the northern region are much flatter. At r $\lesssim$ 250 kpc the pressure profile in the northern region is higher but still compatible with that of the southern region whereas the density is significantly lower. At greater distance, the pressure and density profiles in the North become significantly higher (larger than a factor 2 at r$_{500}$). While the pressure profile slopes in the outskirts are quite similar, that of the density in the northern region is much flatter. 

The temperature and entropy index profiles are higher in the northern region but become compatible with that of the southern region around $r_{500}$, considering the errors. 
The integrated Compton parameter and gas mass profiles behave as the pressure and density profiles, respectively. At $r_{500}$ and $r_{200}$, $Y_{\rm sph}$ is 64 \% and 84\% higher in the northern region, while the density is 35\% and 70\% higher.
These differences are not translated to the hydrostatic mass profiles, which are fully compatible at all radii. Considering the uncertainties, the gas fraction profiles are also consistent. 

The differences observed between the profiles in the two regions strongly support the post merger scenario discussed in Section \ref{sec:Morphology}. The ICM in the northern region present higher temperature, pressure and entropy likely caused by the merger. The density and gas mass profiles in the northern region point towards a redistribution of the gas from the inner part to the outskirts. This may also be the sign of the presence of a clump of gas in the direction of the galaxy density peak. We can note that the shapes of the profiles in the southern region, which is supposedly less affected by the merger, are also compatible with that of a perturbed cluster (especially the presence of a high entropy floor). This indicates that the gas had time to mix after the merger, and supports the post-merger scenario \citep[see e.g.][]{2011ApJ...728...54Z}.
The fact that the hydrostatic mass profiles are equivalent in the two regions derives from the fact that the changes in density and pressure balance each other. However, we can expect the hydrostatic mass bias to be different in the two regions, due to non thermal pressure support and turbulence caused by the merger, and thus the inferred total mass in the two sectors to differ. The impact of the inner structure on the scaling laws will be investigated in Section \ref{subsec:scaling_relations}.

		\subsubsection{Impact of the centre definition}

	\begin{figure*}
	\center
	\begin{tabular}{rr}
	\includegraphics[trim=0cm 2cm 0cm 0cm, width=0.45\textwidth]{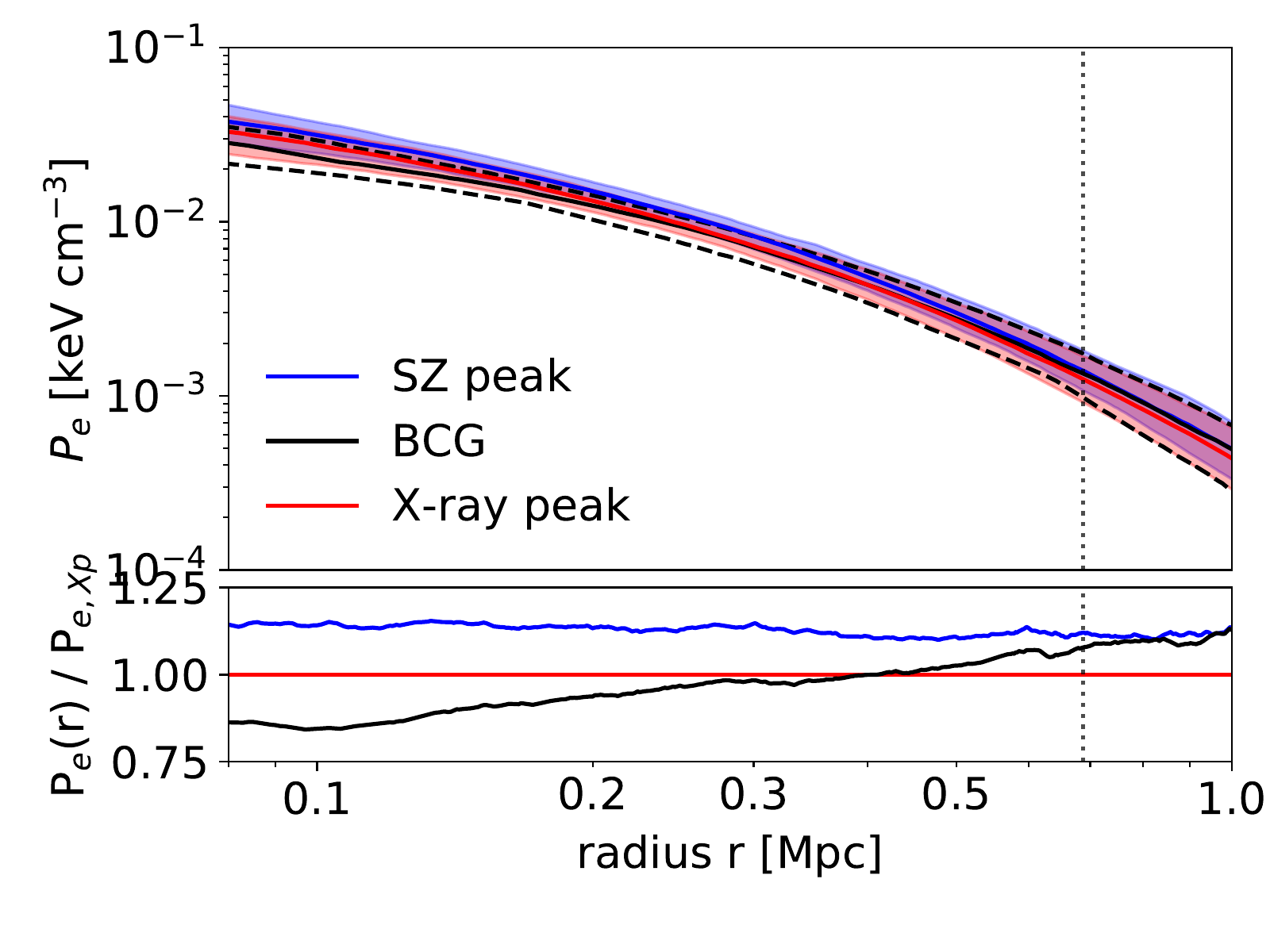} & 
	\includegraphics[trim=0cm 2cm 0cm 0cm, width=0.45\textwidth]{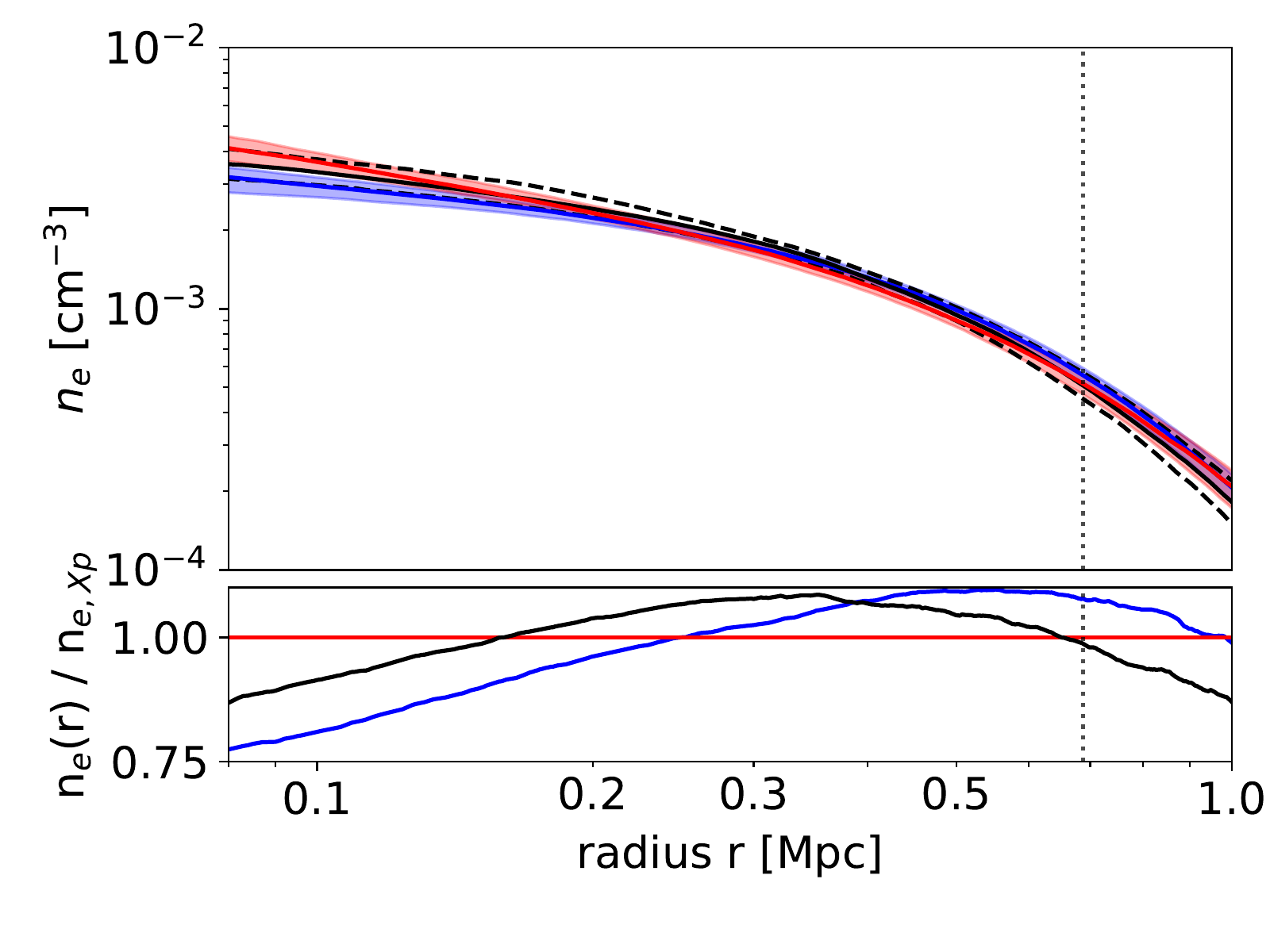} \\
	\includegraphics[trim=0cm 2cm 0cm 0cm, width=0.45\textwidth]{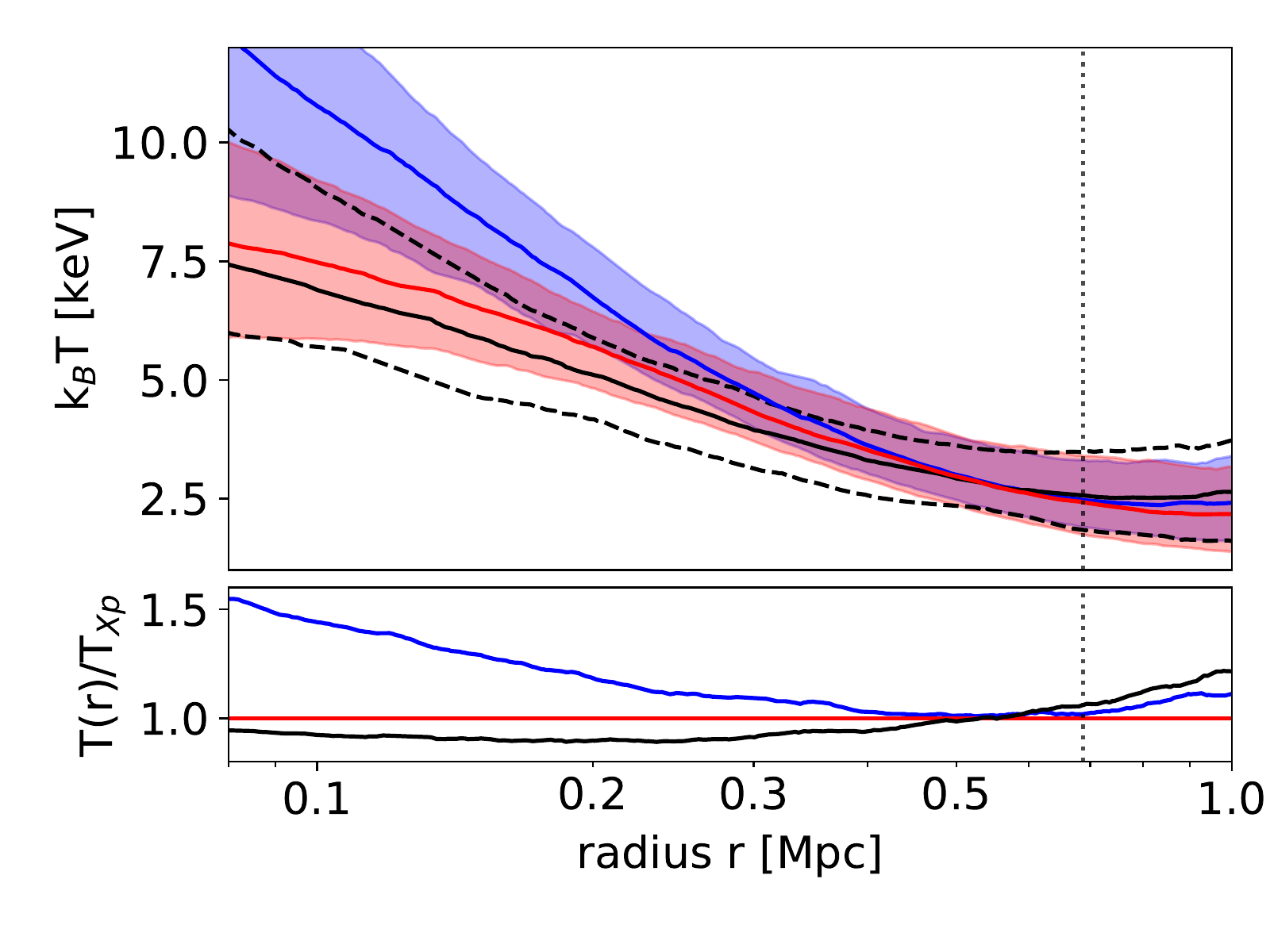} &
	\includegraphics[trim=0cm 2cm 0cm 0cm, width=0.45\textwidth]{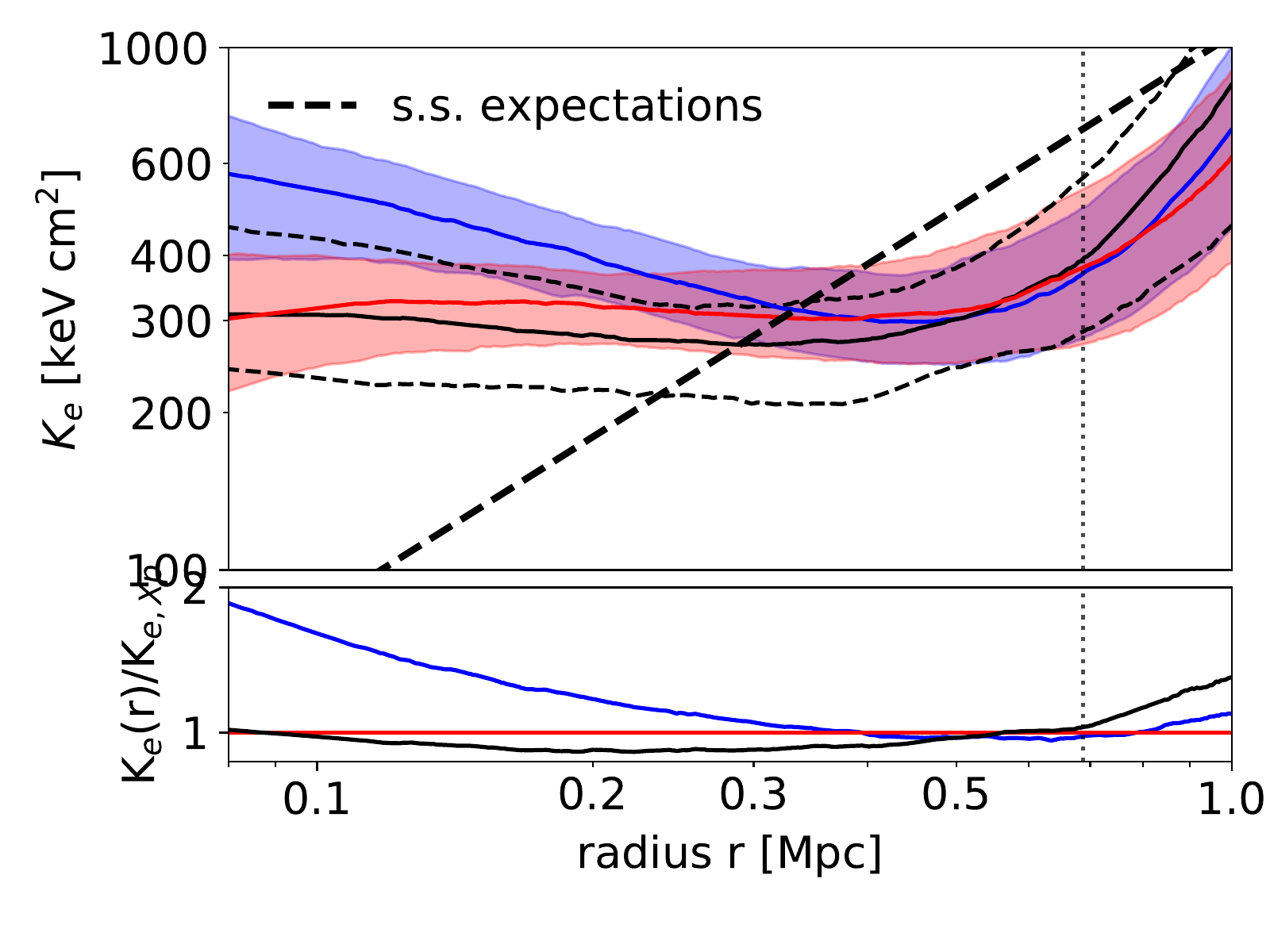} \\
	\includegraphics[trim=0cm 2cm 0cm 0cm, width=0.45\textwidth]{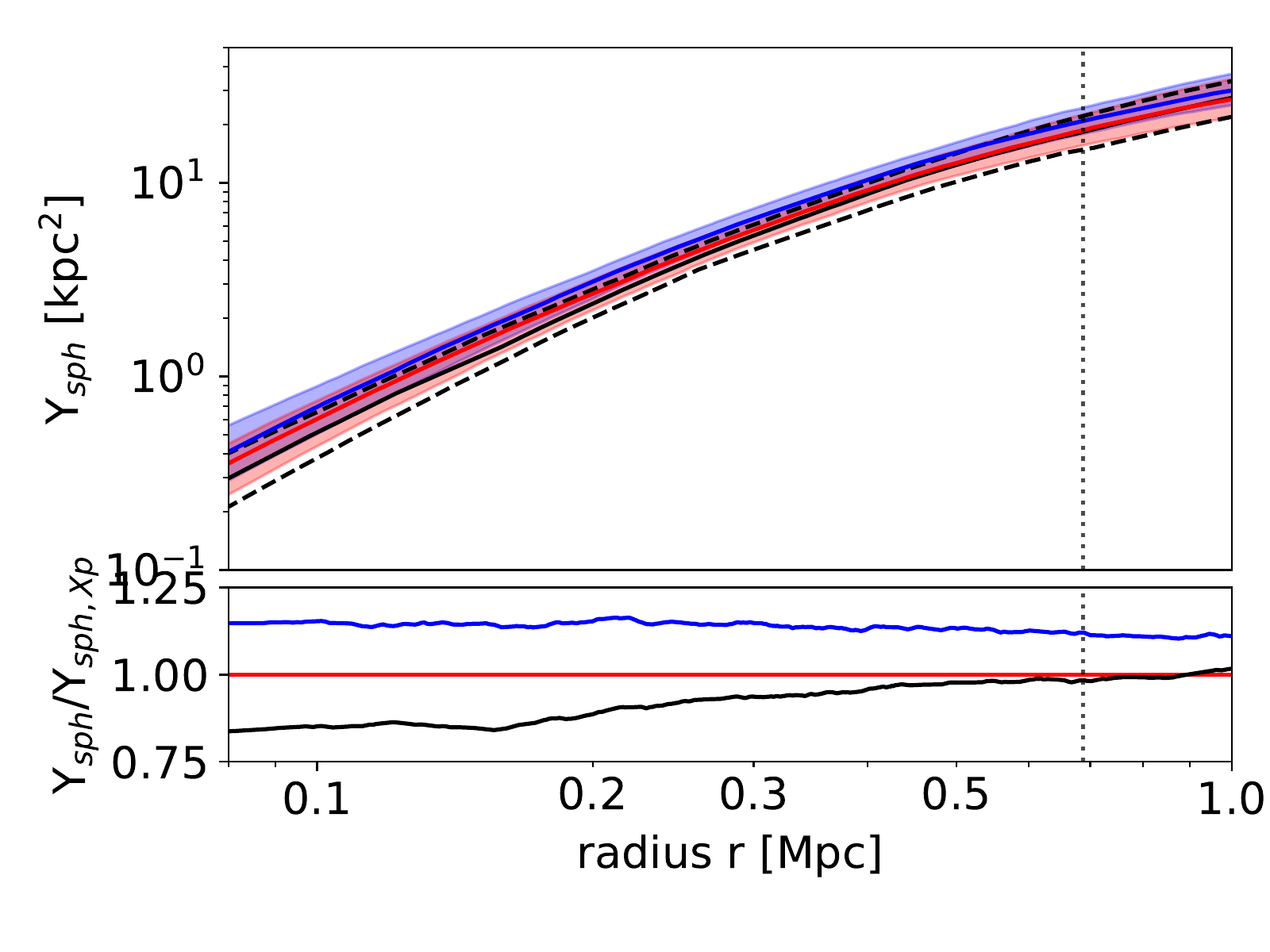} &
	\includegraphics[trim=0cm 2cm 0cm 0cm, width=0.45\textwidth]{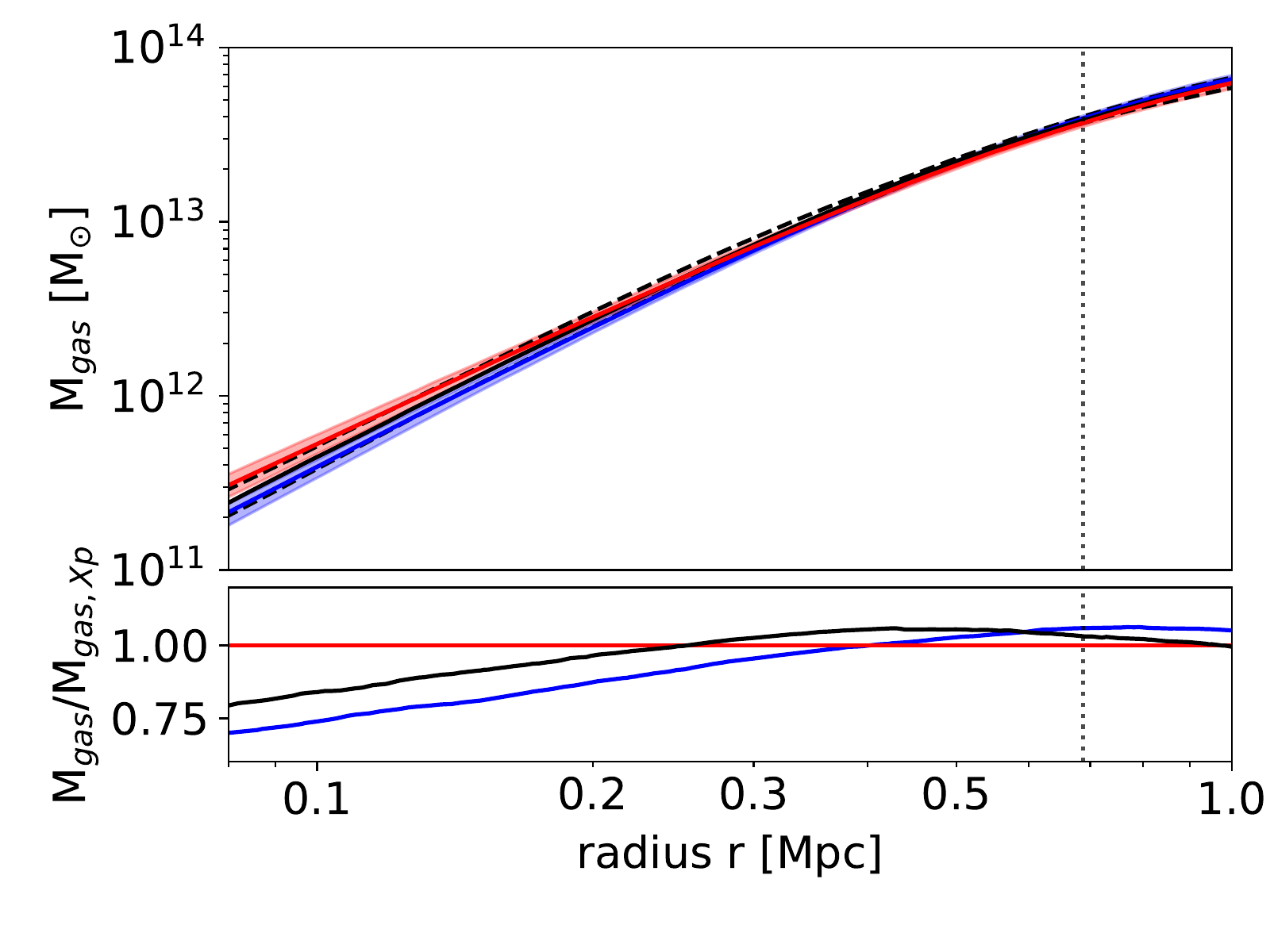} \\
	\includegraphics[trim=0cm 0cm 0cm 0cm, width=0.45\textwidth]{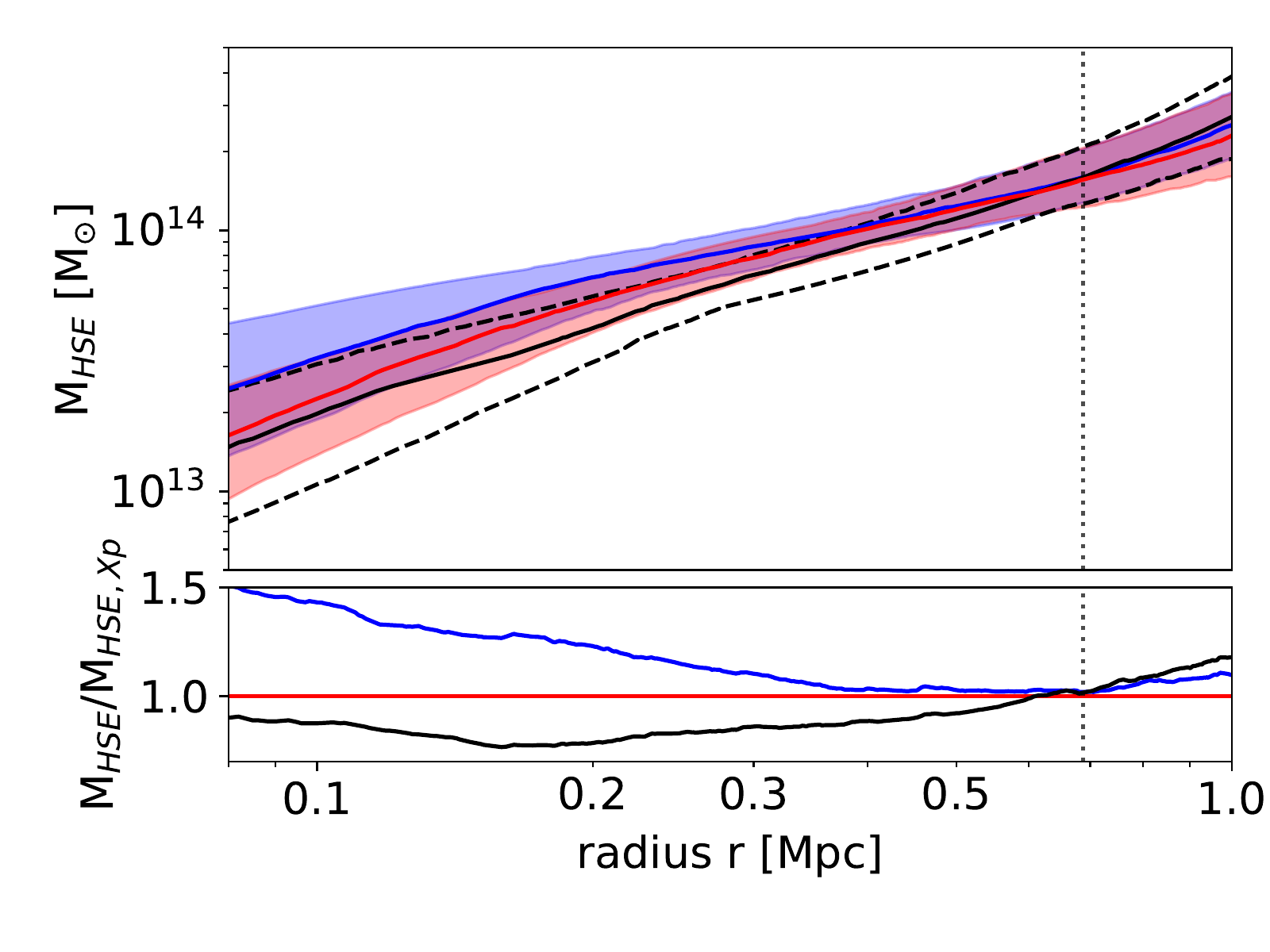} & 
	\includegraphics[trim=0cm 0cm 0cm 0cm, width=0.45\textwidth]{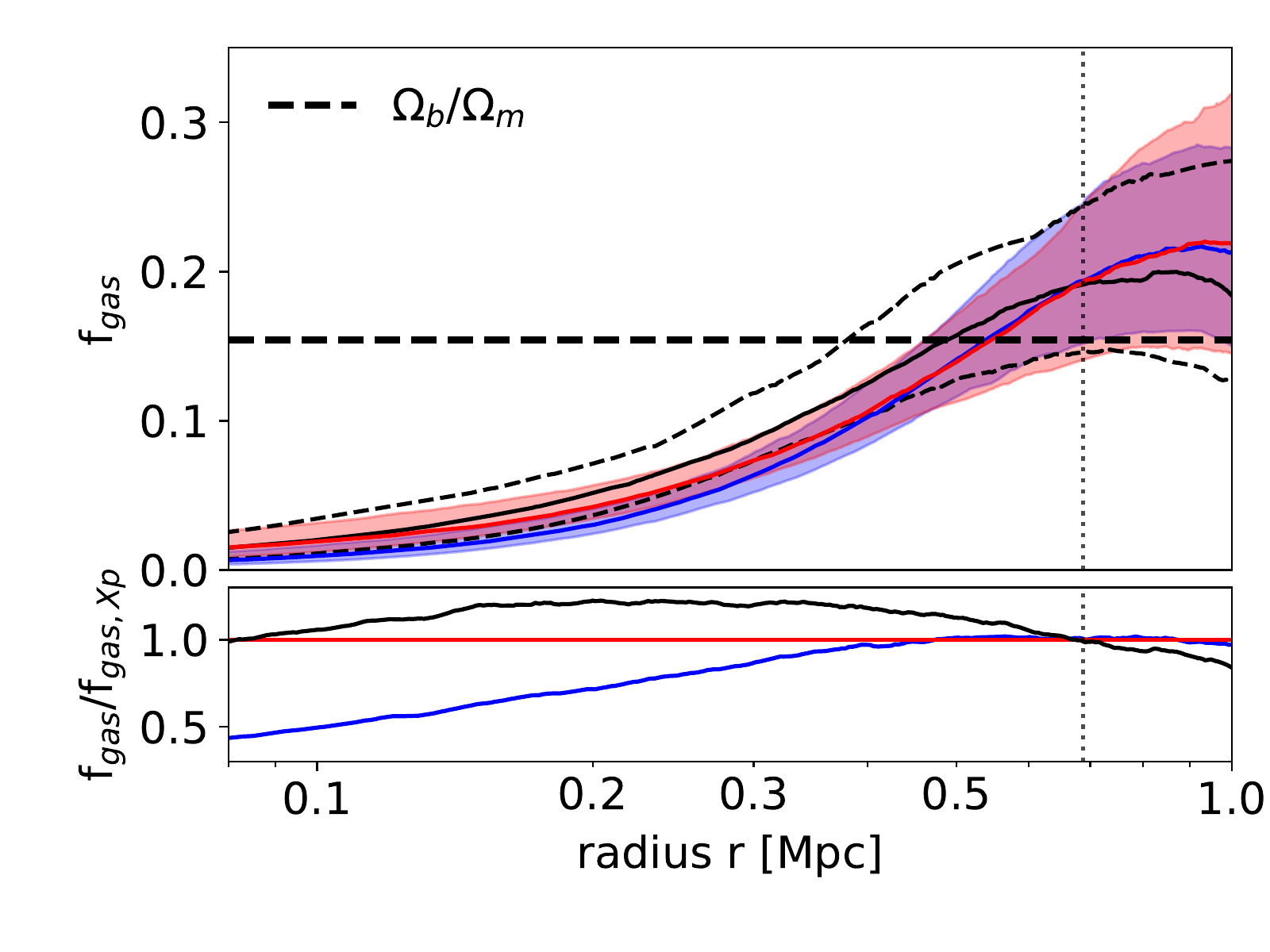}
	\end{tabular}
	\caption{\label{fig:thermo_profiles_centers}\footnotesize{Radial physical profiles of XLSSC~102 computed at the X-ray peak (red), SZ peak (blue) and BCG positions (black). The legend is the same as in Figure \ref{fig:thermo_profiles_Xp}. The bottom inset panel in each plot shows the ratio of the profiles computed at the SZ peak and the BCG position to that computed at the X-ray peak. }}
	\end{figure*}

So far, we used the X-ray peak as our reference position to construct the different profiles. However, as shown in Section \ref{sec:Morphology}, XLSSC~102 presents significant offsets between its different tracers and thus the definition of the cluster centre is not obvious. In the following we investigate the impact of taking the X-ray peak, the SZ peak or the BCG as indicators of the cluster centre on the derived profiles. We did not use the optical density peak as it does not coincide with the gas tracers.

The thermodynamic and mass profiles computed using the three centres are presented in Figure \ref{fig:thermo_profiles_centers}. The order of the plots is the same as in Figure  \ref{fig:thermo_profiles_Xp}. The profiles computed at the X-ray and SZ peaks are shown in red and blue respectively, while that computed at the BCG position is shown by the black lines. 
The bottom inset panel in each plot shows the ratio of the profiles computed at the SZ peak and BCG position to that computed at the X-ray peak position. 

All profiles are compatible with each other considering the error bars. However, as the profiles are derived from the same data the differences between them arise from purely systematic effects. 

We remark that in the inner part ($r \lesssim 200$ kpc), as expected, the pressure and integrated Compton parameter profiles computed at the SZ peak are higher than at other positions, while the density and gas mass profiles are higher when computed at the X-ray peak. Consistent with the merger scenario, the temperature and entropy near the SZ peak are much higher than around the X-ray peak and the BCG. This translates to the hydrostatic mass being higher and the gas fraction lower around the SZ peak.

At larger radii ($r \gtrsim 300$ kpc) the profiles are in better agreement but some differences remain. In particular, at $r = 0.7$ Mpc ($\sim r_{500}$) there are still differences of 11 \% between $Y_{\rm sph}$ computed at the SZ peak and at the X-ray peak or BCG position. Similarly the differences are 6 and 3 \% between $M_{\rm gas}$ computed at the SZ or X-ray peaks or BCG position, respectively. By comparison, the differences in hydrostatic masses are only $\sim$ 2\% at the same radius, but increase to 10 and 17 \% at $r = 1$ Mpc ($\sim r_{200}$). The impact of the different cluster centre choice on the scaling laws will be investigated in Section \ref{subsec:scaling_relations}.

\section{Global properties}
\label{sec:global_prop}

In the previous sections, we studied the morphology and thermodynamic profiles of XLSSC~102. Here we derive the integrated properties of the cluster and compare them to scaling relations calibrated from local samples.
As discussed in Section \ref{sec:intro}, the properties of high redshift low-mass clusters are poorly known. In this Section we thus test if the integrated quantities of XLSSC~102 deviate from what is expected for low redshift more massive clusters.
Moreover, we can see in Sections \ref{sec:Morphology} and \ref{sec:thermo_profiles} that XLSSC~102 appears to be in a post-merger state and highly disturbed. We thus derive different estimates of its mass under different assumptions, and compare them to what is expected from the XXL scaling relations (see Table \ref{tab:prop_summary}). 

	\subsection{Location of XLSSC~102 on $Y_{\rm SZ}$ - Mass scaling relation}
	\label{subsec:scaling_relations}

	\begin{figure*}[ht]
	\centering
	\includegraphics[width=0.7\textwidth]{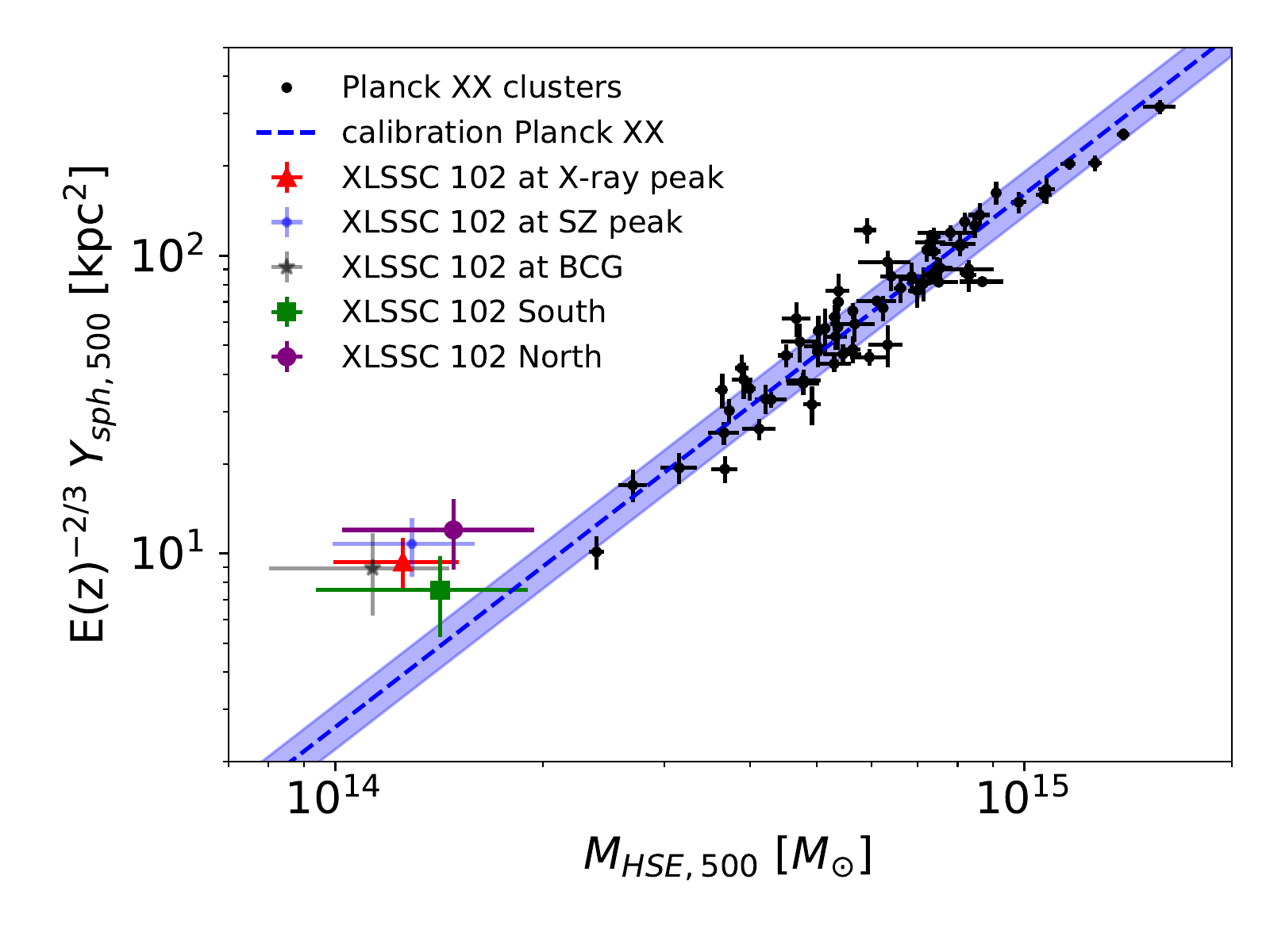}
	\caption{\footnotesize{Location of XLSSC~102 on the scaling relation between the spherically integrated Compton parameter $Y_{500}$ and the hydrostatic mass $M_{{\rm HSE}, 500}$ (i.e the total mass assuming $b_{\rm HSE}=0$). The different markers indicate the values from the combined SZ+X-ray analysis of XLSSC~102 evaluated on the entire cluster at different centres and from the southern and northern regions at the X-ray centre, as indicated in the legend. The values and error bars represent the median and the 68\% c.i. The black data points show the \textit{Planck} calibration cluster sample corrected from Malmquist bias. The blue line and region indicate the fitted relation and its intrinsic scatter \citepalias{Planck2014XX}.}}
		\label{fig:planck_scaling}
	\end{figure*}

	\begin{table}
	\begin{center}
	\caption{Global quantities evaluated at a radius corresponding to spherical overdensity $\Delta$ = 500, computed by assuming no hydrostatic mass bias (upper block) or a value $b_{\rm HSE}=0.2$ (lower block).}
	\label{tab:SO_quantities}

	\begin{tabular}{r | rrrrr}
	\hline
			      & $r_{500}$     &  $M_{500}$                         &  $M_{{\rm gas},500}$                        &   $Y_{{\rm sph},500}$ &   $Y_{{\rm X},500}$  \\
				& Mpc		    &  10$^{14}$ M$_{\odot}$ &   10$^{13}$ M$_{\odot}$	   &   kpc$^2$        &   10$^{14}$ keV M$_{\odot}$  \\

	\hline
\multicolumn{6}{c}{$b_{\rm HSE}=0.0$}  \\
\hline
           X-ray  		&	0.53$^{+ 0.06 }_{- 0.04 }$  &  1.25$^{+ 0.47 }_{- 0.26 }$ &  2.32$^{+ 0.49 }_{- 0.30 }$ & 13.51 $^{+ 4.43 }_{- 2.71 }$ & 0.87 $^{+ 0.30 }_{- 0.18 }$ \\
           SZ  			&  	0.53$^{+ 0.06 }_{- 0.05 }$  &  1.29$^{+ 0.42 }_{- 0.30 }$ &  2.48$^{+ 0.49 }_{- 0.42 }$ & 15.51 $^{+ 4.37 }_{- 3.46 }$ & 0.99 $^{+ 0.29 }_{- 0.22 }$ \\
           BCG			&   0.51$^{+ 0.07 }_{- 0.06 }$  &  1.13$^{+ 0.54 }_{- 0.33 }$ &  2.31$^{+ 0.66 }_{- 0.41 }$ & 12.86 $^{+ 4.23 }_{- 3.97 }$ & 0.82 $^{+ 0.31 }_{- 0.24 }$ \\
           South 		&   0.54$^{+ 0.16 }_{- 0.07 }$  &  1.42$^{+ 0.94 }_{- 0.48 }$ &  2.10$^{+ 0.88 }_{- 0.41 }$ & 10.85 $^{+ 3.47 }_{- 3.31 }$ & 0.68 $^{+ 0.25 }_{- 0.22 }$ \\
           North		&   0.56$^{+ 0.07 }_{- 0.07 }$  &  1.48$^{+ 0.61 }_{- 0.46 }$ &  2.66$^{+ 0.68 }_{- 0.54 }$ & 17.29 $^{+ 5.53 }_{- 4.61 }$ & 1.11 $^{+ 0.43 }_{- 0.31 }$ \\
	\hline

\multicolumn{6}{c}{$b_{\rm HSE}=0.2$}\\
\hline 
	       X-ray  		&	0.58$^{+ 0.08 }_{- 0.04 }$  &  1.69$^{+ 0.72 }_{- 0.34 }$ &  2.80$^{+ 0.66 }_{- 0.37 }$ & $15.43 ^{+ 5.48 }_{- 3.20 }$ & 0.96 $^{+ 0.38 }_{- 0.21 }$ \\
           SZ  			&  	0.59$^{+ 0.07 }_{- 0.05 }$  &  1.75$^{+ 0.67 }_{- 0.43 }$ &  3.00$^{+ 0.63 }_{- 0.52 }$ & $17.57 ^{+ 5.16 }_{- 4.02 }$ & 1.10 $^{+ 0.35 }_{- 0.25 }$ \\
           BCG			&    0.58$^{+ 0.09 }_{- 0.07 }$  &  1.62$^{+ 0.74 }_{- 0.47 }$ &  2.84$^{+ 0.80 }_{- 0.49 }$ & $14.88 ^{+ 4.65 }_{- 4.25 }$ & 0.95 $^{+ 0.33 }_{- 0.27 }$ \\
           South 		&    0.59$^{+ 0.18 }_{- 0.07 }$  &  1.92$^{+ 1.20 }_{- 0.66 }$ &  2.37$^{+ 0.86 }_{- 0.48 }$ & $11.99 ^{+ 3.53 }_{- 3.73 }$ & 0.70 $^{+ 0.31 }_{- 0.22 }$ \\
           North		&    0.62$^{+ 0.09 }_{- 0.08 }$  &  2.04$^{+ 0.98 }_{- 0.65 }$ &  3.21$^{+ 0.82 }_{- 0.61 }$ & $19.60 ^{+ 7.77 }_{- 5.32 }$ & 1.24 $^{+ 0.54 }_{- 0.34 }$ \\
           
	\end{tabular}
	\end{center}
	\end{table}

The SZ flux is known to closely track the total mass \citep[as $ Y_{{\rm SZ},500} \propto M_{500}^{5/3}$ according to self - similar expectations,][]{2013SSRv..177..247G}, and is often used as a low scatter mass proxy \citep[see e.g.][]{2010A&A...517A..92A, Planck2014XX}. Here we compare the integrated quantities $M_{500}$ and $Y_{{\rm SZ},500}$ of XLSSC~102 measured via the SZ+X-ray combination to the scaling relation from \citetalias{Planck2014XX}. This relation was derived from a sample of $z<0.45$ clusters and was used as calibration to estimate the mass of their cosmological cluster sample.

In order to measure the spherically integrated SZ flux $Y_{{\rm SZ},500}$ we first constructed the integrated Compton parameter profile, using the procedure described in Section \ref{sec:thermo_profiles} and the relation from Eq. \ref{eq_icmtool_ysph}. Using the mass profile presented in Figure \ref{fig:thermo_profiles_Xp}, we can derive an overdensity profile (following Eq. \ref{eq_icmtool_rdelta}). This profile is then used to determine the value of $r_{500}$ for each point of the MCMC chain. The values of $Y_{{\rm SZ},500}$ and $M_{500}$ are then also computed for each point and their distributions then encode the uncertainties in the profiles and the $r_{500}$ value. The values of $M_{500}$ and $Y_{{\rm SZ},500}$ are highly correlated, because they depend on the pressure profile and because the value of $r_{500}$ depends on the mass profile. Increasing $r_{500}$ leads to increasing $Y_{{\rm SZ},500}$ for the same $Y_{\rm SZ}(r)$ profile, as well as increasing $M_{500}$ for the same $M(r)$ profile.
The values of the different parameters measured at $r_{500}$ are summarised in Table \ref{tab:SO_quantities}.

Figure \ref{fig:planck_scaling} shows the location of XLSSC~102 on the $Y_{{\rm SZ},500} - M_{{\rm HSE},500} $ relation from \citetalias{Planck2014XX}.
The different coloured markers indicate the values from the combined SZ+X-ray analysis of XLSSC~102 evaluated using different centres and from the southern and northern regions at the X-ray centre, as indicated in the legend. The values and error bars represent the median and the 68\% c.i. 
The black points show the individual measurements of \textit{Planck} clusters corrected from the Malmquist bias. The blue line and region show the calibration adopted and the intrinsic scatter. 
The \textit{Planck} masses are derived from X-ray measurements calibrated using the $Y_{{\rm X},500} - M_{{\rm HSE},500}$ relation from \citetalias{2010A&A...517A..92A}, measured using 20 local relaxed clusters. For a meaningful comparison we used hydrostatic mass (or total mass assuming no HSE bias correction). 

As expected, XLSSC~102 has lower mass and integrated Compton parameter than clusters from the \textit{Planck} sample. 
Despite its low mass and high redshift, XLSCC102 does not strongly deviate from the scaling relation. 
The quantities measured at the different centres are all compatible at 68\% or less considering the error ellipses (not shown here for clarity), even if that computed at the SZ peak presents a slightly higher value of $Y_{{\rm SZ}, 500}$. 
These three data points lie above the \textit{Planck} scaling relation and are marginally consistent with it at 95\%, considering the scatter and error ellipses. 
The values of $Y_{{\rm sph},500}$ for the northern and southern halves of the cluster lie respectively above and below the value for the entire cluster and are compatible with each other at 95\%. The data point corresponding to the southern region is closer to the \textit{Planck} relation. This is likely due to the merging event and the boost of pressure discussed in Sections \ref{sec:Morphology} and \ref{sec:thermo_profiles} \citep[also found in a lower redshift massive cluster by][when masking or not an overpressure region]{ruppin_first_2018}. The median mass values are higher when the cluster is split in halves. This is an effect of the widening and skewing of the $M_{{\rm HSE},500}$ distribution towards higher values when using less data, due to increasing uncertainties in both the mass profile and the scaled radius and the correlation between them. We note that the error bars on the measured mass are large.

While no strong conclusions can be drawn from only one cluster, our results indicate that the internal structure and the choice of the cluster centres may have an impact on the scaling relation scatter. The merging event induces a change in the shape of the pressure profile that enhances the SZ flux while having a moderate impact on the inferred mass.
The fact that XLSSC~102 seems to be offset from the \textit{Planck} relation, with excess SZ flux for its mass (albeit still compatible at 95\%), is unlikely to be due to a change in the relation with mass or redshift, as that is expected to cause a reduced SZ effect \citep[see e.g.][]{2015MNRAS.450.3649S, 2017MNRAS.466.4442L}. However, it could be due to an increase of the scatter at lower mass or higher redshift or both \citep{2017MNRAS.466.4442L}. 
Deeper observations of XLSSC~102 and other systems in the same mass and redshift ranges would be necessary to reduce the error bars and measure the scatter of the relation.

	\subsection{Calibration of the $Y_{\rm SZ}$ - $Y_{\rm X}$ relation}
	\label{subsec:yy_scaling}

		\begin{figure*}
		\includegraphics[width=0.5\textwidth]{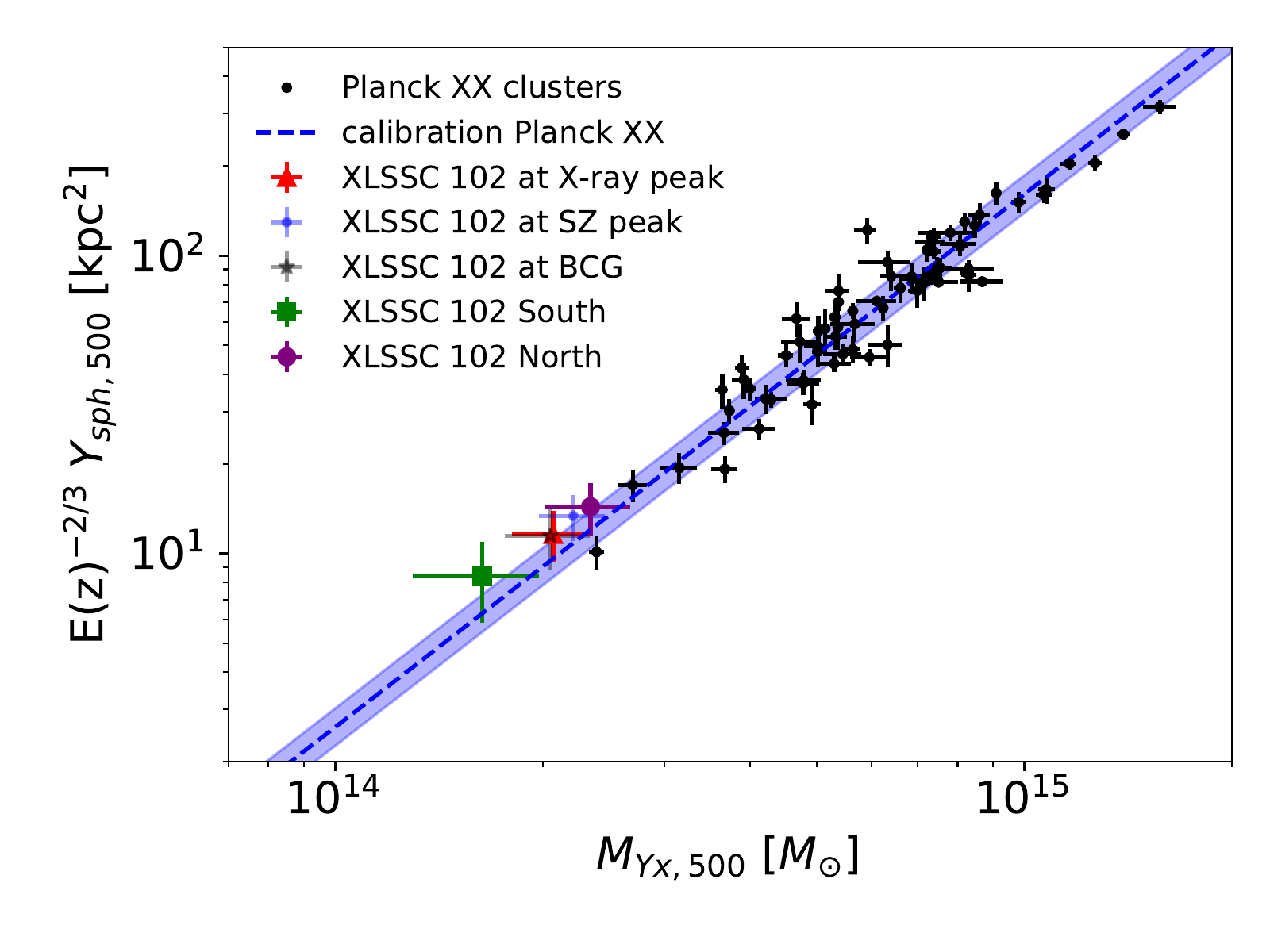}
		  \includegraphics[width=0.5\textwidth]{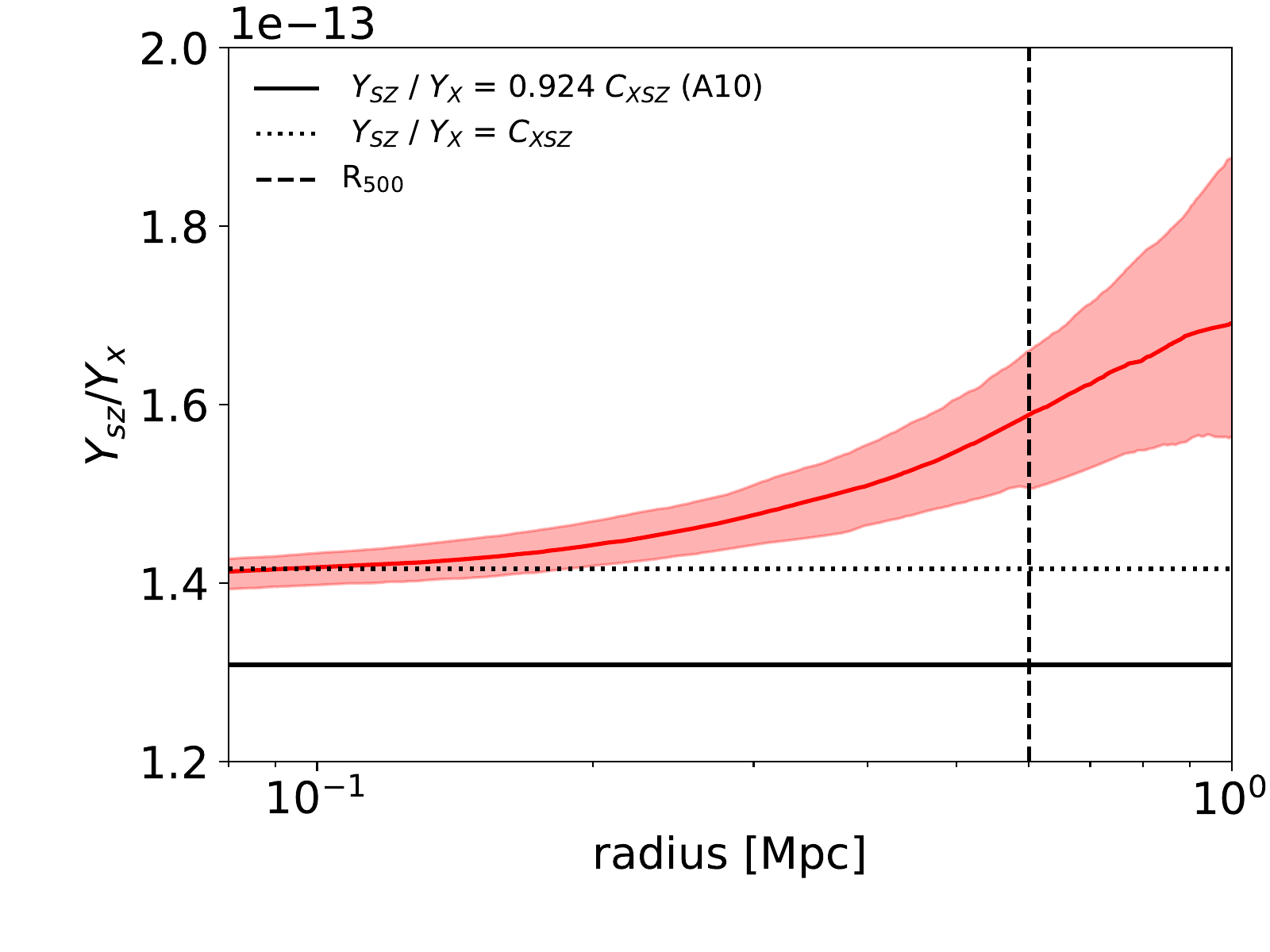}
		 \caption{\label{fig:yy_calib} 
		 \footnotesize{
\textit{Left:} Location of XLSSC~102 on the scaling relation between the spherically integrated Compton parameter $Y_{500}$ and the mass proxy $M_{Y_{\rm X}, 500}$. The legend is the same as in Figure \ref{fig:planck_scaling}. Masses are estimated via the $Y_{\rm X}$ profile and the $Y_{{\rm X},500} - M_{Y_{\rm X}, 500}$ from \citetalias{2010A&A...517A..92A}. As explained in \citetalias{Planck2014XX}, the scaling relations between $Y_{500}$ and $M_{{\rm HSE}, 500}$ or $M_{Y_{\rm X}, 500}$ are the same and only the uncertainties and intrinsic scatter change.
\textit{Right:} Ratio of the spherically integrated Compton parameter $Y_{\rm SZ}$ to its X-ray analogue $Y_{\rm X}$, computed using the X-ray peak. Here $Y_{\rm X}$ is computed using the temperature estimated from the combination of SZ and X-ray data. The solid black line shows the calibration from \citetalias{2010A&A...517A..92A} and the dotted line indicates the value for an isothermal cluster. The dashed black line marks 0.6 Mpc, which is close to the $r_{500}$ values found in Section \ref{subsec:scaling_relations}.}}
		\end{figure*}

Since XLSSC~102 is highly perturbed, our mass estimate may be affected by gas and temperature inhomogeneities (as $M_{\rm HSE}\propto 1/n_e$ , see Eq. \ref{eq_icmtool_masse_hse}). However, the gas mass should be less affected by such effects, as it scales with the integral of the density (see Eq. \ref{eq_icmtool_gas_mass}).
The X-ray analogue of the integrated Compton parameter, $Y_{\rm X}$, defined as $Y_{\rm X} = M_{\rm gas} \times T$ \citep{2006ApJ...650..128K} could thus be more robust than the HSE mass as a mass proxy. 

In order to test the impact of our mass derivation on Figure \ref{fig:planck_scaling}, we thus followed \citetalias{Planck2014XX} and used the $Y_{\rm X} - M$ relation from \citetalias{2010A&A...517A..92A} to estimate masses from our measured integrated Compton parameter profile.
As an X-ray spectroscopic temperature profile is not available for our cluster we used the temperature profile derived from the combination of SZ + X-ray data. We constructed a $Y_{\rm X}$ profile following : 

	\begin{equation}
		Y_{\rm X}(R) = M_{\rm gas}(R) \frac{1}{4/3 \pi R^3} \int_0^R 4 \pi T(r) r^2 dr,
		\label{eq_icmtool_yx}
	\end{equation}

Then, we estimated the hydrostatic mass by iterating about the Y$_{{\rm X}, 500}$ - M$_{Y_{\rm X}, 500}$ scaling relation of \citetalias{2010A&A...517A..92A} (Eq. 2) until convergence. The corresponding value of $R_{{\rm HSE}, 500}$ was used to measure the integrated Compton parameter $Y_{{\rm SZ}, 500}$, as explained in Section \ref{subsec:scaling_relations}.
The new data points are compared to the $Y_{{\rm SZ},500} - M_{{\rm HSE},500} $ relation from \citetalias{Planck2014XX} in the left panel of Figure \ref{fig:yy_calib}. The legend is the same as in Figure \ref{fig:planck_scaling}. As explained in \citetalias{Planck2014XX}, the scaling relation between $Y_{500}$ and $M_{{\rm HSE}, 500}$ is the same as the one between $Y_{500}$ and $M_{Y_{\rm X}, 500}$ albeit with increased uncertainties and intrinsic scatter.

We can see that our data points are much closer to the \textit{Planck} relation when the mass is estimated from $Y_{\rm X}$ and are fully contained within the intrinsic scatter. 
The mass and SZ flux values measured using the northern and southern parts of the cluster are respectively higher and lower than that measured on the whole cluster but each point lies near the scaling law. The values measured based on the X-ray peak and the BCG are the same, whereas that measured using the SZ peak has slightly higher mass and SZ flux. All measurements based on $Y_{\rm X}$ are compatible with each other considering the error bars. However, the mass measurements based on $Y_{\rm X}$ are barely compatible with those from $M_{{\rm HSE}, 500}$, except for the southern region of XLSSC~102.
Our results indicate that the slight tension seen in Figure \ref{fig:planck_scaling} between the values for XLSSC~102 and those of more massive lower redshift clusters is likely due to systematics caused by the merging event and affecting the hydrostatic mass reconstruction.
We note that the error bars are much lower than those of the direct measurements from Section \ref{subsec:scaling_relations}, however, we do not account for the scatter between $M_{Y_{\rm X}, 500}$ and Y$_{{\rm X}, 500}$, which is poorly known for our type of cluster.

As a complementary analysis, we also directly compare the ratio of the spherically integrated Compton parameter $Y_{\rm SZ}$ to its X-ray analogue $Y_{\rm X}$, derived from the combination of X-ray and SZ data. We note however that these two measurements are not totally independent, as $Y_{\rm X}$ depends on the SZ measurements via the temperature.
Following self-similar predictions, we expect $Y_{{\rm SZ}, 500}/Y_{{\rm X}, 500} \propto (M_{Y_{\rm SZ}, 500}/M_{Y_{\rm X}, 500})^{5/3}$ and thus, a ratio constant with mass and no redshift evolution.
The ratio $Y_{\rm SZ}/Y_{\rm X}$ is indeed related to the shape of the temperature and density profiles as : 
	\begin{equation}
	\frac{Y_{\rm SZ}}{Y_{{\rm X}}} (R) = C_{\rm XSZ} \frac{<T_e  n_e>_V}{<n_e>_V} \frac{1}{<T_e>_V} ,
	\label{eq:yy_relation}
	\end{equation}
with $C_{\rm XSZ} = \frac{\sigma_{\rm T}}{ m_e c^2}\frac{1}{\mu_e m_p}  = 1.416 \times 10^{-13}$ kpc$^2$ M$_{\odot}^{-1}$ keV$^{-1}$ and the bracket denoting the volume average in a sphere of radius $R$. For an isothermal cluster, the relation is $\frac{Y_{\rm SZ, 500}}{Y_{{\rm X},500}} = C_{\rm XSZ}$.

Previous studies, using spectroscopically measured temperatures to compute $Y_X$, confirmed the lack of redshift evolution \citep[see e.g.][]{2011ApJ...738...48A}, but found a ratio between $Y_{{\rm SZ}, 500}$ and $Y_{{\rm X}, 500}$ lower than $C_{\rm XSZ}$. This was attributed to the X-ray temperature estimated value being higher than the 'gas mass weighted' temperature, that is, the quantity $\frac{<T_e  n_e>_V}{<n_e>_V}$. In particular, \citetalias{2010A&A...517A..92A} measured $Y_{\rm SZ, 500} =(0.924 \pm 0.004) \cdot Y_{{\rm X},500} \cdot C_{\rm XSZ}$.

As we have computed the profiles of $Y_{\rm SZ}$ and $Y_{\rm X}$ we can calibrate their relation as a function of radius.
The $Y_{\rm SZ}$/$Y_{\rm X}$ ratio is shown in the right part of Figure \ref{fig:yy_calib}, for the profile computed at the X-ray peak. The ratios derived at the other positions or by splitting the cluster in two halves are not shown here, but are consistent.
The isothermal value is indicated by the dotted line and the relation found by \citetalias{2010A&A...517A..92A} is shown by the solid black line. 
We can see that the latter lies below our profile at all sampled radii. This is unlikely to be caused by the overpressure region because the ratio in the two halves of the cluster are compatible. However, the offset may be related to the fact that we do not use a spectroscopically derived temperature profile. The temperatures derived from X-ray and SZ combination or from X-ray spectroscopy may be affected by systematics from the instruments \citep{Mahdavi2013} and from the gas structure \citep{2006ApJ...640..691V, Rasia2014}, leading to differences of the order 10 \% \citep[see e.g.][]{2017A&A...606A..64A, 2017ApJ...838...86R}.

The $Y_{\rm SZ}$/$Y_{\rm X}$ ratio we measure is close to the isothermal value near the centre and increases with the radius, indicating that the temperature computed from the combination of X-ray and SZ data is higher when weighted by the gas mass. 
At R = 0.6 Mpc (which is close to the $r_{500}$ values found in Section \ref{subsec:mass_compare}), our measured ratio is $1.21_{-0.06}^{+0.05}$ times larger than expected from the 
\citetalias{2010A&A...517A..92A} relation. Our value of $Y_{\rm X, 500}$ being lower, this translates to about a 12\% shift in our estimated $M_{Y_{\rm X}, 500}$ with respect to that calibrated with the \citetalias{2010A&A...517A..92A} relation. This might explain the small shift of our data points with respect to the $M_{Y_{\rm X }, 500}$ and $Y_{\rm X, 500}$ scaling relation shown in the left part of Figure \ref{fig:yy_calib}.

	\subsection{Mass estimations comparison}
	\label{subsec:mass_compare}

		\begin{figure}
		  \includegraphics[width=0.5\textwidth]{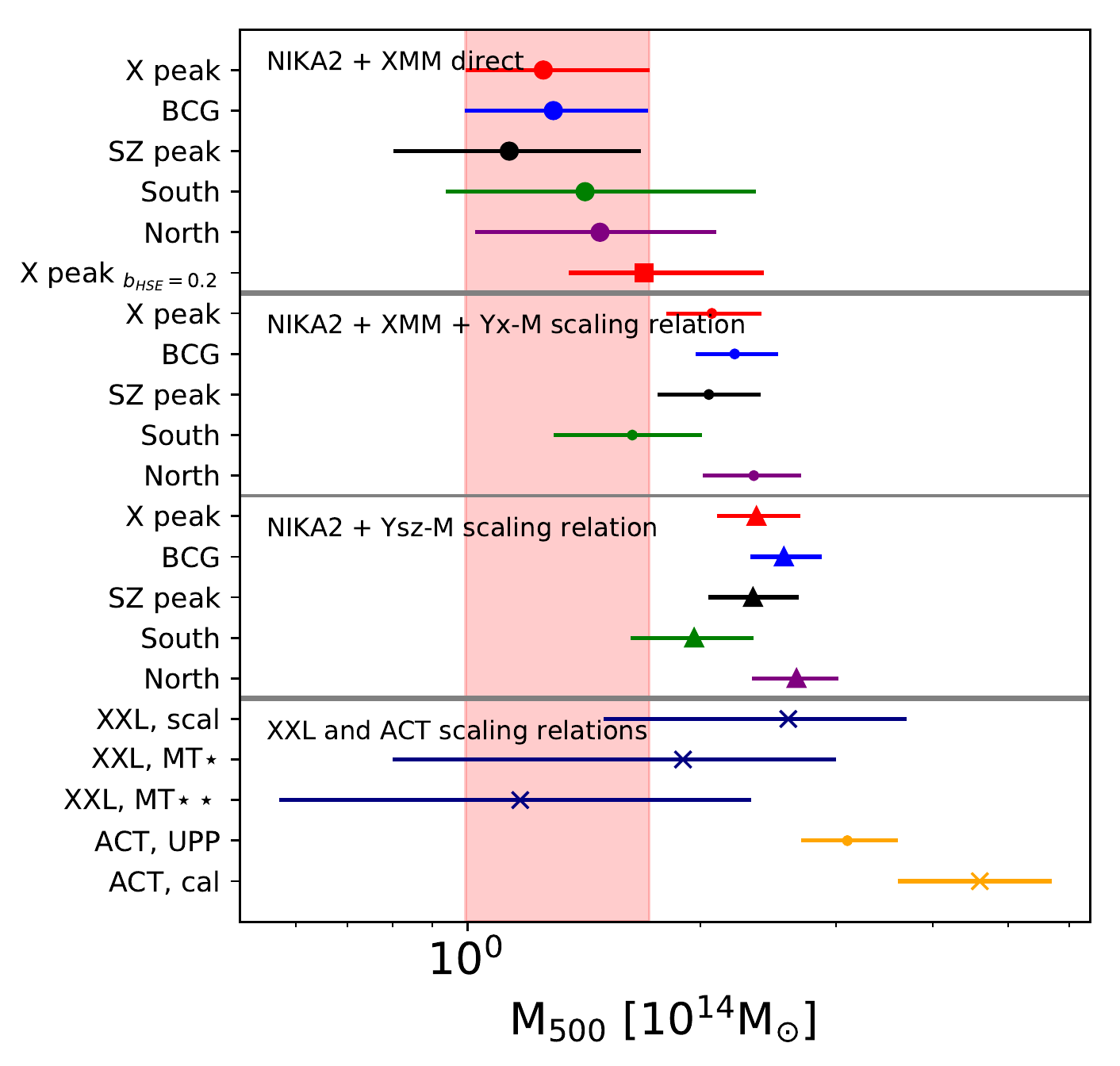}
		 \caption{\footnotesize{Comparison of the different estimates of $M_{500}$ of XLSSC~102. The top part shows the direct measurements from the hydrostatic mass profiles, derived from the combination of NIKA2 and XMM data.
The middle part shows the mass estimates derived from the measurements of $Y_{\rm X}(R)$ and the scaling relation from \citetalias{2010A&A...517A..92A} (dots), and from the measurements of $Y_{\rm SZ}(R)$ and the scaling relation from \citetalias{Planck2014XX} (triangles).
All points and error bars indicate the median and 68\% c.i. around it.
The bottom part shows mass estimation from the XXL survey and from the ACT survey (see Table \ref{tab:prop_summary} and text).
The crosses indicate mass estimates derived from weak lensing calibrated scaling laws.
The pink stripe indicates, for visual reference, the 68\% c.i. around $M_{500}$ measured from the combination of NIKA2 and XMM data, using the X-ray peak and assuming $b_{\rm HSE}=0$.}}
\label{fig:mass_compare}
    \end{figure}

In the previous section we estimated the mass of XLSSC~102 from direct measurements only, and from their combination with the $Y_{\rm X}-M$ scaling relation.
In the absence of complementary X-ray data, the $Y_{\rm SZ, 500} - M_{\rm HSE,500}$ relation from \citetalias{Planck2014XX} can also be used to infer the mass. Similarly to what is done in Section \ref{subsec:yy_scaling}, we directly used our integrated Compton parameter profile to derive the mass. Since $Y_{{\rm SZ},500}$ itself depends on $r_{500}$, which is computed from $M_{500}$, the SZ derived mass is computed by iterating about the scaling relation until convergence.

Having in hand different mass estimates it is now natural to compare them. Figure \ref{fig:mass_compare} shows the different mass measurements for XLSSC~102 and the estimates from the XXL and ACT surveys, derived at a spherical overdensity $\Delta=500$.
The first block of values shows the direct measurements from the combination of NIKA2 and XMM data (see Section \ref{subsec:scaling_relations}). The disks are for the HSE masses at the different centres and splitting the cluster in two halves, with no bias accounted for. The red square shows the mass at the X-ray centre computed by accounting for a bias $b_{\rm HSE}=0.2$ in the HSE mass profile. Because we define the bias with respect to the mass profile (assuming a constant bias value in Eq. \ref{eq_icmtool_hse_bias}), the ratio between $M_{\rm HSE, 500}$ and $M_{b=0.2, 500}$ is not equal to 1 - 0.2 (here we measure $M_{\rm HSE, 500}$/$M_{b=0.2, 500}$= 1 - 0.26).
The second and third blocks show mass estimates derived from the measurements of $Y_{\rm X}(R)$ and the scaling relation from \citetalias{2010A&A...517A..92A} (dots), and from the measurements of $Y_{\rm SZ}(R)$ and the scaling relation from \citetalias{Planck2014XX} (triangles).
The last block shows mass estimates from the XXL survey and from the ACT survey (see Table \ref{tab:prop_summary}). The XXL estimation denoted 'scal' is computed iteratively from X-ray count rates measurements and a set of XXL scaling relations \citepalias{adami_2018}, those denoted 'MT' derives from mass - XXL temperature relations calibrated with weak lensing data \citepalias[see][and \citealt{2016A&A...592A...4L} \citepalias{2016A&A...592A...4L} for 'MT$\star$', and \citealt{2020ApJ...890..148U} for 'MT$\star\star$']{pacaud_xxl_2015}. 
The ACT mass \citep{2018ApJS..235...20H} denoted 'UPP' is computed by assuming that the SZ signal follows the universal pressure profile and the associated scaling relation from \citetalias{2010A&A...517A..92A}. The value denoted 'cal' accounts for a hydrostatic mass bias of $b_{\rm HSE} = 0.32 \pm 0.11 $, measured on external weak lensing data sets.

We can see that the HSE mass values measured directly from the data are in agreement with the XXL expectations, given their large error bars. The ACT masses are larger than the other estimates. As the resolution of ACT is 1.4 arcmin, this could possibly be due to a contamination from a local low temperature CMB fluctuation, artificially boosting the SZ signal. 
The masses estimated from $Y_{\rm SZ}$ and $Y_{\rm X}$ profiles and scaling laws are higher than the one from the direct measurements. However, we can see that the masses measured in the northern part of the cluster are higher than those in the southern. The merging event is thus likely biasing high those values.
We do not see any trend between the HSE masses and those calibrated with weak lensing measurements (indicated by the crosses in Figure \ref{fig:mass_compare}).
We note that since the Subaru HSC survey \citep{Aihara2018,Mandelbaum2018} only partially overlaps with XLSSC~102, the cluster has HSC weak-lensing constraints in the limited radial range, $R\in [0.9, 2]$\,Mpc (comoving), with a very low signal-to-noise ratio of $\mathrm{S/N}= 0.1$ \citep{2020ApJ...890..148U}.

The disparity of values seen in Figure \ref{fig:mass_compare} reflects the difficulty of accurately estimating the mass of a cluster at redshift z$\sim$1 in this mass range, especially with low S/N data and for a disturbed system.
However, it also highlights the strength of resolved SZ observations alone and in combination with (even shallow) X-ray data. 
This is promising for the study of high redshift clusters from the combination of \textit{eROSITA} and high resolution SZ instruments.

\section{Summary and conclusions}
\label{sec:Summary_and_conclusions}

In this study we characterized the morphology, thermodynamic variable profiles and global quantities of XLSSC~102, an X-ray detected cluster at $z=0.97$, with a mass about $M_{\rm{500}}\sim2\times10^{14}$M$_\odot$.
For this purpose we used data recently acquired with NIKA2, a dual band high resolution millimetre camera, to map the SZ signal. 
We compared the NIKA2 observations to X-ray data taken by \textit{XMM-Newton} within the XXL Survey and to optical data from the CFHTLS (Section \ref{sec:Morphology}). This allowed us to study the morphology and inner structure of XLSSC~102, finding that it is likely in a post merger phase.
We then combined the X-ray and SZ data to derive the thermodynamic and mass profiles of XLSSC~102 : pressure, density, temperature, entropy index, gas fraction and HSE mass (Section \ref{sec:thermo_profiles}). We tested the impact of the merger and that of taking different centre definitions. 
Finally, we derived the global properties of XLSSC~102 and compared them to scaling relations of low redshift higher mass clusters and compare our different mass estimates (Section \ref{sec:global_prop}). 

Our main conclusions are as follows : 
\begin{itemize}

\item Our study showed that mapping low-mass clusters with NIKA2 is achievable even with a modest observing time (6.6 hours). With such resolution and sensitivity, it is possible to derive precise thermodynamic and mass profiles up to $\sim r_{500}$ by combining NIKA2 with X-ray survey data, even at high redshift and low mass. 

\item XLSSC~102 is a disturbed cluster that appears to be in a post-merging phase: the galaxy distribution is offset from the gas and presents an elongated morphology, presumably following the merger axis. The BCG, SZ and X-ray peaks are also significantly offset and the gas tracers display elliptical distributions.
This is also seen in the profiles of thermodynamic properties that are typical of disturbed clusters, that is, presenting flat inner temperature and pressure profiles and high entropy floor. When considering only the cluster half containing the overpressure region these features are more pronounced, but they are still visible in the other half, indicating that the gas has had time to mix after the merger.

\item Despite its low mass, high redshift and perturbed dynamical state, XLSSC~102 does not show any strong deviation from standard evolution expectations. Its pressure profile is similar to that found for disturbed clusters in the REXCESS sample and its global quantities are not strongly offset from the local scaling relations. 

\item The choice of centre has a mild effect on the derived profiles but introduces systematics that may broaden the scatter on the scaling relations. The internal structure has a net impact on the shape of the thermodynamic profiles but not much on the reconstructed HSE mass. The values of $Y_{\rm SZ, 500}$ and  $Y_{\rm X, 500}$ are higher in the cluster part that contains the overpressure region such that the mass estimates derived from scaling relations are also higher.

\item There is a disparity in the different estimates of the mass of XLSSC~102. This highlights the difficulty of obtaining accurate mass measurements for high redshift, low-mass systems and the need to combine and compare multi-wavelength and multi-instrument observations.

\end{itemize}

Considering the low S/N of each data set alone, all our findings would not have been possible without combining the different wavelengths. This approach is therefore key to understand cluster physics, especially for the in depth characterisation of low-mass, high-redshift systems.
New observations with the NIKA2 camera and other high-resolution SZ instruments, and their potential combination with \textit{eROSITA} data \cite[see][for its expected cluster selection function]{2012MNRAS.422...44P} will thus likely enhance our understanding of these objects.  The new generation of optical surveys (e.g. LSST and \textit{Euclid}) should also enable precise weak lensing measurements, allowing for a better calibration of the scaling relations. 
Thus, the multi-wavelengths approach is the best route to test the validity limits of our analyses and to extend the mass, redshift and formation stage domains of the cosmological cluster samples, allowing to better exploit the wealth of  information from future surveys.


\begin{acknowledgements}
We are thankful to the anonymous referee for useful comments, which helped improve the quality of the paper. 
We would like to thank the IRAM staff for their support during the campaigns. 
The NIKA dilution cryostat has been designed and built at the Institut N\'eel. In particular, we acknowledge the crucial contribution of the Cryogenics Group, and  in particular Gregory Garde, Henri Rodenas, Jean Paul Leggeri, Philippe Camus. 
This work has been partially funded by the Foundation Nanoscience Grenoble, the LabEx FOCUS ANR-11-LABX-0013 and the ANR under the contracts 'MKIDS' and 'NIKA'. 
This work has benefited from the support of the European Research Council Advanced Grants ORISTARS and M2C under the European Union's Seventh Framework Programme (Grant Agreement nos. 291294 and 340519).

OH acknowledges funding from the European Research Council (ERC) under the European Union's Horizon 2020 research and innovation programme (Grant Agreement No. 679145, project 'COSMO-SIMS').
MS acknowledges financial contribution from contract ASI-INAF n.2017-14-H.0 and INAF `Call per interventi aggiuntivi a sostegno della ricerca di main stream di INAF'.

XXL is an international project based around an XMM Very Large Programme surveying two $25$ deg$^2$ extragalactic fields at a depth of $\sim 6\cdot 10^{-15} {\rm erg}\cdot {\rm cm}^{-2}{\rm s}^{-1}$ in the [0.5--2] keV band for point-like sources. The XXL website is \url{http://irfu.cea.fr/xxl}. Multiband information and spectroscopic follow-up of the X-ray sources are obtained through a number of survey programmes, summarised at \url{http://xxlmultiwave.pbworks.com}. 

This work was supported by the Programme National Cosmology et Galaxies (PNCG) of CNRS/INSU with INP and IN2P3, co-funded by CEA and CNES.

This study is based on observations obtained with MegaPrime/MegaCam, a joint project of CFHT and CEA/IRFU, at the Canada-France-Hawaii Telescope (CFHT) which is operated by the National Research Council (NRC) of Canada, the Institut National des Science de l'Univers of the Centre National de la Recherche Scientifique (CNRS) of France, and the University of Hawaii. This work is based in part on data products produced at Terapix available at the Canadian Astronomy Data Centre as part of the Canada-France-Hawaii Telescope Legacy Survey, a collaborative project of NRC and CNRS. 

This paper present data collected at the Subaru Telescope and retrieved from the HSC data archive system, which is operated by Subaru Telescope and Astronomy Data Center at National Astronomical Observatory of Japan. Data analysis was in part carried out with the cooperation of Center for Computational Astrophysics, National Astronomical Observatory of Japan.

This research made use of Astropy, a community-developed core Python package for Astronomy \citep{Astropy2013}, in addition to NumPy \citep{VanDerWalt2011}, SciPy \citep{Jones2001}, and Ipython \citep{Perez2007}. Figures were generated using Matplotlib \citep{Hunter2007}. 
\end{acknowledgements}

\interlinepenalty=10000
\bibliography{biblio_XLSSC102}

\begin{appendix}

\section{Transfert function}
\label{annexe}
\begin{figure}[h]
	\centering
	\includegraphics[width=0.5\textwidth]{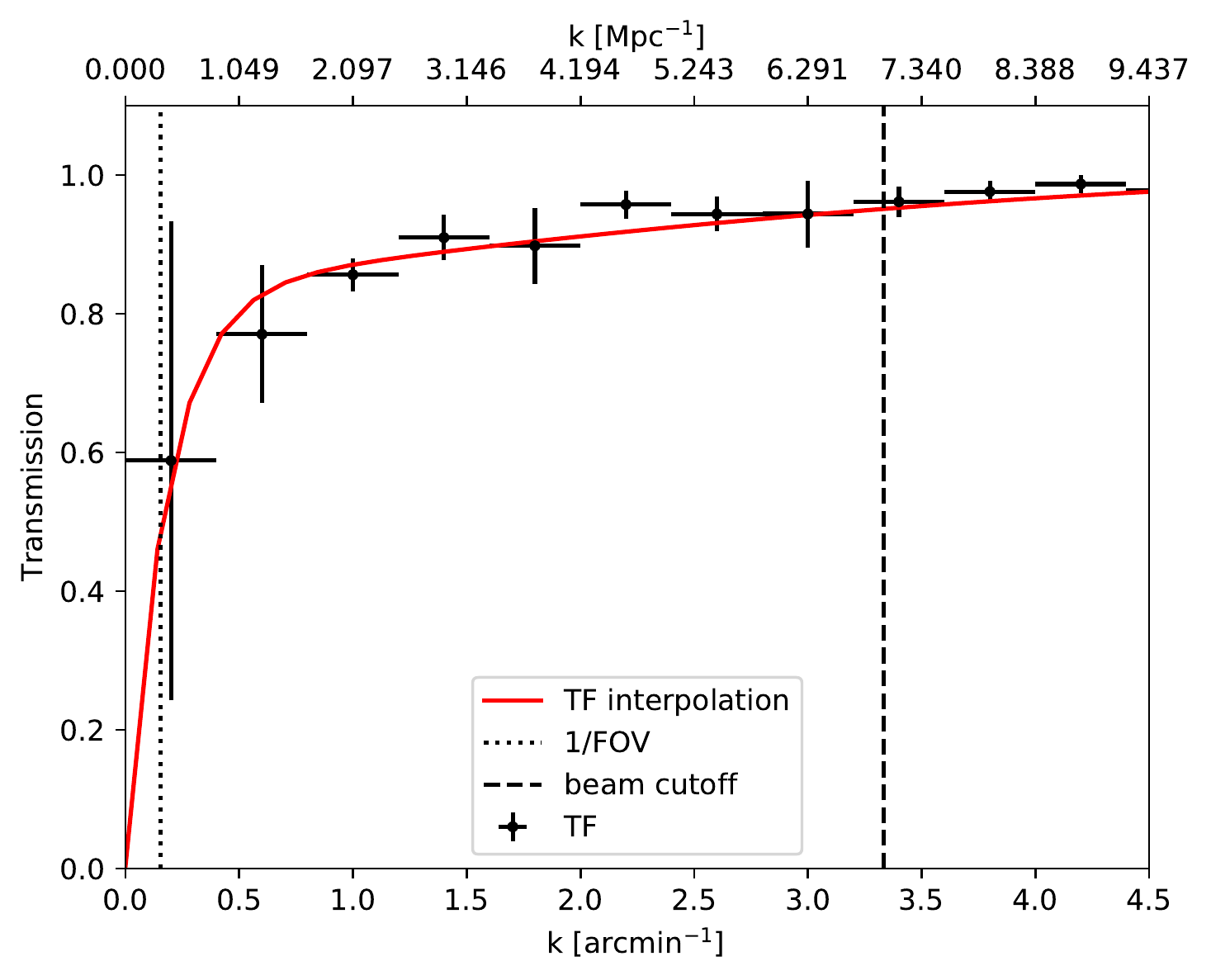}
	\caption{\footnotesize{NIKA2 tranfert function at 150 GHz (transmission as a function of wave number), corresponding to the data reduction employed in this analyse. The black points are measurements and the red line represents the fit of an arbitrary polynomial function used for the interpolation.}}
	\label{fig:TF}
	\end{figure}

Figure \ref{fig:TF} shows the transfert function associated to the reduction of the NIKA2 map of XLSSC102 in the 150 GHz band, as explained in Section \ref{sec:Data_NIKA2}.

\end{appendix}

\end{document}

%% file: listeauthors.tex
\author{
M.~Ricci \inst{\ref{LAPP},\ref{OCA}}\thanks{Corresponding author: Marina Ricci, \url{marina.ricci@lapp.in2p3.fr}}
\and R.~Adam \inst{\ref{CEFCA},\ref{LLR}}
\and D.~Eckert \inst{\ref{geneva}}
\and  P.~Ade \inst{\ref{Cardiff}}
\and  P.~Andr\'e \inst{\ref{CEA}}
\and  A.~Andrianasolo \inst{\ref{IPAG}}
\and  B.~Altieri\inst{\ref{ESAC}}
\and  H.~Aussel \inst{\ref{CEA}}
\and  A.~Beelen \inst{\ref{IAS}}
\and  C.~Benoist\inst{\ref{OCA}}
\and  A.~Beno\^it \inst{\ref{Neel}}
\and  S.~Berta \inst{\ref{IRAMF}}
\and  A.~Bideaud \inst{\ref{Neel}}
\and  M.~Birkinshaw\inst{\ref{Bristol}}
\and  O.~Bourrion \inst{\ref{LPSC}}
\and  D.~Boutigny \inst{\ref{LAPP}}
\and  M.~Bremer\inst{\ref{Bristol}}
\and  M.~Calvo \inst{\ref{Neel}}
\and  A.~Cappi\inst{\ref{OCA},\ref{BolognaObs}}
\and  L.~Chiappetti\inst{\ref{Milan}}
\and  A.~Catalano \inst{\ref{LPSC}}
\and  M.~De~Petris\inst{\ref{Roma}}
\and  F.-X.~D\'esert \inst{\ref{IPAG}}
\and  S.~Doyle \inst{\ref{Cardiff}}
\and  E.~F.~C.~Driessen \inst{\ref{IRAMF}}
\and  L.~Faccioli\inst{\ref{CEA}}
\and  C.~Ferrari\inst{\ref{OCA}}
\and  S.~Fotopoulou\inst{\ref{Durham}}
\and  F.~Gastaldello\inst{\ref{Milan}}
\and  P.~Giles\inst{\ref{Bristol}}
\and  A.~Gomez \inst{\ref{CAB}}
\and  J.~Goupy \inst{\ref{Neel}}
\and  O.~Hahn\inst{\ref{OCA}}
\and  C.~Horellou\inst{\ref{Onsala}}
\and  F.~K\'eruzor\'e \inst{\ref{LPSC}}
\and  E.~Koulouridis\inst{\ref{CEA},\ref{Athene}}
\and  C.~Kramer \inst{\ref{IRAME}}
\and  B.~Ladjelate \inst{\ref{IRAME}}
\and  G.~Lagache \inst{\ref{LAM}}
\and  S.~Leclercq \inst{\ref{IRAMF}}
\and  J.-F.~Lestrade \inst{\ref{LERMA}}
\and  J.F.~Mac\'ias-P\'erez \inst{\ref{LPSC}}
\and  A.~Mantz\inst{\ref{Kavli}}
\and  B.~Maughan\inst{\ref{Bristol}}
\and  S.~Maurogordato\inst{\ref{OCA}}
\and  P.~Mauskopf \inst{\ref{Cardiff},\ref{Arizona}}
\and  A.~Monfardini \inst{\ref{Neel}}
\and  F.~Pacaud\inst{\ref{Bonn}}
\and  L.~Perotto \inst{\ref{LPSC}}
\and  M.~Pierre\inst{\ref{CEA}}
\and  G.~Pisano \inst{\ref{Cardiff}}
\and  E.~Pompei\inst{\ref{ESO}}
\and  N.~Ponthieu \inst{\ref{IPAG}}
\and  V.~Rev\'eret \inst{\ref{CEA}}
\and  A.~Ritacco \inst{\ref{IAS},\ref{ENS}}
\and  C.~Romero \inst{\ref{IRAMF}}
\and  H.~Roussel \inst{\ref{IAP}}
\and  F.~Ruppin \inst{\ref{LPSC}}
\and  M.~S\'{a}nchez Portal \inst{\ref{IRAME}}
\and  K.~Schuster \inst{\ref{IRAMF}}
\and  M.~Sereno\inst{\ref{BolognaObs},\ref{BolognaDepP}}
\and  S.~Shu \inst{\ref{IRAMF}}
\and  A.~Sievers \inst{\ref{IRAME}}
\and  C.~Tucker \inst{\ref{Cardiff}}
\and  K.~Umetsu \inst{\ref{ASIAA}}
}

\institute{
Laboratoire d'Annecy de Physique des Particules, Universit\'e Savoie Mont Blanc, CNRS/IN2P3, F-74941 Annecy, France
  \label{LAPP}
  \and
Laboratoire Lagrange, Universit\'e C\^ote d'Azur, Observatoire de la C\^ote d'Azur, CNRS, Blvd de l'Observatoire, CS 34229, 06304 Nice cedex 4, France
  \label{OCA}
  \and
  Centro de Estudios de F\'isica del Cosmos de Arag\'on (CEFCA), Plaza San Juan, 1, planta 2, E-44001, Teruel, Spain
  \label{CEFCA}
  \and
 LLR, CNRS, \'Ecole Polytechnique, Institut Polytechnique de Paris
  \label{LLR}
  \and
  Department of Astronomy, University of Geneva, ch. d'Ecogia 16, CH-1290 Versoix, Switzerland
   \label{geneva}
   \and
    Astronomy Instrumentation Group, University of Cardiff, UK
  \label{Cardiff}
  \and
AIM, CEA, CNRS, Universit\'e Paris-Saclay, Universit\'e Paris Diderot, Sorbonne Paris Cit\'e, F-91191 Gif-sur-Yvette, France
  \label{CEA}
  \and
Univ. Grenoble Alpes, CNRS, IPAG, F-38000 Grenoble, France 
  \label{IPAG}
 \and
European Space Astronomy Centre (ESA/ESAC), Operations Department, Villanueva de la Can\~{a}da, Madrid, Spain
\label{ESAC}
\and
Institut d'Astrophysique Spatiale (IAS), CNRS and Universit\'e Paris Sud, Orsay, France
  \label{IAS}
\and
Institut N\'eel, CNRS and Universit\'e Grenoble Alpes, France
  \label{Neel}
  \and
Institut de RadioAstronomie Millim\'etrique (IRAM), Grenoble, France
  \label{IRAMF}
\and
HH Wills Physics Laboratory, University of Bristol, Tyndall Avenue, Bristol, BS8 1TL, UK
\label{Bristol} 
  \and
  Laboratoire de Physique Subatomique et de Cosmologie, Universit\'e Grenoble Alpes, CNRS/IN2P3, 53, avenue des Martyrs, Grenoble, France
  \label{LPSC}
  \and
  Istituto Nazionale di Astrofisica (INAF) - Osservatorio di Astrofisica e Scienza dello Spazio (OAS), via Gobetti 93/3, I-40127 Bologna, Italy
\label{BolognaObs}
 \and
INAF - IASF Milan, via A. Corti 12, I-20133 Milano,  Italy
\label{Milan}  
\and
Dipartimento di Fisica, Sapienza Universit\`a di Roma, Piazzale Aldo Moro 5, I-00185 Roma, Italy
  \label{Roma}
\and
Centre for Extragalactic Astronomy, Department of Physics, Durham University, South Road, Durham DH1 3LE, U.K.
\label{Durham}  
\and
Centro de Astrobiolog\'ia (CSIC-INTA), Torrej\'on de Ardoz, 28850 Madrid, Spain
\label{CAB}
\and
Department of Space, Earth and Environment, Chalmers University of Technology, Onsala Space Observatory, SE-439 92 Onsala,
Sweden
\label{Onsala}
\and
Institute for Astronomy \& Astrophysics, Space Applications \& Remote
Sensing, National Observatory of Athens, GR-15236 Palaia Penteli,Greece
  \label{Athene}
\and
Institut de RadioAstronomie Millim\'etrique (IRAM), Granada, Spain 
\label{IRAME}
\and
Aix Marseille Universit\'e, CNRS, LAM (Laboratoire d'Astrophysique de Marseille) UMR 7326, 13388, Marseille, France
  \label{LAM}
\and 
LERMA, Observatoire de Paris, PSL Research University, CNRS, Sorbonne Universités, UPMC Univ. Paris 06, 75014 Paris, France
  \label{LERMA}  
\and
Kavli Institute for Particle Astrophysics \& Cosmology, P. O. Box 2450,
Stanford University, Stanford, CA 94305, USA
\label{Kavli}  
\and
School of Earth and Space Exploration and Department of Physics, Arizona State University, Tempe, AZ 85287
  \label{Arizona}
\and 
Argelander Institut f\"ur Astronomie, Universit\"at Bonn, Auf dem Huegel 71, DE-53121 Bonn, Germany
\label{Bonn}
\and
European Southern Observatory, Alonso de Cordova 3107, Vitacura,
19001 Casilla, Santiago 19, Chile
\label{ESO}
\and
LPENS, Ecole Normale Sup\'erieure, 24 rue Lhomond, 75005, Paris (FR) 
\label{ENS}
\and 
Institut d'Astrophysique de Paris, Sorbonne Universit\'es, UPMC Univ. Paris 06, CNRS UMR 7095, 75014 Paris, France 
  \label{IAP}
\and
INFN, Sezione di Bologna, viale Berti Pichat 6/2, 40127 Bologna, Italy
\label{BolognaDepP}
\and
Academia Sinica Institute of Astronomy and Astrophysics (ASIAA),No. 1, Section 4, Roosevelt Road, Taipei 10617, Taiwan
\label{ASIAA}
}

%% file: paper.bbl
\begin{thebibliography}{112}
\expandafter\ifx\csname natexlab\endcsname\relax\def\natexlab#1{#1}\fi

\bibitem[{Adam(2015)}]{adam:tel-01303736}
Adam, R. 2015, Theses, {Universit{\'e} Grenoble Alpes}

\bibitem[{Adam {et~al.}(2018)Adam, Adane, Ade, André, Andrianasolo, Aussel,
  Beelen, Benoît, Bideaud, Billot, Bourrion, Bracco, Calvo, Catalano,
  Coiffard, Comis, De~Petris, Désert, Doyle, Driessen, Evans, Goupy, Kramer,
  Lagache, Leclercq, Leggeri, Lestrade, Macías-Pérez, Mauskopf, Mayet, Maury,
  Monfardini, Navarro, Pascale, Perotto, Pisano, Ponthieu, Revéret, Rigby,
  Ritacco, Romero, Roussel, Ruppin, Schuster, Sievers, Triqueneaux, Tucker, \&
  Zylka}]{adam_nika2_2018}
Adam, R. {et~al.} 2018, \aap, 609, A115

\bibitem[{{Adam} {et~al.}(2017{\natexlab{a}}){Adam}, {Arnaud}, {Bartalucci},
  {Ade}, {Andr{\'e}}, {Beelen}, {Beno{\^\i}t}, {Bideaud}, {Billot}, {Bourdin},
  {Bourrion}, {Calvo}, {Catalano}, {Coiffard}, {Comis}, {D'Addabbo},
  {D{\'e}sert}, {Doyle}, {Ferrari}, {Goupy}, {Kramer}, {Lagache}, {Leclercq},
  {Mac{\'\i}as-P{\'e}rez}, {Maurogordato}, {Mauskopf}, {Mayet}, {Monfardini},
  {Pajot}, {Pascale}, {Perotto}, {Pisano}, {Pointecouteau}, {Ponthieu},
  {Pratt}, {Rev{\'e}ret}, {Ritacco}, {Rodriguez}, {Romero}, {Ruppin},
  {Schuster}, {Sievers}, {Triqueneaux}, {Tucker}, \&
  {Zylka}}]{2017A&A...606A..64A}
{Adam}, R. {et~al.} 2017{\natexlab{a}}, \aap, 606, A64, 1706.10230

\bibitem[{{Adam} {et~al.}(2017{\natexlab{b}}){Adam}, {Bartalucci}, {Pratt},
  {Ade}, {Andr{\'e}}, {Arnaud}, {Beelen}, {Beno{\^i}t}, {Bideaud}, {Billot},
  {Bourdin}, {Bourrion}, {Calvo}, {Catalano}, {Coiffard}, {Comis}, {D'Addabbo},
  {De Petris}, {D{\'e}mocl{\`e}s}, {D{\'e}sert}, {Doyle}, {Egami}, {Ferrari},
  {Goupy}, {Kramer}, {Lagache}, {Leclercq}, {Mac{\'{\i}}as-P{\'e}rez},
  {Maurogordato}, {Mauskopf}, {Mayet}, {Monfardini}, {Mroczkowski}, {Pajot},
  {Pascale}, {Perotto}, {Pisano}, {Pointecouteau}, {Ponthieu}, {Rev{\'e}ret},
  {Ritacco}, {Rodriguez}, {Romero}, {Ruppin}, {Schuster}, {Sievers},
  {Triqueneaux}, {Tucker}, {Zemcov}, \& {Zylka}}]{Adam2016b}
------. 2017{\natexlab{b}}, \aap, 598, A115, 1606.07721

\bibitem[{Adam {et~al.}(2016)Adam, Comis, Bartalucci, Adane, Ade, André,
  Arnaud, Beelen, Belier, Benoît, Bideaud, Billot, Bourrion, Calvo, Catalano,
  Coiffard, D'Addabbo, Désert, Doyle, Goupy, Hasnoun, Hermelo, Kramer,
  Lagache, Leclercq, Macías-Pérez, Martino, Mauskopf, Mayet, Monfardini,
  Pajot, Pascale, Perotto, Pointecouteau, Ponthieu, Pratt, Revéret, Ritacco,
  Rodriguez, Savini, Schuster, Sievers, Triqueneaux, Tucker, \&
  Zylka}]{adam_high_2016}
Adam, R. {et~al.} 2016, \aap, 586, A122

\bibitem[{{Adam} {et~al.}(2015){Adam}, {Comis}, {Mac{\'{\i}}as-P{\'e}rez},
  {Adane}, {Ade}, {Andr{\'e}}, {Beelen}, {Belier}, {Beno{\^i}t}, {Bideaud},
  {Billot}, {Blanquer}, {Bourrion}, {Calvo}, {Catalano}, {Coiffard},
  {Cruciani}, {D'Addabbo}, {D{\'e}sert}, {Doyle}, {Goupy}, {Kramer},
  {Leclercq}, {Martino}, {Mauskopf}, {Mayet}, {Monfardini}, {Pajot}, {Pascale},
  {Perotto}, {Pointecouteau}, {Ponthieu}, {Rev{\'e}ret}, {Ritacco},
  {Rodriguez}, {Savini}, {Schuster}, {Sievers}, {Tucker}, \&
  {Zylka}}]{Adam2015}
{Adam}, R. {et~al.} 2015, \aap, 576, A12, 1410.2808

\bibitem[{{Adam} {et~al.}(2014){Adam}, {Comis}, {Mac{\'{\i}}as-P{\'e}rez},
  {Adane}, {Ade}, {Andr{\'e}}, {Beelen}, {Belier}, {Beno{\^i}t}, {Bideaud},
  {Billot}, {Boudou}, {Bourrion}, {Calvo}, {Catalano}, {Coiffard}, {D'Addabbo},
  {D{\'e}sert}, {Doyle}, {Goupy}, {Kramer}, {Leclercq}, {Martino}, {Mauskopf},
  {Mayet}, {Monfardini}, {Pajot}, {Pascale}, {Perotto}, {Pointecouteau},
  {Ponthieu}, {Rev{\'e}ret}, {Rodriguez}, {Savini}, {Schuster}, {Sievers},
  {Tucker}, \& {Zylka}}]{Adam2014}
------. 2014, \aap, 569, A66, 1310.6237

\bibitem[{Adam {et~al.}(2015)Adam, Comis, Macías-Pérez, Adane, Ade, André,
  Beelen, Belier, Benoît, Bideaud, Billot, Blanquer, Bourrion, Calvo,
  Catalano, Coiffard, Cruciani, D’Addabbo, Désert, Doyle, Goupy, Kramer,
  Leclercq, Martino, Mauskopf, Mayet, Monfardini, Pajot, Pascale, Perotto,
  Pointecouteau, Ponthieu, Revéret, Ritacco, Rodriguez, Savini, Schuster,
  Sievers, Tucker, \& Zylka}]{adam_pressure_2015}
Adam, R. {et~al.} 2015, \aap, 576, A12

\bibitem[{{Adami} {et~al.}(2018){Adami}, {Giles}, {Koulouridis}, {Pacaud},
  {Caretta}, {Pierre}, {Eckert}, {Ramos- Ceja}, {Gastaldello}, {Fotopoulou},
  {Guglielmo}, {Lidman}, {Sadibekova}, {Iovino}, {Maughan}, {Chiappetti},
  {Alis}, {Altieri}, {Baldry}, {Bottini}, {Birkinshaw}, {Bremer}, {Brown},
  {Cucciati}, {Driver}, {Elmer}, {Ettori}, {Evrard}, {Faccioli}, {Granett},
  {Grootes}, {Guzzo}, {Hopkins}, {Horellou}, {Lef{\`e}vre}, {Liske}, {Malek},
  {Marulli}, {Maurogordato}, {Owers}, {Paltani}, {Poggianti}, {Polletta},
  {Plionis}, {Pollo}, {Pompei}, {Ponman}, {Rapetti}, {Ricci}, {Robotham},
  {Tuffs}, {Tasca}, {Valtchanov}, {Vergani}, {Wagner}, {Willis}, \& {XXL
  Consortium}}]{adami_2018}
{Adami}, C. {et~al.} 2018, \aap, 620, A5, 1810.03849, (XXL Paper XX)

\bibitem[{{Aihara} {et~al.}(2018{\natexlab{a}}){Aihara}, {Arimoto},
  {Armstrong}, {Arnouts}, {Bahcall}, {Bickerton}, {Bosch}, {Bundy}, {Capak},
  {Chan}, {Chiba}, {Coupon}, {Egami}, {Enoki}, {Finet}, {Fujimori}, {Fujimoto},
  {Furusawa}, {Furusawa}, {Goto}, {Goulding}, {Greco}, {Greene}, {Gunn},
  {Hamana}, {Harikane}, {Hashimoto}, {Hattori}, {Hayashi}, {Hayashi},
  {He{\l}miniak}, {Higuchi}, {Hikage}, {Ho}, {Hsieh}, {Huang}, {Huang},
  {Ikeda}, {Imanishi}, {Inoue}, {Iwasawa}, {Iwata}, {Jaelani}, {Jian},
  {Kamata}, {Karoji}, {Kashikawa}, {Katayama}, {Kawanomoto}, {Kayo}, {Koda},
  {Koike}, {Kojima}, {Komiyama}, {Konno}, {Koshida}, {Koyama}, {Kusakabe},
  {Leauthaud}, {Lee}, {Lin}, {Lin}, {Lupton}, {Mandelbaum}, {Matsuoka},
  {Medezinski}, {Mineo}, {Miyama}, {Miyatake}, {Miyazaki}, {Momose}, {More},
  {More}, {Moritani}, {Moriya}, {Morokuma}, {Mukae}, {Murata}, {Murayama},
  {Nagao}, {Nakata}, {Niida}, {Niikura}, {Nishizawa}, {Obuchi}, {Oguri},
  {Oishi}, {Okabe}, {Okamoto}, {Okura}, {Ono}, {Onodera}, {Onoue}, {Osato},
  {Ouchi}, {Price}, {Pyo}, {Sako}, {Sawicki}, {Shibuya}, {Shimasaku},
  {Shimono}, {Shirasaki}, {Silverman}, {Simet}, {Speagle}, {Spergel},
  {Strauss}, {Sugahara}, {Sugiyama}, {Suto}, {Suyu}, {Suzuki}, {Tait},
  {Takada}, {Takata}, {Tamura}, {Tanaka}, {Tanaka}, {Tanaka}, {Tanaka},
  {Terai}, {Terashima}, {Toba}, {Tominaga}, {Toshikawa}, {Turner}, {Uchida},
  {Uchiyama}, {Umetsu}, {Uraguchi}, {Urata}, {Usuda}, {Utsumi}, {Wang}, {Wang},
  {Wong}, {Yabe}, {Yamada}, {Yamanoi}, {Yasuda}, {Yeh}, {Yonehara}, \&
  {Yuma}}]{2018PASJ...70S...4A}
{Aihara}, H. {et~al.} 2018{\natexlab{a}}, \pasj, 70, S4, 1704.05858

\bibitem[{{Aihara} {et~al.}(2018{\natexlab{b}}){Aihara}, {Armstrong},
  {Bickerton}, {Bosch}, {Coupon}, {Furusawa}, {Hayashi}, {Ikeda}, {Kamata},
  {Karoji}, {Kawanomoto}, {Koike}, {Komiyama}, {Lang}, {Lupton}, {Mineo},
  {Miyatake}, {Miyazaki}, {Morokuma}, {Obuchi}, {Oishi}, {Okura}, {Price},
  {Takata}, {Tanaka}, {Tanaka}, {Tanaka}, {Uchida}, {Uraguchi}, {Utsumi},
  {Wang}, {Yamada}, {Yamanoi}, {Yasuda}, {Arimoto}, {Chiba}, {Finet},
  {Fujimori}, {Fujimoto}, {Furusawa}, {Goto}, {Goulding}, {Gunn}, {Harikane},
  {Hattori}, {Hayashi}, {He{\l}miniak}, {Higuchi}, {Hikage}, {Ho}, {Hsieh},
  {Huang}, {Huang}, {Imanishi}, {Iwata}, {Jaelani}, {Jian}, {Kashikawa},
  {Katayama}, {Kojima}, {Konno}, {Koshida}, {Kusakabe}, {Leauthaud}, {Lee},
  {Lin}, {Lin}, {Mandelbaum}, {Matsuoka}, {Medezinski}, {Miyama}, {Momose},
  {More}, {More}, {Mukae}, {Murata}, {Murayama}, {Nagao}, {Nakata}, {Niida},
  {Niikura}, {Nishizawa}, {Oguri}, {Okabe}, {Ono}, {Onodera}, {Onoue}, {Ouchi},
  {Pyo}, {Shibuya}, {Shimasaku}, {Simet}, {Speagle}, {Spergel}, {Strauss},
  {Sugahara}, {Sugiyama}, {Suto}, {Suzuki}, {Tait}, {Takada}, {Terai}, {Toba},
  {Turner}, {Uchiyama}, {Umetsu}, {Urata}, {Usuda}, {Yeh}, \&
  {Yuma}}]{Aihara2018}
------. 2018{\natexlab{b}}, \pasj, 70, S8, 1702.08449

\bibitem[{{Albrecht} {et~al.}(2006){Albrecht}, {Bernstein}, {Cahn}, {Freedman},
  {Hewitt}, {Hu}, {Huth}, {Kamionkowski}, {Kolb}, {Knox}, {Mather}, {Staggs},
  \& {Suntzeff}}]{2006astro.ph..9591A}
{Albrecht}, A. {et~al.} 2006, ArXiv Astrophysics e-prints, astro-ph/0609591

\bibitem[{{Allen} {et~al.}(2011){Allen}, {Evrard}, \& {Mantz}}]{Allen2011}
{Allen}, S.~W., {Evrard}, A.~E., \& {Mantz}, A.~B. 2011, \araa, 49, 409,
  1103.4829

\bibitem[{{Anders} \& {Grevesse}(1989)}]{1989GeCoA..53..197A}
{Anders}, E., \& {Grevesse}, N. 1989, \gca, 53, 197

\bibitem[{{Andersson} {et~al.}(2011){Andersson}, {Benson}, {Ade}, {Aird},
  {Armstrong}, {Bautz}, {Bleem}, {Brodwin}, {Carlstrom}, {Chang}, {Crawford},
  {Crites}, {de Haan}, {Desai}, {Dobbs}, {Dudley}, {Foley}, {Forman},
  {Garmire}, {George}, {Gladders}, {Halverson}, {High}, {Holder}, {Holzapfel},
  {Hrubes}, {Jones}, {Joy}, {Keisler}, {Knox}, {Lee}, {Leitch}, {Lueker},
  {Marrone}, {McMahon}, {Mehl}, {Meyer}, {Mohr}, {Montroy}, {Murray}, {Padin},
  {Plagge}, {Pryke}, {Reichardt}, {Rest}, {Ruel}, {Ruhl}, {Schaffer}, {Shaw},
  {Shirokoff}, {Song}, {Spieler}, {Stalder}, {Staniszewski}, {Stark}, {Stubbs},
  {Vand erlinde}, {Vieira}, {Vikhlinin}, {Williamson}, {Yang}, {Zahn}, \&
  {Zenteno}}]{2011ApJ...738...48A}
{Andersson}, K. {et~al.} 2011, \apj, 738, 48, 1006.3068

\bibitem[{{Arnaud} {et~al.}(2010){Arnaud}, {Pratt}, {Piffaretti},
  {B{\"o}hringer}, {Croston}, \& {Pointecouteau}}]{2010A&A...517A..92A}
{Arnaud}, M., {Pratt}, G.~W., {Piffaretti}, R., {B{\"o}hringer}, H., {Croston},
  J.~H., \& {Pointecouteau}, E. 2010, \aap, 517, A92, 0910.1234

\bibitem[{{Astropy Collaboration} {et~al.}(2013){Astropy Collaboration},
  {Robitaille}, {Tollerud}, {Greenfield}, {Droettboom}, {Bray}, {Aldcroft},
  {Davis}, {Ginsburg}, {Price-Whelan}, {Kerzendorf}, {Conley}, {Crighton},
  {Barbary}, {Muna}, {Ferguson}, {Grollier}, {Parikh}, {Nair}, {Unther},
  {Deil}, {Woillez}, {Conseil}, {Kramer}, {Turner}, {Singer}, {Fox}, {Weaver},
  {Zabalza}, {Edwards}, {Azalee Bostroem}, {Burke}, {Casey}, {Crawford},
  {Dencheva}, {Ely}, {Jenness}, {Labrie}, {Lim}, {Pierfederici}, {Pontzen},
  {Ptak}, {Refsdal}, {Servillat}, \& {Streicher}}]{Astropy2013}
{Astropy Collaboration} {et~al.} 2013, \aap, 558, A33, 1307.6212

\bibitem[{{Becker} {et~al.}(1995){Becker}, {White}, \&
  {Helfand}}]{1995ApJ...450..559B}
{Becker}, R.~H., {White}, R.~L., \& {Helfand}, D.~J. 1995, \apj, 450, 559

\bibitem[{{Benoist}(2014)}]{2014becs.confE...8B}
{Benoist}, C. 2014, in Building the Euclid Cluster Survey - Scientific Program,
  proceedings of a conference held July 6-11 2014 at the Sexten Center for
  Astrophysics., 8

\bibitem[{{Birkinshaw}(1999)}]{Birkinshaw1999}
{Birkinshaw}, M. 1999, \physrep, 310, 97, arXiv:astro-ph/9808050

\bibitem[{{Bocquet} {et~al.}(2016){Bocquet}, {Saro}, {Dolag}, \&
  {Mohr}}]{2016MNRAS.456.2361B}
{Bocquet}, S., {Saro}, A., {Dolag}, K., \& {Mohr}, J.~J. 2016, \mnras, 456,
  2361, 1502.07357

\bibitem[{{B{\"o}hringer} \& {Werner}(2010)}]{bohringer2010}
{B{\"o}hringer}, H., \& {Werner}, N. 2010, \aapr, 18, 127

\bibitem[{{Bourrion} {et~al.}(2016){Bourrion}, {Benoit}, {Bouly}, {Bouvier},
  {Bosson}, {Calvo}, {Catalano}, {Goupy}, {Li}, {Mac{\'\i}as-P{\'e}rez},
  {Monfardini}, {Tourres}, {Ponchant}, \& {Vescovi}}]{Bourrion2016}
{Bourrion}, O. {et~al.} 2016, Journal of Instrumentation, 11, P11001,
  1602.01288

\bibitem[{Calvo {et~al.}(2016)Calvo, Benoit, Catalano, Goupy, Monfardini,
  Ponthieu, Barria, Bres, Grollier, Garde, Leggeri, Pont, Triqueneaux, Adam,
  Bourrion, Macías-Pérez, Rebolo, Ritacco, Scordilis, Tourres, Vescovi,
  Désert, Adane, Coiffard, Leclercq, Doyle, Mauskopf, Tucker, Ade, André,
  Beelen, Belier, Bideaud, Billot, Comis, D'Addabbo, Kramer, Martino, Mayet,
  Pajot, Pascale, Perotto, Revéret, Rodriguez, Savini, Schuster, Sievers, \&
  Zylka}]{calvo_nika2_2016}
Calvo, M. {et~al.} 2016, Journal of Low Temperature Physics, 184, 816, arXiv:
  1601.02774

\bibitem[{{Catalano} {et~al.}(2014){Catalano}, {Calvo}, {Ponthieu}, {Adam},
  {Adane}, {Ade}, {Andr{\'e}}, {Beelen}, {Belier}, {Beno{\^i}t}, {Bideaud},
  {Billot}, {Boudou}, {Bourrion}, {Coiffard}, {Comis}, {D'Addabbo},
  {D{\'e}sert}, {Doyle}, {Goupy}, {Kramer}, {Leclercq},
  {Mac{\'{\i}}as-P{\'e}rez}, {Martino}, {Mauskopf}, {Mayet}, {Monfardini},
  {Pajot}, {Pascale}, {Perotto}, {Rev{\'e}ret}, {Rodriguez}, {Savini},
  {Schuster}, {Sievers}, {Tucker}, \& {Zylka}}]{Catalano2014}
{Catalano}, A. {et~al.} 2014, \aap, 569, A9, 1402.0260

\bibitem[{{Chiappetti} {et~al.}(2018){Chiappetti}, {Fotopoulou}, {Lidman},
  {Faccioli}, {Pacaud}, {Elyiv}, {Paltani}, {Pierre}, {Plionis}, {Adami},
  {Alis}, {Altieri}, {Baldry}, {Bolzonella}, {Bongiorno}, {Brown}, {Driver},
  {Elmer}, {Franzetti}, {Grootes}, {Guglielmo}, {Iovino}, {Koulouridis},
  {Lef{\`e}vre}, {Liske}, {Maurogordato}, {Melnyk}, {Owers}, {Poggianti},
  {Polletta}, {Pompei}, {Ponman}, {Robotham}, {Sadibekova}, {Tuffs},
  {Valtchanov}, {Vignali}, \& {Wagner}}]{2018A&A...620A..12C}
{Chiappetti}, L. {et~al.} 2018, \aap, 620, A12, 1810.03929, (XXL Paper XXVII)

\bibitem[{{Cialone} {et~al.}(2018){Cialone}, {De Petris}, {Sembolini}, {Yepes},
  {Baldi}, \& {Rasia}}]{2018MNRAS.477..139C}
{Cialone}, G., {De Petris}, M., {Sembolini}, F., {Yepes}, G., {Baldi}, A.~S.,
  \& {Rasia}, E. 2018, \mnras, 477, 139, 1708.03325

\bibitem[{{Coleman} {et~al.}(1980){Coleman}, {Wu}, \&
  {Weedman}}]{coleman_colors_1980}
{Coleman}, G.~D., {Wu}, C.-C., \& {Weedman}, D.~W. 1980, \apjs, 43, 393

\bibitem[{Coupon {et~al.}(2009)Coupon, Ilbert, Kilbinger, McCracken, Mellier,
  Arnouts, Bertin, Hudelot, Schultheis, Le~Fèvre, Le~Brun, Guzzo, Bardelli,
  Zucca, Bolzonella, Garilli, Zamorani, Zanichelli, Tresse, \&
  Aussel}]{coupon_photometric_2009}
Coupon, J. {et~al.} 2009, \aap, 500, 981

\bibitem[{{Donahue} {et~al.}(2016){Donahue}, {Ettori}, {Rasia}, {Sayers},
  {Zitrin}, {Meneghetti}, {Voit}, {Golwala}, {Czakon}, {Yepes}, {Baldi},
  {Koekemoer}, \& {Postman}}]{2016ApJ...819...36D}
{Donahue}, M. {et~al.} 2016, \apj, 819, 36, 1601.04947

\bibitem[{{Eckert} {et~al.}(2016){Eckert}, {Ettori}, {Coupon}, {Gastaldello},
  {Pierre}, {Melin}, {Le Brun}, {McCarthy}, {Adami}, {Chiappetti}, {Faccioli},
  {Giles}, {Lavoie}, {Lef{\`e}vre}, {Lieu}, {Mantz}, {Maughan}, {McGee},
  {Pacaud}, {Paltani}, {Sadibekova}, {Smith}, \&
  {Ziparo}}]{2016A&A...592A..12E}
{Eckert}, D. {et~al.} 2016, \aap, 592, A12, 1512.03814, (XXL Paper XIII)

\bibitem[{{Eckert} {et~al.}(2013{\natexlab{a}}){Eckert}, {Ettori}, {Molendi},
  {Vazza}, \& {Paltani}}]{2013A&A...551A..23E}
{Eckert}, D., {Ettori}, S., {Molendi}, S., {Vazza}, F., \& {Paltani}, S.
  2013{\natexlab{a}}, \aap, 551, A23, 1301.0624

\bibitem[{{Eckert} {et~al.}(2017){Eckert}, {Ettori}, {Pointecouteau},
  {Molendi}, {Paltani}, \& {Tchernin}}]{Eckert2017}
{Eckert}, D., {Ettori}, S., {Pointecouteau}, E., {Molendi}, S., {Paltani}, S.,
  \& {Tchernin}, C. 2017, Astronomische Nachrichten, 338, 293, 1611.05051

\bibitem[{{Eckert} {et~al.}(2011){Eckert}, {Molendi}, \&
  {Paltani}}]{Eckert2011}
{Eckert}, D., {Molendi}, S., \& {Paltani}, S. 2011, \aap, 526, A79, 1011.3302

\bibitem[{{Eckert} {et~al.}(2013{\natexlab{b}}){Eckert}, {Molendi}, {Vazza},
  {Ettori}, \& {Paltani}}]{2013A&A...551A..22E}
{Eckert}, D., {Molendi}, S., {Vazza}, F., {Ettori}, S., \& {Paltani}, S.
  2013{\natexlab{b}}, \aap, 551, A22, 1301.0617

\bibitem[{{Ettori} {et~al.}(2013){Ettori}, {Donnarumma}, {Pointecouteau},
  {Reiprich}, {Giodini}, {Lovisari}, \& {Schmidt}}]{2013SSRv..177..119E}
{Ettori}, S., {Donnarumma}, A., {Pointecouteau}, E., {Reiprich}, T.~H.,
  {Giodini}, S., {Lovisari}, L., \& {Schmidt}, R.~W. 2013, \ssr, 177, 119,
  1303.3530

\bibitem[{{Fioc} \& {Rocca-Volmerange}(1997)}]{1997A&A...326..950F}
{Fioc}, M., \& {Rocca-Volmerange}, B. 1997, \aap, 326, 950, astro-ph/9707017

\bibitem[{{Ghirardini} {et~al.}(2019){Ghirardini}, {Eckert}, {Ettori},
  {Pointecouteau}, {Molendi}, {Gaspari}, {Rossetti}, {De Grandi}, {Roncarelli},
  {Bourdin}, {Mazzotta}, {Rasia}, \& {Vazza}}]{Ghirardini2019}
{Ghirardini}, V. {et~al.} 2019, \aap, 621, A41, 1805.00042

\bibitem[{{Giles} {et~al.}(2016){Giles}, {Maughan}, {Pacaud}, {Lieu}, {Clerc},
  {Pierre}, {Adami}, {Chiappetti}, {D{\'e}mocl{\'e}s}, {Ettori}, {Le
  F{\'e}vre}, {Ponman}, {Sadibekova}, {Smith}, {Willis}, \&
  {Ziparo}}]{giles_xxl_2015}
{Giles}, P.~A. {et~al.} 2016, \aap, 592, A3, 1512.03833, (XXL Paper III)

\bibitem[{{Giodini} {et~al.}(2013){Giodini}, {Lovisari}, {Pointecouteau},
  {Ettori}, {Reiprich}, \& {Hoekstra}}]{2013SSRv..177..247G}
{Giodini}, S., {Lovisari}, L., {Pointecouteau}, E., {Ettori}, S., {Reiprich},
  T.~H., \& {Hoekstra}, H. 2013, \ssr, 177, 247, 1305.3286

\bibitem[{{Gwyn}(2012)}]{2012AJ....143...38G}
{Gwyn}, S.~D.~J. 2012, \aj, 143, 38, 1101.1084

\bibitem[{{Hilton} {et~al.}(2018){Hilton}, {Hasselfield}, {Sif{\'o}n},
  {Battaglia}, {Aiola}, {Bharadwaj}, {Bond}, {Choi}, {Crichton}, {Datta},
  {Devlin}, {Dunkley}, {D{\"u}nner}, {Gallardo}, {Gralla}, {Hincks}, {Ho},
  {Hubmayr}, {Huffenberger}, {Hughes}, {Koopman}, {Kosowsky}, {Louis},
  {Madhavacheril}, {Marriage}, {Maurin}, {McMahon}, {Miyatake}, {Moodley},
  {N{\ae}ss}, {Nati}, {Newburgh}, {Niemack}, {Oguri}, {Page}, {Partridge},
  {Schmitt}, {Sievers}, {Spergel}, {Staggs}, {Trac}, {van Engelen},
  {Vavagiakis}, \& {Wollack}}]{2018ApJS..235...20H}
{Hilton}, M. {et~al.} 2018, \apjs, 235, 20, 1709.05600

\bibitem[{Hudelot {et~al.}(2012)Hudelot, Cuillandre, Withington, Goranova,
  McCracken, Magnard, Mellier, Regnault, Betoule, Aussel, Kavelaars, Fernique,
  Bonnarel, Ochsenbein, \& Ilbert}]{hudelot_vizier_2012}
Hudelot, P. {et~al.} 2012, VizieR Online Data Catalog, 2317

\bibitem[{{Hudson} {et~al.}(2010){Hudson}, {Mittal}, {Reiprich}, {Nulsen},
  {Andernach}, \& {Sarazin}}]{2010A&A...513A..37H}
{Hudson}, D.~S., {Mittal}, R., {Reiprich}, T.~H., {Nulsen}, P.~E.~J.,
  {Andernach}, H., \& {Sarazin}, C.~L. 2010, \aap, 513, A37, 0911.0409

\bibitem[{Hunter(2007)}]{Hunter2007}
Hunter, J.~D. 2007, Computing In Science \& Engineering, 9, 90

\bibitem[{{Hurier} {et~al.}(2013){Hurier}, {Mac{\'{\i}}as-P{\'e}rez}, \&
  {Hildebrandt}}]{2013A&A...558A.118H}
{Hurier}, G., {Mac{\'{\i}}as-P{\'e}rez}, J.~F., \& {Hildebrandt}, S. 2013,
  \aap, 558, A118, 1007.1149

\bibitem[{{Huterer} {et~al.}(2015){Huterer}, {Kirkby}, {Bean}, {Connolly},
  {Dawson}, {Dodelson}, {Evrard}, {Jain}, {Jarvis}, {Linder}, {Mandelbaum},
  {May}, {Raccanelli}, {Reid}, {Rozo}, {Schmidt}, {Sehgal}, {Slosar}, {van
  Engelen}, {Wu}, \& {Zhao}}]{2015APh....63...23H}
{Huterer}, D. {et~al.} 2015, Astroparticle Physics, 63, 23, 1309.5385

\bibitem[{Ilbert {et~al.}(2006)Ilbert, Arnouts, McCracken, Bolzonella, Bertin,
  Fevre, Mellier, Zamorani, Pello, Iovino, Tresse, Bottini, Garilli, Brun,
  Maccagni, Picat, Scaramella, Scodeggio, Vettolani, Zanichelli, Adami,
  Bardelli, Cappi, Charlot, Ciliegi, Contini, Cucciati, Foucaud, Franzetti,
  Gavignaud, Guzzo, Marano, Marinoni, Mazure, Meneux, Merighi, Paltani, Pollo,
  Pozzetti, Radovich, Zucca, Bondi, Bongiorno, Busarello, De~La~Torre,
  Gregorini, Lamareille, Mathez, Merluzzi, Ripepi, Rizzo, \&
  Vergani}]{ilbert_accurate_2006}
Ilbert, O. {et~al.} 2006, \aap, 457, 841, arXiv: astro-ph/0603217

\bibitem[{{Itoh} {et~al.}(1998){Itoh}, {Kohyama}, \&
  {Nozawa}}]{1998ApJ...502....7I}
{Itoh}, N., {Kohyama}, Y., \& {Nozawa}, S. 1998, \apj, 502, 7, astro-ph/9712289

\bibitem[{Jones {et~al.}(2001)Jones, Oliphant, Peterson, {et~al.}}]{Jones2001}
Jones, E., Oliphant, T., Peterson, P., {et~al.} 2001, {SciPy}: Open source
  scientific tools for {Python}

\bibitem[{{Kaiser}(1986)}]{1986MNRAS.222..323K}
{Kaiser}, N. 1986, \mnras, 222, 323

\bibitem[{{Katayama} {et~al.}(2003){Katayama}, {Hayashida}, {Takahara}, \&
  {Fujita}}]{2003ApJ...585..687K}
{Katayama}, H., {Hayashida}, K., {Takahara}, F., \& {Fujita}, Y. 2003, \apj,
  585, 687

\bibitem[{Kinney {et~al.}(1996)Kinney, Calzetti, Bohlin, McQuade,
  Storchi-Bergmann, \& Schmitt}]{kinney_template_1996}
Kinney, A.~L., Calzetti, D., Bohlin, R.~C., McQuade, K., Storchi-Bergmann, T.,
  \& Schmitt, H.~R. 1996, \apj, 467, 38

\bibitem[{{Komatsu} {et~al.}(2001){Komatsu}, {Matsuo}, {Kitayama}, {Hattori},
  {Kawabe}, {Kohno}, {Kuno}, {Schindler}, {Suto}, \&
  {Yoshikawa}}]{2001PASJ...53...57K}
{Komatsu}, E. {et~al.} 2001, \pasj, 53, 57, astro-ph/0006293

\bibitem[{{Koulouridis} \& {Bartalucci}(2019)}]{2019A&A...623L..10K}
{Koulouridis}, E., \& {Bartalucci}, I. 2019, \aap, 623, L10, 1903.02919

\bibitem[{{Koulouridis} {et~al.}(2018){Koulouridis}, {Ricci}, {Giles}, {Adami},
  {Ramos-Ceja}, {Pierre}, {Plionis}, {Lidman}, {Georgantopoulos}, {Chiappetti},
  {Elyiv}, {Ettori}, {Faccioli}, {Fotopoulou}, {Gastaldello}, {Pacaud},
  {Paltani}, \& {Vignali}}]{2018A&A...620A..20K}
{Koulouridis}, E. {et~al.} 2018, \aap, 620, A20, 1809.00683, (XXL Paper XXXV)

\bibitem[{{Kravtsov} {et~al.}(2006){Kravtsov}, {Vikhlinin}, \&
  {Nagai}}]{2006ApJ...650..128K}
{Kravtsov}, A.~V., {Vikhlinin}, A., \& {Nagai}, D. 2006, \apj, 650, 128,
  astro-ph/0603205

\bibitem[{{Le Brun} {et~al.}(2017){Le Brun}, {McCarthy}, {Schaye}, \&
  {Ponman}}]{2017MNRAS.466.4442L}
{Le Brun}, A.~M.~C., {McCarthy}, I.~G., {Schaye}, J., \& {Ponman}, T.~J. 2017,
  \mnras, 466, 4442, 1606.04545

\bibitem[{{Lieu} {et~al.}(2016){Lieu}, {Smith}, {Giles}, {Ziparo}, {Maughan},
  {D{\'e}mocl{\`e}s}, {Pacaud}, {Pierre}, {Adami}, {Bah{\'e}}, {Clerc},
  {Chiappetti}, {Eckert}, {Ettori}, {Lavoie}, {Le Fevre}, {McCarthy},
  {Kilbinger}, {Ponman}, {Sadibekova}, \& {Willis}}]{2016A&A...592A...4L}
{Lieu}, M. {et~al.} 2016, \aap, 592, A4, 1512.03857, (XXL Paper IV)

\bibitem[{Lin \& Mohr(2004)}]{0004-637X-617-2-879}
Lin, Y.-T., \& Mohr, J.~J. 2004, \apj, 617, 879

\bibitem[{Lin {et~al.}(2006)Lin, Mohr, Gonzalez, \&
  Stanford}]{lin_evolution_2006}
Lin, Y.-T., Mohr, J.~J., Gonzalez, A.~H., \& Stanford, S.~A. 2006, \apjl, 650,
  L99

\bibitem[{{Mahdavi} {et~al.}(2013){Mahdavi}, {Hoekstra}, {Babul}, {Bildfell},
  {Jeltema}, \& {Henry}}]{Mahdavi2013}
{Mahdavi}, A., {Hoekstra}, H., {Babul}, A., {Bildfell}, C., {Jeltema}, T., \&
  {Henry}, J.~P. 2013, \apj, 767, 116, 1210.3689

\bibitem[{{Mandelbaum} {et~al.}(2018){Mandelbaum}, {Miyatake}, {Hamana},
  {Oguri}, {Simet}, {Armstrong}, {Bosch}, {Murata}, {Lanusse}, {Leauthaud},
  {Coupon}, {More}, {Takada}, {Miyazaki}, {Speagle}, {Shirasaki}, {Sif{\'o}n},
  {Huang}, {Nishizawa}, {Medezinski}, {Okura}, {Okabe}, {Czakon}, {Takahashi},
  {Coulton}, {Hikage}, {Komiyama}, {Lupton}, {Strauss}, {Tanaka}, \&
  {Utsumi}}]{Mandelbaum2018}
{Mandelbaum}, R. {et~al.} 2018, \pasj, 70, S25, 1705.06745

\bibitem[{{Mantz} {et~al.}(2018){Mantz}, {Abdulla}, {Allen}, {Carlstrom},
  {Logan}, {Marrone}, {Maughan}, {Willis}, {Pacaud}, \&
  {Pierre}}]{2018A&A...620A...2M}
{Mantz}, A.~B. {et~al.} 2018, \aap, 620, A2, 1703.08221, (XXL Paper XVII)

\bibitem[{{Maughan} {et~al.}(2007){Maughan}, {Jones}, {Jones}, \& {Van
  Speybroeck}}]{Maughan2007}
{Maughan}, B.~J., {Jones}, C., {Jones}, L.~R., \& {Van Speybroeck}, L. 2007,
  \apj, 659, 1125, astro-ph/0609690

\bibitem[{{McCarthy} {et~al.}(2011){McCarthy}, {Schaye}, {Bower}, {Ponman},
  {Booth}, {Dalla Vecchia}, \& {Springel}}]{2011MNRAS.412.1965M}
{McCarthy}, I.~G., {Schaye}, J., {Bower}, R.~G., {Ponman}, T.~J., {Booth},
  C.~M., {Dalla Vecchia}, C., \& {Springel}, V. 2011, \mnras, 412, 1965,
  1008.4799

\bibitem[{Monfardini {et~al.}(2011)Monfardini, Benoit, Bideaud, Swenson,
  Cruciani, Camus, Hoffmann, Désert, {S. Doyle}, Ade, Mauskopf, Tucker,
  Roesch, Leclercq, Schuster, Endo, Baryshev, Baselmans, Ferrari, Yates,
  Bourrion, Macias-Perez, Vescovi, Calvo, \&
  Giordano}]{monfardini_dual-band_2011}
Monfardini, A. {et~al.} 2011, ApJS, 194, 24

\bibitem[{{Mroczkowski} {et~al.}(2009){Mroczkowski}, {Bonamente}, {Carlstrom},
  {Culverhouse}, {Greer}, {Hawkins}, {Hennessy}, {Joy}, {Lamb}, {Leitch},
  {Loh}, {Maughan}, {Marrone}, {Miller}, {Muchovej}, {Nagai}, {Pryke}, {Sharp},
  \& {Woody}}]{2009ApJ...694.1034M}
{Mroczkowski}, T. {et~al.} 2009, \apj, 694, 1034, 0809.5077

\bibitem[{{Mroczkowski} {et~al.}(2012){Mroczkowski}, {Dicker}, {Sayers},
  {Reese}, {Mason}, {Czakon}, {Romero}, {Young}, {Devlin}, {Golwala},
  {Korngut}, {Sarazin}, {Bock}, {Koch}, {Lin}, {Molnar}, {Pierpaoli}, {Umetsu},
  \& {Zemcov}}]{Mroczkowski2012}
------. 2012, \apj, 761, 47, 1205.0052

\bibitem[{{Nagai} {et~al.}(2007){Nagai}, {Kravtsov}, \&
  {Vikhlinin}}]{2007ApJ...668....1N}
{Nagai}, D., {Kravtsov}, A.~V., \& {Vikhlinin}, A. 2007, \apj, 668, 1,
  astro-ph/0703661

\bibitem[{{Oguri} {et~al.}(2018){Oguri}, {Lin}, {Lin}, {Nishizawa}, {More},
  {More}, {Hsieh}, {Medezinski}, {Miyatake}, {Jian}, {Lin}, {Takada}, {Okabe},
  {Speagle}, {Coupon}, {Leauthaud}, {Lupton}, {Miyazaki}, {Price}, {Tanaka},
  {Chiu}, {Komiyama}, {Okura}, {Tanaka}, \& {Usuda}}]{2018PASJ...70S..20O}
{Oguri}, M. {et~al.} 2018, \pasj, 70, S20, 1701.00818

\bibitem[{{Pacaud} {et~al.}(2016){Pacaud}, {Clerc}, {Giles}, {Adami},
  {Sadibekova}, {Pierre}, {Maughan}, {Lieu}, {Le F{\`e}vre}, {Alis}, {Altieri},
  {Ardila}, {Baldry}, {Benoist}, {Birkinshaw}, {Chiappetti},
  {D{\'e}mocl{\`e}s}, {Eckert}, {Evrard}, {Faccioli}, {Gastaldello}, {Guennou},
  {Horellou}, {Iovino}, {Koulouridis}, {Le Brun}, {Lidman}, {Liske},
  {Maurogordato}, {Menanteau}, {Owers}, {Poggianti}, {Pomar{\`e}de}, {Pompei},
  {Ponman}, {Rapetti}, {Reiprich}, {Smith}, {Tuffs}, {Valageas}, {Valtchanov},
  {Willis}, \& {Ziparo}}]{pacaud_xxl_2015}
{Pacaud}, F. {et~al.} 2016, \aap, 592, A2, 1512.04264, (XXL Paper II)

\bibitem[{P\'erez \& Granger(2007)}]{Perez2007}
P\'erez, F., \& Granger, B.~E. 2007, Computing in Science and Engineering, 9,
  21

\bibitem[{{Perotto} {et~al.}(2019){Perotto}, {Ponthieu},
  {Mac{\'\i}as-P{\'e}rez}, {Adam}, {Ade}, {Andr{\'e}}, {Andrianasolo},
  {Aussel}, {Beelen}, {Beno{\^\i}t}, {Berta}, {Bideaud}, {Bourrion}, {Calvo},
  {Catalano}, {Comis}, {De Petris}, {D{\'e}sert}, {Doyle}, {Driessen},
  {Garc{\'\i}a}, {Gomez}, {Goupy}, {John}, {K{\'e}ruzor{\'e}}, {Kramer},
  {Ladjelate}, {Lagache}, {Leclercq}, {Lestrade}, {Maury}, {Mauskopf}, {Mayet},
  {Monfardini}, {Navarro}, {Pe{\~n}alver}, {Pierfederici}, {Pisano},
  {Rev{\'e}ret}, {Ritacco}, {Romero}, {Roussel}, {Ruppin}, {Schuster}, {Shu},
  {Sievers}, {Tucker}, \& {Zylka}}]{2019arXiv191002038P}
{Perotto}, L. {et~al.} 2019, \aap, in press, arXiv:1910.02038, 1910.02038

\bibitem[{{Pierre} {et~al.}(2016){Pierre}, {Pacaud}, {Adami}, {Alis},
  {Altieri}, {Baran}, {Benoist}, {Birkinshaw}, {Bongiorno}, {Bremer}, {Brusa},
  {Butler}, {Ciliegi}, {Chiappetti}, {Clerc}, {Corasaniti}, {Coupon}, {De
  Breuck}, {Democles}, {Desai}, {Delhaize}, {Devriendt}, {Dubois}, {Eckert},
  {Elyiv}, {Ettori}, {Evrard}, {Faccioli}, {Farahi}, {Ferrari}, {Finet},
  {Fotopoulou}, {Fourmanoit}, {Gandhi}, {Gastaldello}, {Gastaud},
  {Georgantopoulos}, {Giles}, {Guennou}, {Guglielmo}, {Horellou}, {Husband},
  {Huynh}, {Iovino}, {Kilbinger}, {Koulouridis}, {Lavoie}, {Le Brun}, {Le
  Fevre}, {Lidman}, {Lieu}, {Lin}, {Mantz}, {Maughan}, {Maurogordato},
  {McCarthy}, {McGee}, {Melin}, {Melnyk}, {Menanteau}, {Novak}, {Paltani},
  {Plionis}, {Poggianti}, {Pomarede}, {Pompei}, {Ponman}, {Ramos-Ceja},
  {Ranalli}, {Rapetti}, {Raychaudury}, {Reiprich}, {Rottgering}, {Rozo},
  {Rykoff}, {Sadibekova}, {Santos}, {Sauvageot}, {Schimd}, {Sereno}, {Smith},
  {Smol{\v c}i{\'c}}, {Snowden}, {Spergel}, {Stanford}, {Surdej}, {Valageas},
  {Valotti}, {Valtchanov}, {Vignali}, {Willis}, \& {Ziparo}}]{pierre_xxl_2015}
{Pierre}, M. {et~al.} 2016, \aap, 592, A1, 1512.04317, (XXL Paper I)

\bibitem[{Pierre {et~al.}(2011)Pierre, Pacaud, Juin, Melin, Clerc, \&
  Corasaniti}]{pierre_precision_2011}
Pierre, M., Pacaud, F., Juin, J.~B., Melin, J.~B., Clerc, N., \& Corasaniti,
  P.~S. 2011, \mnras, 414, 1732, arXiv: 1009.3182

\bibitem[{{Pillepich} {et~al.}(2012){Pillepich}, {Porciani}, \&
  {Reiprich}}]{2012MNRAS.422...44P}
{Pillepich}, A., {Porciani}, C., \& {Reiprich}, T.~H. 2012, \mnras, 422, 44,
  1111.6587

\bibitem[{{Planck Collaboration} {et~al.}(2014){Planck Collaboration}, {Ade},
  {Aghanim}, {Armitage-Caplan}, {Arnaud}, {Ashdown}, {Atrio-Barandela},
  {Aumont}, {Baccigalupi}, {Banday}, \& et~al.}]{Planck2014XX}
{Planck Collaboration} {et~al.} 2014, \aap, 571, A20, 1303.5080

\bibitem[{{Planck Collaboration} {et~al.}(2013){Planck Collaboration}, {Ade},
  {Aghanim}, {Arnaud}, {Ashdown}, {Atrio-Barandela}, {Aumont}, {Baccigalupi},
  {Balbi}, {Banday}, \& et~al.}]{2013A&A...550A.131P}
------. 2013, \aap, 550, A131, 1207.4061

\bibitem[{{Planck Collaboration} {et~al.}(2016{\natexlab{a}}){Planck
  Collaboration}, {Ade}, {Aghanim}, {Arnaud}, {Ashdown}, {Aumont},
  {Baccigalupi}, {Banday}, {Barreiro}, {Barrena}, {Bartlett}, {Bartolo},
  {Battaner}, {Battye}, {Benabed}, {Beno{\^\i}t}, {Benoit-L{\'e}vy}, {Bernard},
  {Bersanelli}, {Bielewicz}, {Bikmaev}, {B{\"o}hringer}, {Bonaldi}, {Bonavera},
  {Bond}, {Borrill}, {Bouchet}, {Bucher}, {Burenin}, {Burigana}, {Butler},
  {Calabrese}, {Cardoso}, {Carvalho}, {Catalano}, {Challinor}, {Chamballu},
  {Chary}, {Chiang}, {Chon}, {Christensen}, {Clements}, {Colombi}, {Colombo},
  {Combet}, {Comis}, {Couchot}, {Coulais}, {Crill}, {Curto}, {Cuttaia},
  {Dahle}, {Danese}, {Davies}, {Davis}, {De Bernardis}, {de Rosa}, {de Zotti},
  {Delabrouille}, {D{\'e}sert}, {Dickinson}, {Diego}, {Dolag}, {Dole},
  {Donzelli}, {Dor{\'e}}, {Douspis}, {Ducout}, {Dupac}, {Efstathiou},
  {Eisenhardt}, {Elsner}, {En{\ss}lin}, {Eriksen}, {Falgarone}, {Fergusson},
  {Feroz}, {Ferragamo}, {Finelli}, {Forni}, {Frailis}, {Fraisse}, {Franceschi},
  {Frejsel}, {Galeotta}, {Galli}, {Ganga}, {G{\'e}nova-Santos}, {Giard},
  {Giraud-H{\'e}raud}, {Gjerl{\o}w}, {Gonz{\'a}lez-Nuevo}, {G{\'o}rski},
  {Grainge}, {Gratton}, {Gregorio}, {Gruppuso}, {Gudmundsson}, {Hansen},
  {Hanson}, {Harrison}, {Hempel}, {Henrot- Versill{\'e}},
  {Hern{\'a}ndez-Monteagudo}, {Herranz}, {Hildebrandt}, {Hivon}, {Hobson},
  {Holmes}, {Hornstrup}, {Hovest}, {Huffenberger}, {Hurier}, {Jaffe}, {Jaffe},
  {Jin}, {Jones}, {Juvela}, {Keih{\"a}nen}, {Keskitalo}, {Khamitov}, {Kisner},
  {Kneissl}, {Knoche}, {Kunz}, {Kurki-Suonio}, {Lagache}, {Lamarre}, {Lasenby},
  {Lattanzi}, {Lawrence}, {Leonardi}, {Lesgourgues}, {Levrier}, {Liguori},
  {Lilje}, {Linden-V{\o}rnle}, {L{\'o}pez- Caniego}, {Lubin},
  {Mac{\'\i}as-P{\'e}rez}, {Maggio}, {Maino}, {Mak}, {Mandolesi}, {Mangilli},
  {Martin}, {Mart{\'\i}nez-Gonz{\'a}lez}, {Masi}, {Matarrese}, {Mazzotta},
  {McGehee}, {Mei}, {Melchiorri}, {Melin}, {Mendes}, {Mennella}, {Migliaccio},
  {Mitra}, {Miville-Desch{\^e}nes}, {Moneti}, {Montier}, {Morgante},
  {Mortlock}, {Moss}, {Munshi}, {Murphy}, {Naselsky}, {Nastasi}, {Nati},
  {Natoli}, {Netterfield}, {N{\o}rgaard-Nielsen}, {Noviello}, {Novikov},
  {Novikov}, {Olamaie}, {Oxborrow}, {Paci}, {Pagano}, {Pajot}, {Paoletti},
  {Pasian}, {Patanchon}, {Pearson}, {Perdereau}, {Perotto}, {Perrott},
  {Perrotta}, {Pettorino}, {Piacentini}, {Piat}, {Pierpaoli}, {Pietrobon},
  {Plaszczynski}, {Pointecouteau}, {Polenta}, {Pratt}, {Pr{\'e}zeau}, {Prunet},
  {Puget}, {Rachen}, {Reach}, {Rebolo}, {Reinecke}, {Remazeilles}, {Renault},
  {Renzi}, {Ristorcelli}, {Rocha}, {Rosset}, {Rossetti}, {Roudier}, {Rozo},
  {Rubi{\~n}o-Mart{\'\i}n}, {Rumsey}, {Rusholme}, {Rykoff}, {Sandri}, {Santos},
  {Saunders}, {Savelainen}, {Savini}, {Schammel}, {Scott}, {Seiffert},
  {Shellard}, {Shimwell}, {Spencer}, {Stanford}, {Stern}, {Stolyarov},
  {Stompor}, {Streblyanska}, {Sudiwala}, {Sunyaev}, {Sutton}, {Suur-Uski},
  {Sygnet}, {Tauber}, {Terenzi}, {Toffolatti}, {Tomasi}, {Tramonte},
  {Tristram}, {Tucci}, {Tuovinen}, {Umana}, {Valenziano}, {Valiviita}, {Van
  Tent}, {Vielva}, {Villa}, {Wade}, {Wandelt}, {Wehus}, {White}, {Wright},
  {Yvon}, {Zacchei}, \& {Zonca}}]{2016A&A...594A..27P}
------. 2016{\natexlab{a}}, \aap, 594, A27

\bibitem[{{Planck Collaboration} {et~al.}(2016{\natexlab{b}}){Planck
  Collaboration}, {Ade}, {Aghanim}, {Arnaud}, {Ashdown}, {Aumont},
  {Baccigalupi}, {Banday}, {Barreiro}, {Barrena}, \& et~al.}]{PlanckXXVII2015}
------. 2016{\natexlab{b}}, \aap, 594, A27, 1502.01598

\bibitem[{{Planck Collaboration} {et~al.}(2016{\natexlab{c}}){Planck
  Collaboration}, {Ade}, {Aghanim}, {Arnaud}, {Ashdown}, {Aumont},
  {Baccigalupi}, {Banday}, {Barreiro}, {Bartlett}, {Bartolo}, {Battaner},
  {Battye}, {Benabed}, {Beno{\^\i}t}, {Benoit-L{\'e}vy}, {Bernard},
  {Bersanelli}, {Bielewicz}, {Bock}, {Bonaldi}, {Bonavera}, {Bond}, {Borrill},
  {Bouchet}, {Bucher}, {Burigana}, {Butler}, {Calabrese}, {Cardoso},
  {Catalano}, {Challinor}, {Chamballu}, {Chary}, {Chiang}, {Christensen},
  {Church}, {Clements}, {Colombi}, {Colombo}, {Combet}, {Comis}, {Couchot},
  {Coulais}, {Crill}, {Curto}, {Cuttaia}, {Danese}, {Davies}, {Davis}, {De
  Bernardis}, {de Rosa}, {de Zotti}, {Delabrouille}, {D{\'e}sert}, {Diego},
  {Dolag}, {Dole}, {Donzelli}, {Dor{\'e}}, {Douspis}, {Ducout}, {Dupac},
  {Efstathiou}, {Elsner}, {En{\ss}lin}, {Eriksen}, {Falgarone}, {Fergusson},
  {Finelli}, {Forni}, {Frailis}, {Fraisse}, {Franceschi}, {Frejsel},
  {Galeotta}, {Galli}, {Ganga}, {Giard}, {Giraud-H{\'e}raud}, {Gjerl{\o}w},
  {Gonz{\'a}lez-Nuevo}, {G{\'o}rski}, {Gratton}, {Gregorio}, {Gruppuso},
  {Gudmundsson}, {Hansen}, {Hanson}, {Harrison}, {Henrot-Versill{\'e}},
  {Hern{\'a}ndez-Monteagudo}, {Herranz}, {Hildebrandt}, {Hivon}, {Hobson},
  {Holmes}, {Hornstrup}, {Hovest}, {Huffenberger}, {Hurier}, {Jaffe}, {Jaffe},
  {Jones}, {Juvela}, {Keih{\"a}nen}, {Keskitalo}, {Kisner}, {Kneissl},
  {Knoche}, {Kunz}, {Kurki-Suonio}, {Lagache}, {L{\"a}hteenm{\"a}ki},
  {Lamarre}, {Lasenby}, {Lattanzi}, {Lawrence}, {Leonardi}, {Lesgourgues},
  {Levrier}, {Liguori}, {Lilje}, {Linden-V{\o}rnle}, {L{\'o}pez- Caniego},
  {Lubin}, {Mac{\'\i}as-P{\'e}rez}, {Maggio}, {Maino}, {Mandolesi}, {Mangilli},
  {Maris}, {Martin}, {Mart{\'\i}nez-Gonz{\'a}lez}, {Masi}, {Matarrese},
  {McGehee}, {Meinhold}, {Melchiorri}, {Melin}, {Mendes}, {Mennella},
  {Migliaccio}, {Mitra}, {Miville-Desch{\^e}nes}, {Moneti}, {Montier},
  {Morgante}, {Mortlock}, {Moss}, {Munshi}, {Murphy}, {Naselsky}, {Nati},
  {Natoli}, {Netterfield}, {N{\o}rgaard-Nielsen}, {Noviello}, {Novikov},
  {Novikov}, {Oxborrow}, {Paci}, {Pagano}, {Pajot}, {Paoletti}, {Partridge},
  {Pasian}, {Patanchon}, {Pearson}, {Perdereau}, {Perotto}, {Perrotta},
  {Pettorino}, {Piacentini}, {Piat}, {Pierpaoli}, {Pietrobon}, {Plaszczynski},
  {Pointecouteau}, {Polenta}, {Popa}, {Pratt}, {Pr{\'e}zeau}, {Prunet},
  {Puget}, {Rachen}, {Rebolo}, {Reinecke}, {Remazeilles}, {Renault}, {Renzi},
  {Ristorcelli}, {Rocha}, {Roman}, {Rosset}, {Rossetti}, {Roudier},
  {Rubi{\~n}o-Mart{\'\i}n}, {Rusholme}, {Sandri}, {Santos}, {Savelainen},
  {Savini}, {Scott}, {Seiffert}, {Shellard}, {Spencer}, {Stolyarov}, {Stompor},
  {Sudiwala}, {Sunyaev}, {Sutton}, {Suur- Uski}, {Sygnet}, {Tauber}, {Terenzi},
  {Toffolatti}, {Tomasi}, {Tristram}, {Tucci}, {Tuovinen}, {T{\"u}rler},
  {Umana}, {Valenziano}, {Valiviita}, {Van Tent}, {Vielva}, {Villa}, {Wade},
  {Wandelt}, {Wehus}, {Weller}, {White}, {Yvon}, {Zacchei}, \&
  {Zonca}}]{2016A&A...594A..24P}
------. 2016{\natexlab{c}}, \aap, 594, A24

\bibitem[{{Planck Collaboration} {et~al.}(2018){Planck Collaboration},
  {Aghanim}, {Akrami}, {Ashdown}, {Aumont}, {Baccigalupi}, {Ballardini},
  {Banday}, {Barreiro}, \& {Bartolo}}]{2018arXiv180706209P}
------. 2018, arXiv e-prints, arXiv:1807.06209, 1807.06209

\bibitem[{{Planck Collaboration} {et~al.}(2016{\natexlab{d}}){Planck
  Collaboration}, {Aghanim}, {Arnaud}, {Ashdown}, {Aumont}, {Baccigalupi},
  {Banday}, {Barreiro}, {Bartlett}, {Bartolo}, \& et~al.}]{2016A&A...594A..22P}
------. 2016{\natexlab{d}}, \aap, 594, A22, 1502.01596

\bibitem[{{Pointecouteau} {et~al.}(1999){Pointecouteau}, {Giard}, {Benoit},
  {D{\'e}sert}, {Aghanim}, {Coron}, {Lamarre}, \&
  {Delabrouille}}]{1999ApJ...519L.115P}
{Pointecouteau}, E., {Giard}, M., {Benoit}, A., {D{\'e}sert}, F.~X., {Aghanim},
  N., {Coron}, N., {Lamarre}, J.~M., \& {Delabrouille}, J. 1999, \apjl, 519,
  L115

\bibitem[{{Pratt} {et~al.}(2019){Pratt}, {Arnaud}, {Biviano}, {Eckert},
  {Ettori}, {Nagai}, {Okabe}, \& {Reiprich}}]{2019SSRv..215...25P}
{Pratt}, G.~W., {Arnaud}, M., {Biviano}, A., {Eckert}, D., {Ettori}, S.,
  {Nagai}, D., {Okabe}, N., \& {Reiprich}, T.~H. 2019, \ssr, 215, 25,
  1902.10837

\bibitem[{{Pratt} {et~al.}(2010){Pratt}, {Arnaud}, {Piffaretti},
  {B{\"o}hringer}, {Ponman}, {Croston}, {Voit}, {Borgani}, \&
  {Bower}}]{2010A&A...511A..85P}
{Pratt}, G.~W. {et~al.} 2010, \aap, 511, A85, 0909.3776

\bibitem[{{Rasia} {et~al.}(2014){Rasia}, {Lau}, {Borgani}, {Nagai}, {Dolag},
  {Avestruz}, {Granato}, {Mazzotta}, {Murante}, {Nelson}, \&
  {Ragone-Figueroa}}]{Rasia2014}
{Rasia}, E. {et~al.} 2014, \apj, 791, 96

\bibitem[{Reiprich {et~al.}(2013)Reiprich, Basu, Ettori, Israel, Lovisari,
  Molendi, Pointecouteau, \& Roncarelli}]{Reiprich2013}
Reiprich, T.~H., Basu, K., Ettori, S., Israel, H., Lovisari, L., Molendi, S.,
  Pointecouteau, E., \& Roncarelli, M. 2013, Space Science Reviews, 177, 195

\bibitem[{{Ricci} {et~al.}(2018){Ricci}, {Benoist}, {Maurogordato}, {Adami},
  {Chiappetti}, {Gastaldello}, {Guglielmo}, {Poggianti}, {Sereno}, {Adam},
  {Arnouts}, {Cappi}, {Koulouridis}, {Pacaud}, {Pierre}, \&
  {Ramos-Ceja}}]{Ricci_LF}
{Ricci}, M. {et~al.} 2018, \aap, 620, A13, 1807.03207, (XXL Paper XXVIII)

\bibitem[{{Romero} {et~al.}(2017){Romero}, {Mason}, {Sayers}, {Mroczkowski},
  {Sarazin}, {Donahue}, {Baldi}, {Clarke}, {Young}, {Sievers}, {Dicker},
  {Reese}, {Czakon}, {Devlin}, {Korngut}, \& {Golwala}}]{2017ApJ...838...86R}
{Romero}, C.~E. {et~al.} 2017, \apj, 838, 86, 1608.03980

\bibitem[{{Rossetti} {et~al.}(2017){Rossetti}, {Gastaldello}, {Eckert}, {Della
  Torre}, {Pantiri}, {Cazzoletti}, \& {Molendi}}]{2017MNRAS.468.1917R}
{Rossetti}, M., {Gastaldello}, F., {Eckert}, D., {Della Torre}, M., {Pantiri},
  G., {Cazzoletti}, P., \& {Molendi}, S. 2017, \mnras, 468, 1917, 1702.06961

\bibitem[{{Rossetti} {et~al.}(2016){Rossetti}, {Gastaldello}, {Ferioli},
  {Bersanelli}, {De Grandi}, {Eckert}, {Ghizzardi}, {Maino}, \&
  {Molendi}}]{2016MNRAS.457.4515R}
{Rossetti}, M. {et~al.} 2016, \mnras, 457, 4515, 1512.00410

\bibitem[{{Ruppin} {et~al.}(2018{\natexlab{a}}){Ruppin}, {Mayet}, {Pratt},
  {Adam}, {Ade}, {Andr{\'e}}, {Arnaud}, {Aussel}, {Bartalucci}, {Beelen},
  {Beno{\^i}t}, {Bideaud}, {Bourrion}, {Calvo}, {Catalano}, {Comis}, {De
  Petris}, {D{\'e}sert}, {Doyle}, {Driessen}, {Goupy}, {Kramer}, {Lagache},
  {Leclercq}, {Lestrade}, {Mac{\'{\i}}as-P{\'e}rez}, {Mauskopf}, {Monfardini},
  {Perotto}, {Pisano}, {Pointecouteau}, {Ponthieu}, {Rev{\'e}ret}, {Ritacco},
  {Romero}, {Roussel}, {Schuster}, {Sievers}, {Tucker}, \&
  {Zylka}}]{ruppin_first_2017}
{Ruppin}, F. {et~al.} 2018{\natexlab{a}}, \aap, 615, A112, 1712.09587

\bibitem[{{Ruppin} {et~al.}(2018{\natexlab{b}}){Ruppin}, {Mayet}, {Pratt},
  {Adam}, {Ade}, {Andr{\'e}}, {Arnaud}, {Aussel}, {Bartalucci}, {Beelen},
  {Beno{\^i}t}, {Bideaud}, {Bourrion}, {Calvo}, {Catalano}, {Comis}, {De
  Petris}, {D{\'e}sert}, {Doyle}, {Driessen}, {Goupy}, {Kramer}, {Lagache},
  {Leclercq}, {Lestrade}, {Mac{\'{\i}}as-P{\'e}rez}, {Mauskopf}, {Monfardini},
  {Perotto}, {Pisano}, {Pointecouteau}, {Ponthieu}, {Rev{\'e}ret}, {Ritacco},
  {Romero}, {Roussel}, {Schuster}, {Sievers}, {Tucker}, \&
  {Zylka}}]{ruppin_first_2018}
------. 2018{\natexlab{b}}, \aap, 615, A112, 1712.09587

\bibitem[{{Salvatier} {et~al.}(2016){Salvatier}, {Wiecki}, \&
  {Fonnesbeck}}]{pymc3}
{Salvatier}, J., {Wiecki}, T.~V., \& {Fonnesbeck}, C. 2016, PeerJ Computer
  Science, 2:e55

\bibitem[{{Santos} {et~al.}(2008){Santos}, {Rosati}, {Tozzi}, {B{\"o}hringer},
  {Ettori}, \& {Bignamini}}]{2008A&A...483...35S}
{Santos}, J.~S., {Rosati}, P., {Tozzi}, P., {B{\"o}hringer}, H., {Ettori}, S.,
  \& {Bignamini}, A. 2008, \aap, 483, 35, 0802.1445

\bibitem[{{Sayers} {et~al.}(2013){Sayers}, {Mroczkowski}, {Zemcov}, {Korngut},
  {Bock}, {Bulbul}, {Czakon}, {Egami}, {Golwala}, {Koch}, {Lin}, {Mantz},
  {Molnar}, {Moustakas}, {Pierpaoli}, {Rawle}, {Reese}, {Rex}, {Shitanishi},
  {Siegel}, \& {Umetsu}}]{Sayers2013}
{Sayers}, J. {et~al.} 2013, \apj, 778, 52, 1312.3680

\bibitem[{{Sereno} {et~al.}(2015){Sereno}, {Ettori}, \&
  {Moscardini}}]{2015MNRAS.450.3649S}
{Sereno}, M., {Ettori}, S., \& {Moscardini}, L. 2015, \mnras, 450, 3649,
  1407.7869

\bibitem[{{Sunyaev} \& {Zel'dovich}(1972)}]{Sunyaev1972}
{Sunyaev}, R.~A., \& {Zel'dovich}, Y.~B. 1972, \apspr, 4, 173

\bibitem[{{Sunyaev} \& {Zel'dovich}(1980)}]{Sunyaev1980}
------. 1980, \araa, 18, 537

\bibitem[{{Tanaka} {et~al.}(2018){Tanaka}, {Coupon}, {Hsieh}, {Mineo},
  {Nishizawa}, {Speagle}, {Furusawa}, {Miyazaki}, \&
  {Murayama}}]{2018PASJ...70S...9T}
{Tanaka}, M. {et~al.} 2018, \pasj, 70, S9, 1704.05988

\bibitem[{{Tchernin} {et~al.}(2016){Tchernin}, {Eckert}, {Ettori},
  {Pointecouteau}, {Paltani}, {Molendi}, {Hurier}, {Gastaldello}, {Lau},
  {Nagai}, {Roncarelli}, \& {Rossetti}}]{2016A&A...595A..42T}
{Tchernin}, C. {et~al.} 2016, \aap, 595, A42, 1606.05657

\bibitem[{{Umetsu} {et~al.}(2020){Umetsu}, {Sereno}, {Lieu}, {Miyatake},
  {Medezinski}, {Nishizawa}, {Giles}, {Gastaldello}, {McCarthy}, {Kilbinger},
  {Birkinshaw}, {Ettori}, {Okabe}, {Chiu}, {Coupon}, {Eckert}, {Fujita},
  {Higuchi}, {Koulouridis}, {Maughan}, {Miyazaki}, {Oguri}, {Pacaud}, {Pierre},
  {Rapetti}, \& {Smith}}]{2020ApJ...890..148U}
{Umetsu}, K. {et~al.} 2020, \apj, 890, 148, 1909.10524

\bibitem[{{Van Der Walt} {et~al.}(2011){Van Der Walt}, {Colbert}, \&
  {Varoquaux}}]{VanDerWalt2011}
{Van Der Walt}, S., {Colbert}, S.~C., \& {Varoquaux}, G. 2011, ArXiv e-prints,
  1102.1523

\bibitem[{{Vikhlinin} {et~al.}(2006){Vikhlinin}, {Kravtsov}, {Forman}, {Jones},
  {Markevitch}, {Murray}, \& {Van Speybroeck}}]{2006ApJ...640..691V}
{Vikhlinin}, A., {Kravtsov}, A., {Forman}, W., {Jones}, C., {Markevitch}, M.,
  {Murray}, S.~S., \& {Van Speybroeck}, L. 2006, \apj, 640, 691,
  astro-ph/0507092

\bibitem[{{Voit}(2005)}]{2005RvMP...77..207V}
{Voit}, G.~M. 2005, Reviews of Modern Physics, 77, 207, astro-ph/0410173

\bibitem[{{Weinberg} {et~al.}(2013){Weinberg}, {Mortonson}, {Eisenstein},
  {Hirata}, {Riess}, \& {Rozo}}]{2013PhR...530...87W}
{Weinberg}, D.~H., {Mortonson}, M.~J., {Eisenstein}, D.~J., {Hirata}, C.,
  {Riess}, A.~G., \& {Rozo}, E. 2013, \physrep, 530, 87, 1201.2434

\bibitem[{{Witzel} {et~al.}(1979){Witzel}, {Schmidt}, {Pauliny-Toth}, \&
  {Nauber}}]{1979AJ.....84..942W}
{Witzel}, A., {Schmidt}, J., {Pauliny-Toth}, I.~I.~K., \& {Nauber}, U. 1979,
  \aj, 84, 942

\bibitem[{{Zhang} {et~al.}(2014){Zhang}, {Yu}, \& {Lu}}]{2014ApJ...796..138Z}
{Zhang}, C., {Yu}, Q., \& {Lu}, Y. 2014, \apj, 796, 138, 1406.4019

\bibitem[{{Zhang} {et~al.}(2017){Zhang}, {Reiprich}, {Schneider}, {Clerc},
  {Merloni}, {Schwope}, {Borm}, {Andernach}, {Caretta}, \&
  {Wu}}]{2017A&A...599A.138Z}
{Zhang}, Y.-Y. {et~al.} 2017, \aap, 599, A138

\bibitem[{{ZuHone}(2011)}]{2011ApJ...728...54Z}
{ZuHone}, J.~A. 2011, \apj, 728, 54, 1004.3820

\end{thebibliography}
